\documentclass[smallcondensed]{svjour3}    
\smartqed  
\usepackage{graphicx}
\usepackage{amssymb}
\usepackage{rotating}
\usepackage{color}
\usepackage{ulem}

\begin{document}

\title{Formation of Terrestrial Planets in Disks with Different Surface Density Profiles}

\titlerunning{Terrestrial Planet Formation} 

\author{Nader Haghighipour \\
        Othon C. Winter
}

\authorrunning{Haghighipour \& Winter} 

\institute{N. Haghighipour \at
           Institute for Astronomy,
           University of Hawaii, Honolulu, HI 96822, USA,
           Tel: +1-808-956-6098,
           Fax: +1-808-956-4532,
           \email{nader@ifa.hawaii.edu}\\           
           O. C. Winter \at
           UNESP - Univ. Estadual Paulista, Grupo de Din\'amica Orbital \& Planetologia, 
           Guaratinguet\'a, CEP 12516-410, SP, Brazil, \email{ocwinter@pq.cnpq.br}}

\maketitle

\begin{abstract}

We present the results of an extensive study of the final stage of terrestrial planet formation
in disks with different surface density profiles and for different orbital configurations of
Jupiter and Saturn. We carried out simulations in the context of the classical model with disk
surface densities proportional to ${r^{-0.5}}, {r^{-1}}$ and ${r^{-1.5}}$, and also using partially
depleted, non-uniform disks as in the recent model of Mars formation by Izidoro et al (2014). The
purpose of our study is to determine how the final assembly of planets and their physical properties 
are affected by the total mass of the disk and its radial profile. Because as a result of the 
interactions of giant planets with the protoplanetary disk, secular resonances will also play important 
roles in the orbital assembly and properties of the final terrestrial planets, we will study the effect
of these resonances as well. In that respect, we divide this study into two parts.
When using a partially depleted disk (Part 1), we are particularly interested in 
examining the effect of secular resonances on the formation of Mars and orbital stability of terrestrial 
planets. When using the disk in the classical model (Part 2), our goal is to determine trends that
may exist between the disk surface density profile and the final properties of terrestrial planets.
In the context of the depleted disk model,
results of our study show that in general, the $\nu_5$ resonance does not have 
a significant effect on the dynamics of planetesimals and planetary embryos, and the final orbits 
of terrestrial planets. However, $\nu_6$ and $\nu_{16}$ resonances play important roles in
clearing their affecting areas. While these resonances do not alter the orbits of Mars and other 
terrestrial planets, they strongly deplete the region of the asteroid belt ensuring that no additional 
mass will be scattered into the accretion zone of Mars so that it can maintain its mass and orbital stability.
In the context of the classical model, the effects of these resonances 
are stronger in disks with less steep surface density profiles. Our results indicate that 
when considering the classical model (Part 2), the final planetary systems
do not seem to show a trend between the disk surface density profile and the mean number of the final 
planets, their masses, time of formation, and distances to the central star. Some small
correlations were observed where, for instance, in disks with steeper surface density profiles, the final planets
were drier, or their water contents decreased when Saturn was added to the simulations. However, in general,
the final orbital and physical properties of terrestrial planets seem to vary from one system to another 
and depend on the mass of the disk, the spatial distribution of protoplanetary bodies (i.e., disk surface density 
profile), and the initial orbital configuration of giant planets. We present results of our simulations
and discuss their implications for the formation of Mars and other terrestrial planets,
as well as the physical properties of these objects such as their masses and water contents.


\end{abstract}

\section{Introduction}
\label{intro}

Recent efforts in explaining the origin of Mars (its small mass and short time of formation compared to Earth) 
have profoundly altered our views of the formation and evolution of terrestrial planets in our solar system. 
Motivated by the work of Hansen (2009), the two current models
that have successfully accounted for the formation of this planet (Izidoro et al 2014, Walsh et a. 2011)
suggest that Mars formed in a protoplanetary disk where, unlike the traditional models of terrestrial planet formation,
the distribution of solid material did not follow a uniform functional form of the radial distance, but instead 
it contained non-uniformities [See Chambers (2014) for a detailed review of these two models]. 
Simulations by these authors indicate that
the accretion of Mars completed in a {\it mass-rich} part of the disk and soon after its accretion, 
Mars was scattered into a region of the disk with low mass and low surface density. In this region, 
the feeding zone of Mars was devoid of solid material which prevented Mars from growing larger.
The local planetary embryos in this region were also much smaller than Mars
and their perturbation was not strong enough to alter Mars orbital dynamics. As a result, Mars maintained 
its mass and orbit for the duration of the evolution of the solar system.

Although the models by Izidoro et al (2014) and Walsh et al (2011) are based on the same
underlying physics (i.e. Mars formed in a mass-rich region of a disk with a non-uniform surface density
profile, and was scattered into a local mass-depleted region), 
the non-uniformities in the distribution of protoplanetary material in these two models 
have different origins. Walsh et al (2011) considered the initial distribution of the solid material in the
protoplanetary disk (i.e. planetesimals and planetary embryos) to follow the classical model in which
objects are distributed according to a specific disk surface density profile and are placed at certain
distances from one another given by a number of their mutual Hill's radii. Because in this model, the disk
contains no non-uniformity, these authors considered an inward and then outward migration for Jupiter 
and Saturn to induce variations in the local distribution of the disk material. 
Izidoro et al (2014), on the other hand, assumed that at the onset of the Runaway growth and prior to the
end of the Oligarchic growth when the disk will self-adjust, the protoplanetary disks are naturally inhomogeneous
and they inherit their non-uniformities during their evolution, from the initial non-uniformities existed 
in the disk of planetesimals. The implications of these assumptions is that Walsh et al presents Mars as the
evidence to the migration of giant planets, whereas in the model by Izidoro et al, Mars is a natural
product of the evolution of the protoplanetary disk. However, despite these differences,
an important common feature of these two models is that their success in forming Mars (and other terrestrial planets) 
strongly depends on the spatial distribution of the disk material, and the interactions 
of giant planets (specifically, Jupiter and Saturn) with planetesimals and protoplanetary bodies.
 
The strong effects of Jupiter and Saturn on the formation of terrestrial planets has been known for over three decades.
As shown by many authors, the final assembly of terrestrial planets, their masses, and their water contents are 
strongly correlated with the dynamical properties and orbital architecture of these two giant planets
(Wetherill 1990a\&b, 1994, 1996, 1998; Agnor et al 1999; Chambers and Wetherill 1998; 
Chambers 2001; Chambers and Cassen 2002; Levison and Agnor 2003; Raymond et al 2004, 2005a\&b, 2006, 2007, 2009; 
Agnor and Lin 2012; Izidoro et al 2013; Lykawka and Ito 2013; Quintana and Lissauer 2014; Kaib and Cowan 2015;
Kaib and Chambers 2015). For instance, within the context of the classical model,
Raymond et al (2005b) explored the correlation between the final planetary systems and the slope of the disk surface density profile
in disks with equal masses and showed that in systems with one Jupiter-mass giant planet in a circular orbit at 5.5 AU,
the number of the final terrestrial planets will be larger, they will be more massive, form more quickly, have lower water contents,
and will be closer to the central star when the disk has a steeper surface density profile. 
Recently, Quintana and Lissauer (2014) also studied the formation of terrestrial planets and the water contents
of the final planets in the inner part of the solar system, and showed that although giant planets can have
profound effects on the formation of these bodies, they are not necessary for radial mixing of disk material
and the delivery of water and volatiles to the accretion zone of Earth.
In a recent study, Izidoro et al (2015) also considered the effect of Jupiter and Saturn
on the formation of terrestrial planets in disks with very steep radial profiles (e.g., ${r^{-2.5}}$  to ${r^{-5.5}}$), 
and showed that although the accumulation of solid material in the inner part of such disks may provide a 
favorable condition for the formation of Mars-analogs, they cannot reproduce the depletion and dynamical state 
of the asteroid belt. These authors suggested that the depletion and dynamical excitation of the asteroid belt 
must have happened because of an external mechanism such as the migration of Jupiter and Saturn.

The mean-motion (Nesvorny and Morbidelli 1998) and secular resonances (Milani and Kne\u znevi\'c 1990; Minton and Malhotra 2009, 2011; 
Brasser et al 2013; Lykawka and Ito 2013; {Kaib and Chambers 2015) associated with giant planets also have strong effects on the
formation and evolution of terrestrial planets. These resonances cause the orbital eccentricities of 
small bodies interior to their orbits to reach high values which in turn results in their ejection from the system or their collisions with the 
Sun or other bodies. As such, giant planets have played a major role in sculpting the
asteroid belt, depleting its mass, constraining the orbital eccentricities and inclinations 
of asteroids, and the accretion of planetesimals and planetary embryos in the late stage
of terrestrial planet formation.

Among the secular resonances $\nu_5$, $\nu_6$, and $\nu_{16}$ have been shown to have significant effects
on the dynamics of the protoplanetary disk. In the classical model of the formation of terrestrial
planets, as first noted by Chambers and Wetherill (1998) and subsequently shown by Levison and 
Agnor (2003), Raymond et al. (2009), Minton and Malhotra (2009, 2011), Agnor and Lin (2012), Brasser et al (2013),
and Kaib and Chambers (2015), these resonances significantly alter the orbits of planetesimals, 
planetary embryos, and the final terrestrial planets, causing many of these objects 
to be ejected from the system. 
Agnor and Lin (2012) studied these resonances during the migration of Jupiter and Saturn within the
context of the Nice model (Gomes et al. 2005) and showed that they impose stringent conditions on the 
timescales of terrestrial planet formation as well as the growth and migration of giant planets.
These authors suggested that in order for terrestrial planet formation to proceed constructively,
these planets have to form after the migration of giant planets is completed.
Terrestrial planet formation during planetesimal-driven migration of Jupiter and Saturn
was also studied by Lykawka and Ito (2013). These authors showed that if the orbits of Jupiter and Saturn are
considered to be more eccentric than their current orbits, their secular resonances strongly deplete the 
region between 1.5 AU and 2 AU. However, as shown by these authors, this depletion did not prevent their simulations
from forming a massive Mars, a problem that existed with the classical model as well.
More recently Malhotra (2014) and Kaib and Chambers (2015) also studied the effect of
the secular resonances of the giant planets on the final orbital assembly of terrestrial planets.
Malhotra (2014) showed that the $\nu_5$ secular resonance during the migration
of giant planets, as discussed in Minton and Malhotra (2009, 2011), excites the orbital eccentricities of
terrestrial planets causing many orbital crossings among these objects. Results of the simulations
by this author point to a low probability for constructive formation and subsequent stability
of terrestrial planets during giant planets migration, even when this migration is rapid as in
the jumping Jupiter scenario (Morbidelli et al 2009, 2010). Kaib and Chambers (2015) have found
that in order for the terrestrial planets to settle in their current orbits and maintain their orbital architecture
for the age of the solar system, giant planets' instability (e.g., as in the Nice model) must have occurred prior
to the formation of terrestrial planets suggesting that such instability is not the source of the Late Heavy Bombardment.

The significance of the results of the above-mentioned studies motivated us to examine the effects 
of different disk surface density profiles and secular resonances both in our model of Mars and terrestrial planet formation 
(Izidoro et al. 2014) and in the classical model. 
Given that similar to the Grand Tack scenario (Walsh et al 2011), in our model, Mars forms as a planetary embryo and 
is scattered into its current orbit through interaction with other protoplanetary bodies, it would 
be important to determine how this scenario would be affected by the secular resonances of giant planets and the 
initial distribution of planetesimals and planetary embryos, and also what conditions they impose on the 
ranges of the initial semimajor axes and eccentricities
of Jupiter and Saturn in order for the formation of Mars and other terrestrial planets to proceed efficiently. 
In the context of the classical model, the goal of our study is to expand on the work of
Raymond et al (2005b) and explore the possibility of trends between the final orbital architecture and properties of 
terrestrial planets, and the distribution of solid material in the disk,
by considering two giant planets, varying their initial orbital elements, and also considering disks with different
initial masses and radial profiles.

We divide our paper into two parts. In Part 1,
we begin by examining the effects of secular resonances. As mentioned earlier, in both models of Mars 
formation by Izidoro et al (2014) and Walsh et al (2011), Mars forms as a stranded planetary embryo. Because
as suggested by cosmochemical data, the time of the formation of Mars is within the first 
10 Myr after the formation of the first solids (Nimmo and Kleine 2007, Dauphas and Pourmand 2011), 
we focus our study of the effects of secular resonances on 
the first few million years of the last stage of terrestrial planet formation.
As we are interested in comparing our results with those in previous
studies, we begin by studying $\nu_5$, $\nu_6$ and $\nu_{16}$ resonances in the classical model, and continue 
to systems with non-uniform protoplanetary disks.

During the formation of terrestrial planets, the dynamics of 
protoplanetary bodies, in addition to the perturbation of giant planets, is also affected by 
their own mutual interactions. The latter is the subject of Part 2.
To study the effects of these interactions, 
we assume different spatial distributions for these objects and consider
disks with surface densities proportional to $r^{-0.5}$, $r^{-1}$ and $r^{-1.5}$. 
 We continue our study by
focusing on the correlation between the initial distribution of material in protoplanetary disks and
the number, masses, water contents and orbital elements of the final terrestrial planets.

The outline of this paper is as follows. We begin in section 2 by studying the effects of secular resonances in the 
classical model of terrestrial planet formation, and for disks with different surface density profiles. These results
will be used as the basis for comparison when in section 3, we carry out similar analysis for disks with 
non-uniform distributions of protoplanetary material. Section 4 presents Part 2 of our study and has to do with the 
analysis of the final planetary systems for different protoplanetary disks in the classical model. Section 5 
concludes this study by presenting a summary, and discussing the implications of our analysis for the formation of
terrestrial planets.

\vskip 35pt
\noindent
{\bf PART 1: SECULAR RESONANCES}
\vskip 10pt
\noindent
In this part of the paper, we present the results of our study of the effects of 
secular resonances on the formation of terrestrial planets in disks with different surface
density profiles. The goal of our study is to explore how these resonances affect the final mass and
orbital architecture of terrestrial planets, in particular Mars, in our model of Mars formation in a
non-uniform and partially depleted disk (Izidoro et al 2014). For the sake of comparison, we begin 
by studying secular resonances in the classical model. Because during the evolution of the system, the locations
of secular resonances change with the orbits of giant planets and the mass of the disk, we
present a complete study of these effects in section 2, and use the results as a basis for comparing with similar
analysis in partially depleted disks in section 3.

\section{Secular Resonances $\nu_5$, $\nu_6$, and $\nu_{16}$ in the Classical Model}

{\bf \subsection{The Protoplanetary Disk and Initial Set Up}}

We consider a system consisting of the Sun, a bi-modal disk of planetesimals and planetary embryos extending 
from 0.5 AU to 4 AU, and two giant planets with masses equal to those of Jupiter and Saturn. 
The bi-modality of the disk is to ensure that our models would be consistent 
with the outcomes of Runaway and Oligarchic growths where, as shown by Kokubo and Ida (1998, 2000), simulations 
result in a bi-modal distribution of mass in a protoplanetary disk. Also, the submergence of planetary embryos 
in a swarm of planetesimals subjects these objects to dynamical friction which as shown by O'Brien et al. (2006) 
and Morishima et al. (2008) prevents the eccentricities and inclinations of planetary embryos to reach high values.

We distribute planetesimals and planetary embryos according to a surface density proportional to 
8 (g/cm$^2$) $r^{-\alpha}$ where $\alpha=0.5, 1$, and 1.5. The distances between planetary embryos are 
randomly chosen to vary between 5 and 10 mutual Hill radii. The masses of these objects increase with their semimajor axes 
$(a)$ and the number of their mutual Hill radii $(\Delta)$ as ${a^{3/4}}\,{\Delta^{3/2}}$ (Kokubo and Ida 2000; 
Raymond et al 2005, 2009; Izidoro et al. 2013, 2014). In the disk with $\alpha = 0.5$, we choose protoplanetary masses 
from the range of 0.005-0.5 Earth-masses. When $\alpha=1$, this range is 0.009-0.2 Earth-masses, and for $\alpha=1.5$, it 
is chosen to be 0.015-0.07 Earth-masses. The total mass of the disk in these three models will then be equal to $\sim 9.6$, 
6.6, and 4.9 Earth-masses, respectively.

For each value of $\alpha$, we divide the mass of the disk almost equally between the
total mass of the planetesimals and the total mass of planetary embryos. Due to computational limitations, we 
follow Raymond et al (2009) and consider the embryo-to-planetesimal mass-ratio to be $\sim 8$ at 1.5 AU.
With this mass-ratio, the number of embryos and planetesimals in disk model $\alpha=0.5$ is 70 and 1015, in the 
$\alpha=1$ disk model, it is 66 and 1008, and in the disk with $\alpha=1.5$, it is 73 and 1000, respectively. 
We note that as stated above, for each value of $\alpha$, the masses of embryos scale with their
orbital distances. This causes in disks with small values of  $\alpha$ (e.g., 0.5), where the disk is quite massive, 
very large  embryos to exist in the outer parts.

We also follow Izidoro et al (2013, 2014) and consider all planetesimals and planetary embryos to be initially 
in circular orbits. The orbital inclinations of these objects are randomly chosen from the range of $10^{-4}$ to 
$10^{-3}$ deg., and their mean-anomalies are taken randomly from the range of 0 to 360$^\circ$. The initial values 
of the arguments of periastrons and longitudes of ascending nodes of these bodies are set to zero.

Finally, we consider three different orbital configurations for the giant planets. In the first configuration, 
we consider both planets to be initially in the current orbits of Jupiter and Saturn (classical model). 
In the second configuration, we change the initial values of their orbital eccentricities to 0.1 
(Raymond et al. 2009), and in the third configuration, we consider
the giant planets to be initially in circular orbits and choose their orbital elements to be similar 
to those in the Nice model (${a_{\rm Jup}}=5.45$ AU,
${a_{\rm Sat}}=8.18$ AU) (Tsiganis et al. 2005; Gomes et al. 2005; Morbidelli et al. 2005).

{\bf \subsection{Numerical Simulations and Results}}

{\bf \subsubsection{Locations of $\nu_5$ and $\nu_6$ Resonances}}

We integrated our systems for different values of $\alpha$ (the radial profile of the disk 
surface density). For each value of $\alpha$,
we ran simulations with and without giant planets. When giant planets were included, we integrated
the system for the three different orbital configurations of these objects. In each of the latter cases, 
we ran simulations for three different initial conditions of planetesimals and planetary embryos. 
In total, 63 simulations were carried out.

We used the hybrid integrator routine in a modified version of the N-body integration package Mercury 
(Chambers 1999; Izidoro et al, in prep.), and carried out simulations for 300 Myr. The time-step of 
integrations were set to 6 days. 
In all our simulations, we set up the integrations such that planetesimals interact with
the giant planets and planetary embryos, but they do not interact with one another. For each set of initial
conditions, we once considered planetesimals to be massless test particles, and once considered
each to have a small mass equal to 0.002 Earth-masses.

To determine the effects of $\nu_5$ and $\nu_6$ resonances, we first
identified the locations of these resonances, analytically. According to the linear secular theory, 
$\nu_5$ and $\nu_6$ appear where the rate of the precession of the longitude of perihelion $(g)$ 
of a small body becomes equal to that of Jupiter and Saturn, respectively. In general, 
in a multiple planet system, $g$ varies with the semimajor axis of the small object as

\begin{equation}
g\,=\,{\sum\limits_{i=1}^N}\, {1\over 4}\,n\,\Big({{M_i}\over {M_\odot}}\Big)\>
{\Big({a\over {a_i}}\Big)^2}\> {b_{3/2}^{(1)}}\,.
\end{equation}

\vskip 5pt
\noindent
In this equation, $n$ is the mean-motion of the object, $a$ is its semimajor axis,
$M_i$ and $a_i$ represent the mass and semimajor axis of the $i$th planet, $M_\odot$ is
the mass of the central star (in this case, the Sun), $b_{3/2}^{(1)}$ is a Laplace coefficient, and $N$ is the number of 
planets in the system. Figure 1 shows the graph of $g$ for a massless particle
in terms of its semimajor axis in the Sun-Jupiter-Saturn system. The initial orbital elements of the two giant 
planets in the top panel are similar to those in their current orbits
and in the bottom panel are equal to those in the Nice model
(note that because locations of secular resonances are functions of the masses and
semimajor axes of giant planets, the top panel represents the case where Jupiter and Saturn
are initially in an $e=0.1$ eccentric orbit, as well). As shown by this figure, $({\nu_5} , {\nu_6})$ are expected to 
appear at (0.62 , 1.82) AU and (0.956 , 3.07) AU from top to bottom, respectively\footnote{Because in the Nice model, 
Jupiter and Saturn are initially in circular orbits, $\nu_6$ does not manifest itself clearly.
In order to make this resonance show, we slightly increased the eccentricities of the two
giant planets to 0.01. This is consistent with the initial orbital eccentricities of these planets in
the Nice model-II (Levison et al 2011).}.

\begin{figure}
\centering{
\includegraphics[width=9cm]{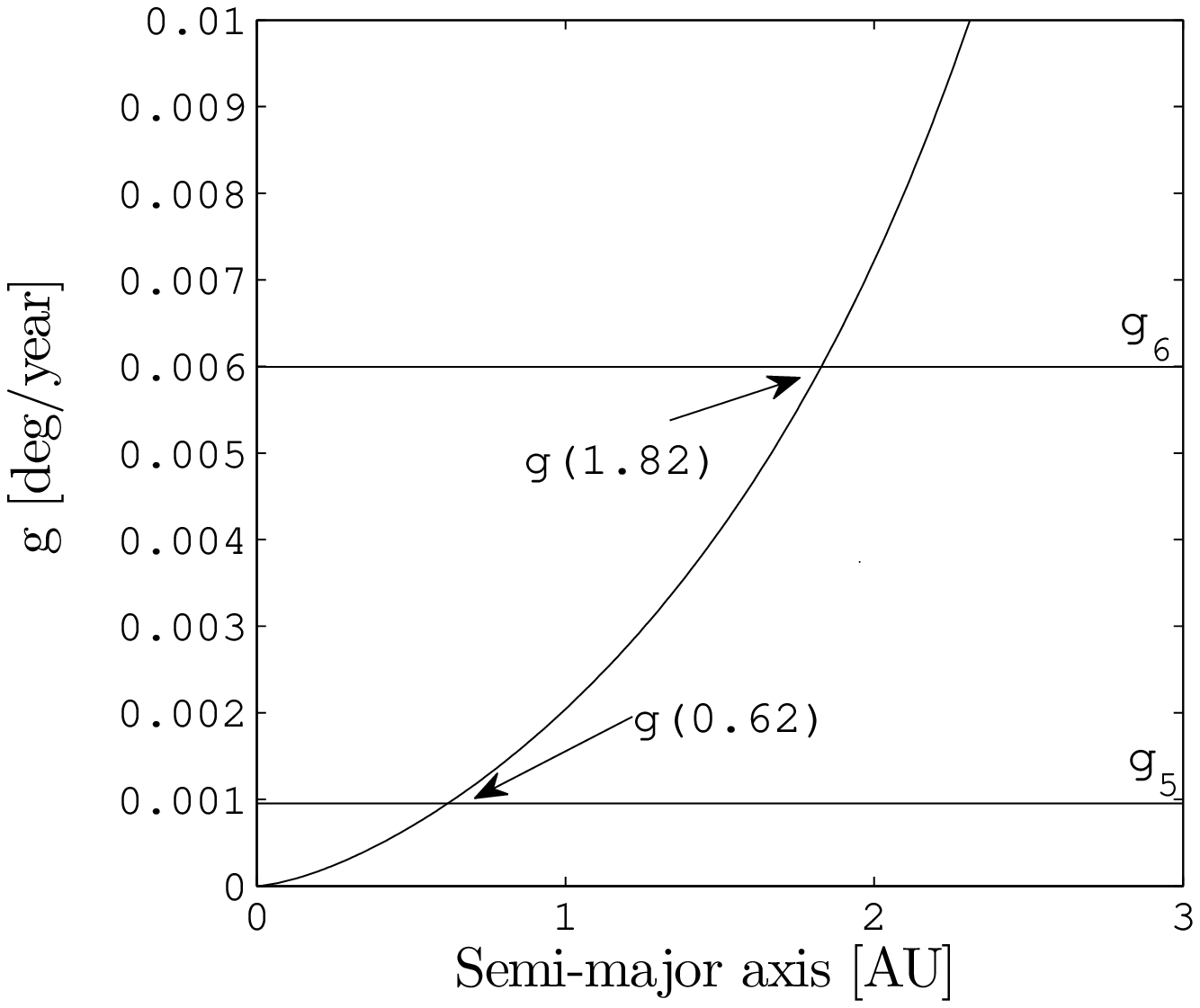}
\vskip 2pt
\includegraphics[width=9cm]{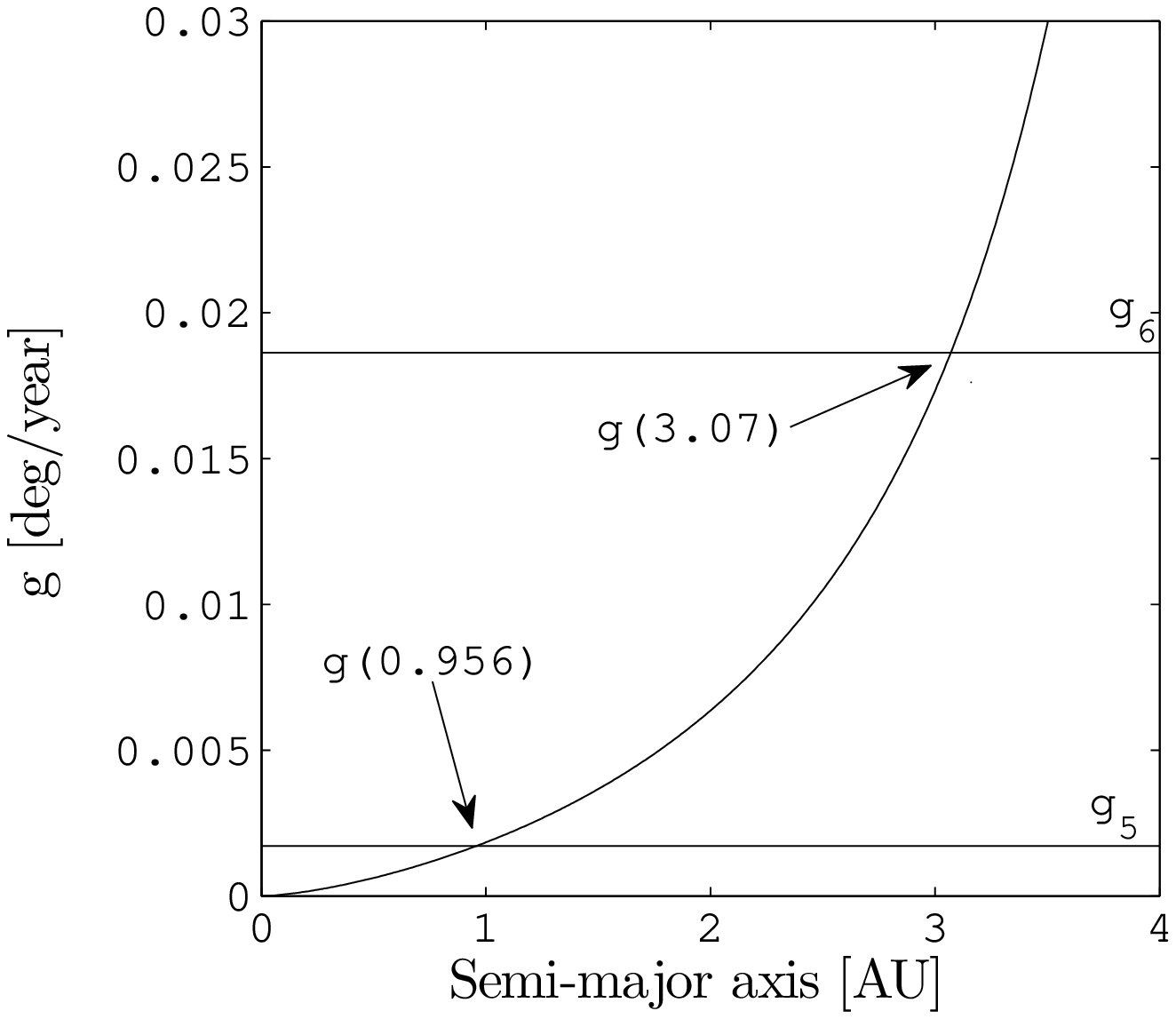}}
\vskip 5pt
\caption{Graphs of the variations of the precession of the longitude of perihelion of 
a test particle (quantity $g$ in equation 1) 
in terms of its semimajor axis in a system with Jupiter and Saturn. The horizontal lines show the 
eccentricity-pericenter eigenfrequencies of the two giant planets. The top panel
corresponds to a system where the initial orbital elements of Jupiter and Saturn are similar
to their current values. The bottom panel corresponds to Jupiter and Saturn initially in the
Nice model. Note that because the value of $g$ depends only on 
the semimajor axes of the giant planets and their masses, the top panel also shows the 
variations of $g$ in a system where the orbits of Jupiter and Saturn are initially eccentric.} 
\end{figure}

\begin{figure}
\centering{
\includegraphics[width=7.2cm]{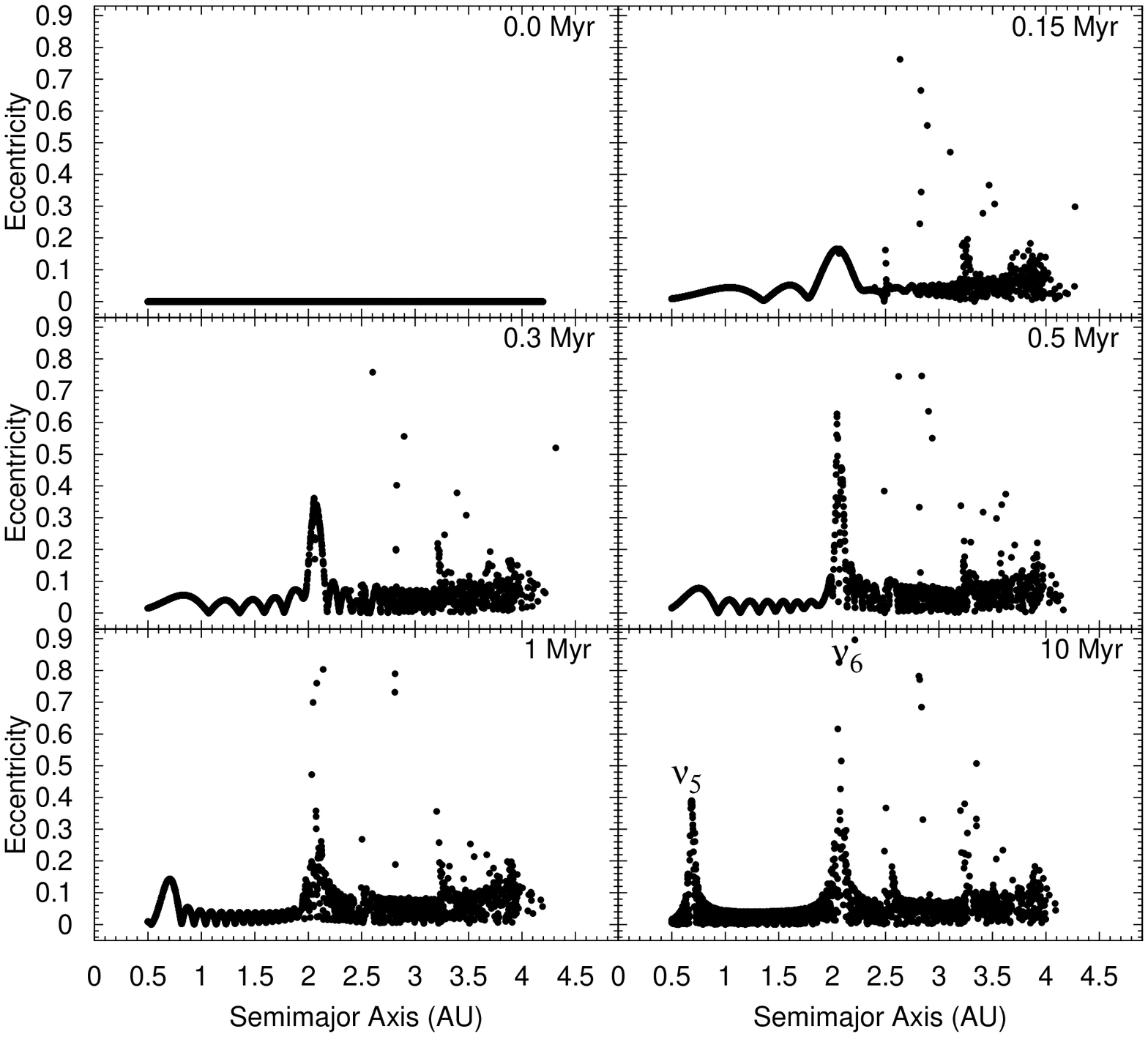}
\vskip -5pt
\includegraphics[width=7.2cm]{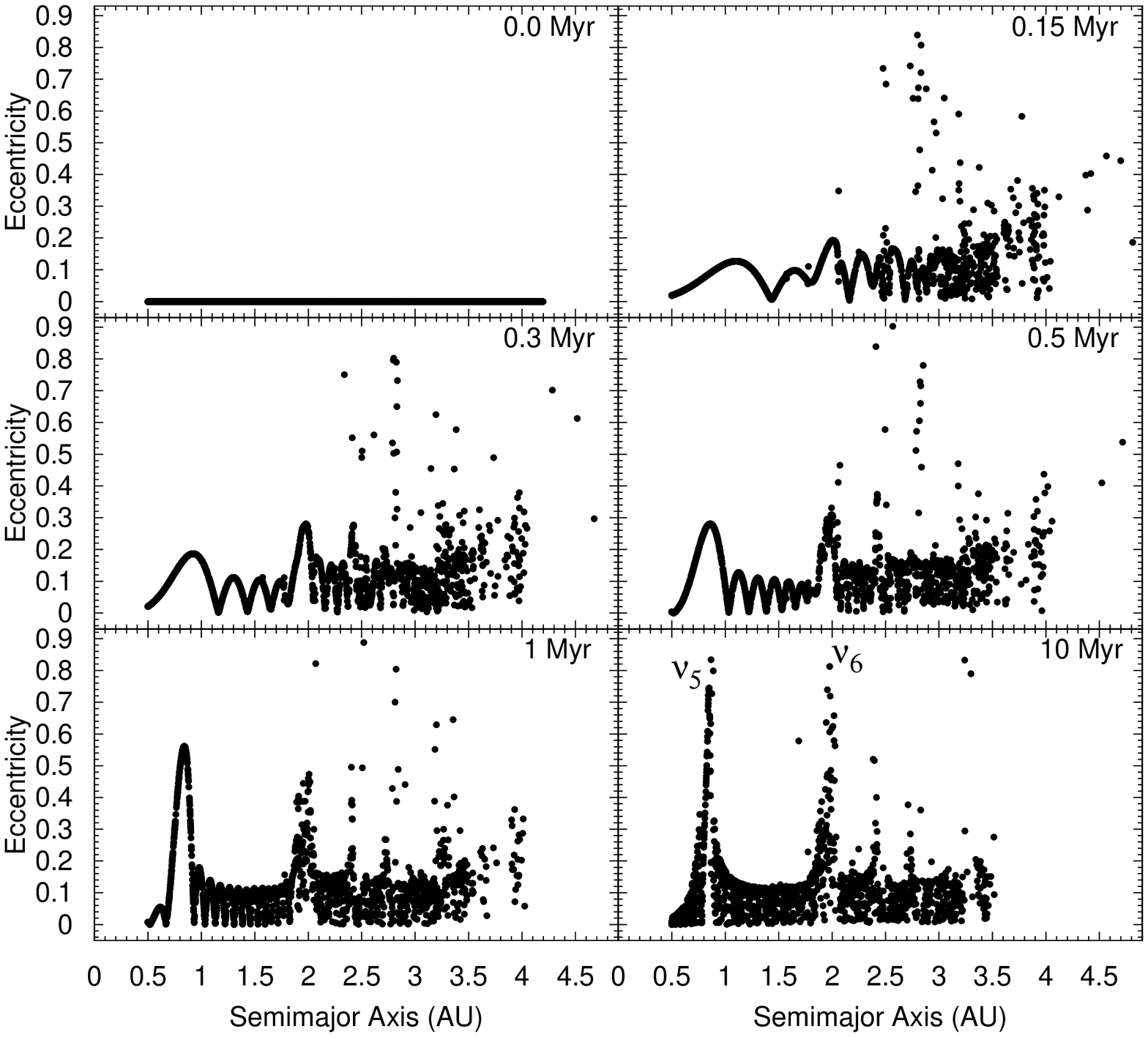}
\vskip -5pt
\includegraphics[width=7.2cm]{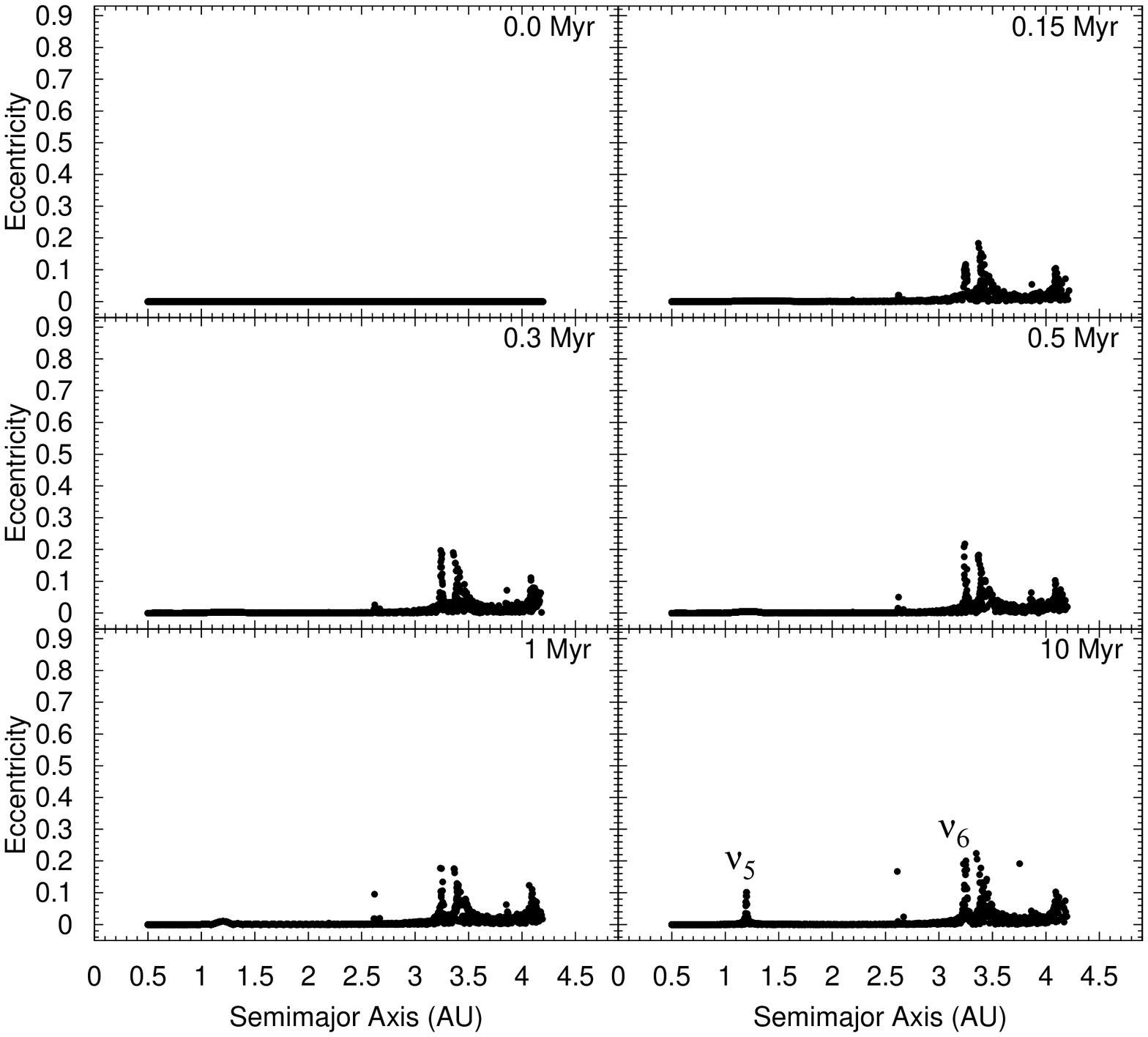}
}
\caption{Snapshots of the evolution of a disk of massless test particles in an $(a,e)$ space 
in a system where Jupiter and Saturn are initially in their current orbits (top), have an initial 
eccentricity of 0.1 (middle), and are initially as in the Nice model (bottom). The locations of $\nu_5$ and
$\nu_6$ resonances have been marked on each panel.} 
\end{figure}

\vskip 30pt
\noindent
{\it {\bf Effect of the Mutual Interactions of Giant Planets}}

\vskip 5pt

It is important to note that in the derivation of equation (1), it has been 
assumed that during the dynamical evolution of the system, the semimajor axes of the giant 
planets will stay unchanged. However, in reality, and in particular during the formation of terrestrial
planets, these objects interact with one another and with the protoplanetary disk interior to the orbit
of Jupiter, and their semimajor axes undergo secular changes. As a result, the locations 
of secular resonances for a test particle will vary during the evolution of the system, and may be slightly 
different from those shown in Figure 1 when the system reaches a stable dynamical state. 
This suggests that a more precise way of identifying the locations of secular resonances would be 
to integrate the orbit of a test particle at different distances interior to the orbits of 
the giant planets. Such simulation have shown that for instance, when the effects of all four 
giant planets in the solar system are 
taken into account, for a massless  particle initially in a circular orbit, $\nu_5$ and $\nu_6$ 
appear at approximately 0.7 AU and 2.1 AU, respectively.

To determine the more precise locations of these resonances in our systems,
we considered a disk consisting of only test particles initially in circular and coplanar orbits, 
and numerically integrated the orbit of 
each object for the same orbital architectures of Jupiter and Saturn as in Figure 1. 
Figure 2 shows the snapshots of the simulations in the $(a,e)$ space for 10 Myr.
As shown here, at the end of the integrations,
$({\nu_5} , {\nu_6})$ appear at (0.7 , 2.0-2.1) AU for a system 
with Jupiter and Saturn initially in their current orbits (top panel), (0.9 , 2.0) AU when the two giant planets 
have an initial eccentricity of 0.1 (middle panel), and (1.25 , 3.25) AU when they are initially in the same configuration 
as in the Nice model (bottom panel)\footnote{In all our numerical integrations throughout this study, we determine 
the locations of secular resonances by monitoring the precession rates of protoplanetary bodies and planetesimals,
and comparing them with those of Jupiter and Saturn. Secular resonances increase the orbital eccentricities and
inclinations of their affected bodies.  We also monitor these quantities and mark the time and semimajor axis
at which eccentricity and inclination are maximally enhanced.} 
. The results shown in the middle panel are in good agreement
with the results of similar analysis by Raymond et al (2009) who also considered an initial eccentricity of 0.1
for Jupiter and Saturn. In their simulations, these authors found
the $\nu_6$ resonance to be at 2.2 AU. The slight difference between our results and those by Raymond et al (2009) 
is due to slight
differences in our initial conditions. In their simulations, Raymond et al (2009) considered Jupiter and 
Saturn to be initially at 5.25 AU and 9.54 AU, respectively. However, in our simulations, we considered 
the initial semimajor axes of these planets to be 5.204 AU and 9.581 AU. For the sake of completeness, we 
performed simulations with  identical initial conditions as those in Raymond et al (2009) and found that
the $\nu_6$ resonance appears at 2.2 AU.

\vskip 20pt

{\bf \subsubsection{The Location of $\nu_{16}$ Resonance}}

We also examined the position of secular resonance $\nu_{16}$ in the $(a,i)$ space. 
This secular resonance appears when the precession of the nodal longitude of a body becomes equal to
the orbital precession of Saturn. Figure 3 shows the snapshots
of the evolution of the inclinations of test particles for the same orbital configurations of Jupiter and Saturn
as in Figure 2. As shown here, after 10 Myr of integrations,
$\nu_{16}$ appears at 1.9 AU in simulations where Jupiter and Saturn are initially in 
their current orbits (top panel), at 2.0 AU when they have an initial eccentricity of 0.1 (middle panel), and at 
3.2 AU when the giant planets are taken to be as in the Nice model (bottom panel).

\begin{figure}
\centering{
\includegraphics[width=7.2cm]{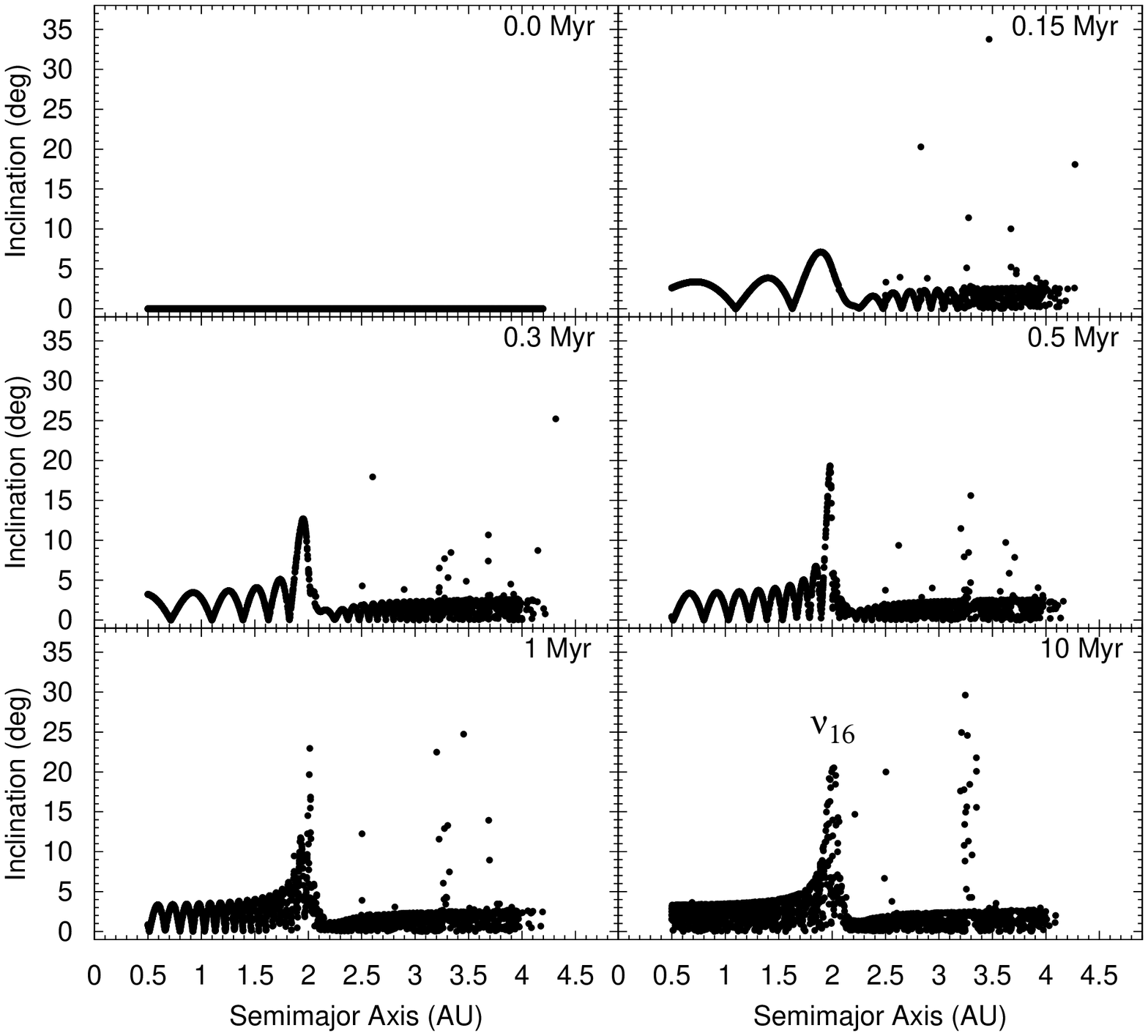}
\vskip -5pt
\includegraphics[width=7.2cm]{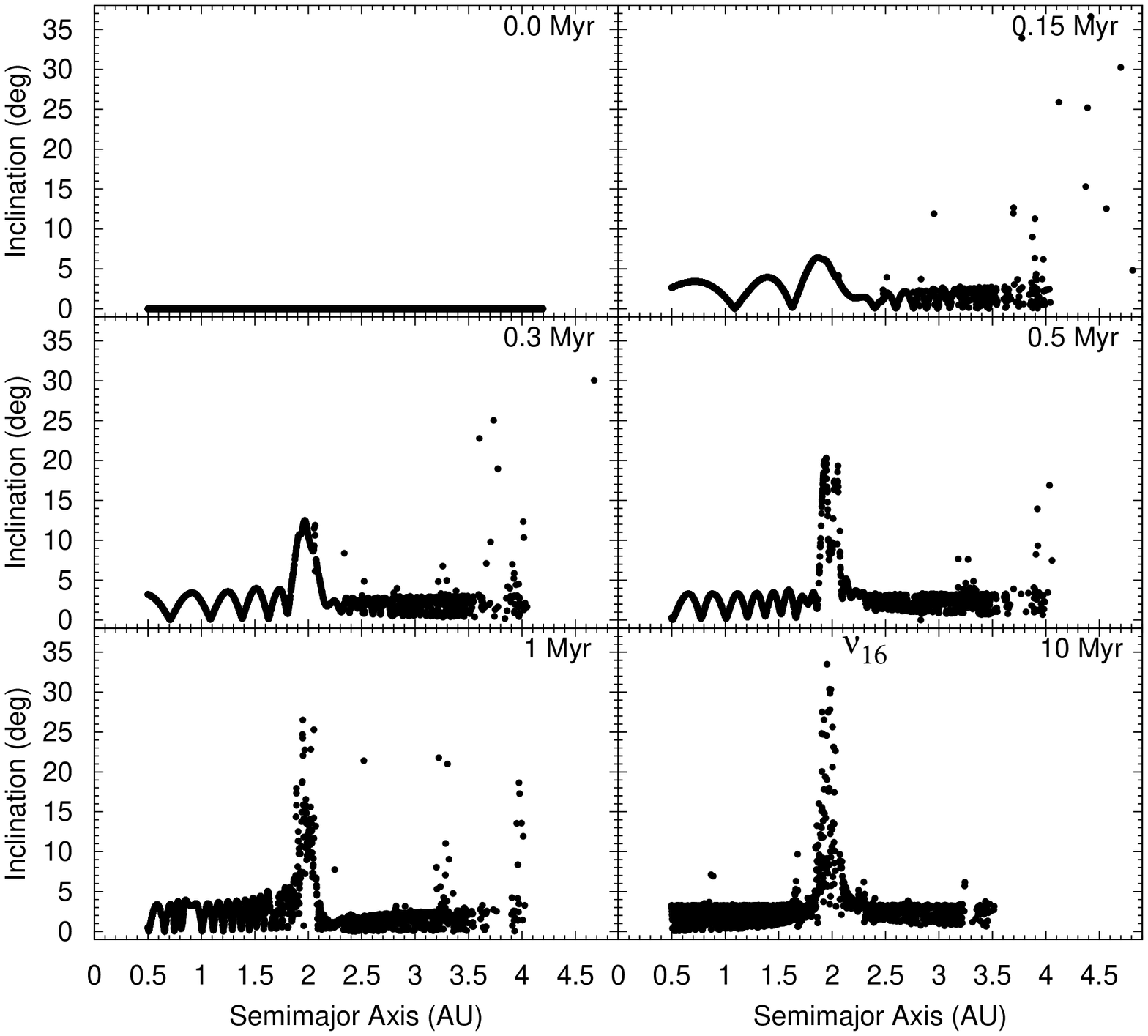}
\vskip -5pt
\includegraphics[width=7.2cm]{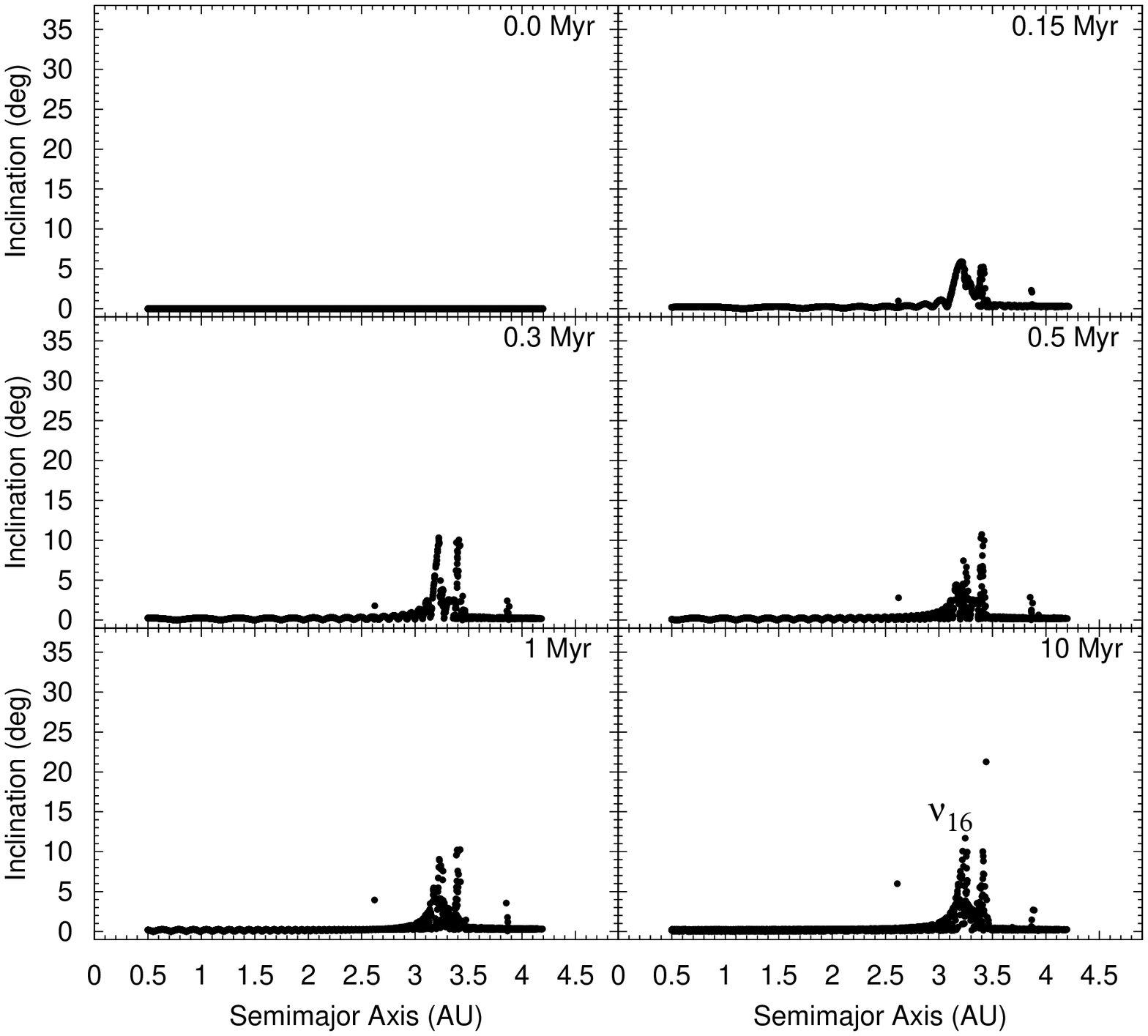}
}
\caption{Snapshots of the evolution of a disk of massless test particles in an $(a,i)$ space
in a system where Jupiter and Saturn are in their current orbits (top), have an eccentricity of 0.1
(middle), and are as in the Nice model (bottom). The location of $\nu_{16}$ resonance has been marked 
on each panel.} 
\end{figure}

An inspection of the results indicates that in simulations where Jupiter and Saturn are considered 
to be initially in their current orbits, the inclinations of objects at the location of the $\nu_{16}$ resonance raise to 
$\sim {20^\circ}$ within the first million years. Outside the resonance, inclinations were no larger than $5^\circ$. 
When the giant planets are considered to have an initial orbital 
eccentricity of 0.1, the inclinations of the objects affected by the $\nu_{16}$ resonance raise to a larger value
$(\sim {35^\circ})$ and the spatial region affect by this resonance becomes larger (middle panel of Figure 3). This
expansion of the affected region of Saturn's secular resonance
occurs for both $\nu_6$ and $\nu_{16}$ resonances and can be attributed to the variations of the semimajor axis
of Saturn due to the perturbation of Jupiter. As shown by equation (1), in a disk of massless test particles, 
positions of secular resonances vary as a function of giant planets' semimajor axes. When during the evolution
of a system, the semimajor axis of a giant planet oscillates around a certain value, the position of its corresponding
secular resonances also vary within a certain region. In the model considered here, the semimajor axis of Saturn
varies within a range around its initial value due to the perturbation of Jupiter. Our simulations show that 
when Jupiter and Saturn were considered to
be initially in their current orbits, these variations were $\sim 0.08$ AU. When the initial eccentricities of these planets
were increased to 0.1, the range of the variation of Saturn's semimajor axis increased to $\sim 0.2$ AU. It is
important to note that as a result of the perturbation of Saturn, the semimajor axis of Jupiter also undergoes 
variations. However, the range of these variations is smaller than 0.01 AU.

In the rest of this study, we will use 
the results of Figures 2 and 3 as the basis to which we will compare the results of our simulations with 
massive disks.

\vskip 20pt
{\bf \subsubsection{Effect of Disk-Mass: Outward Displacement of Secular Resonances}}

Unlike the disks in the simulations of Figures 2 and 3, protoplanetary disks are massive.
As shown by Ward (1981), in a massive disk, the interaction between an object and the disk's
material results in an apsidal 
regression in the orbit of the body which counteracts the effect of an outer giant planet and 
reduces the rate of the object's orbital precession. Since from equation (1), $g$ is an 
increasing function of the object's semimajor axis, a decrease in the rate of the
orbital precession of the object implies that in order for $g$ to approach the rate of the 
orbital precession of a giant planet, the location of an apsidal secular resonance has
to shift to a larger semimajor axis. Such shifts in the locations of $\nu_5$ and $\nu_6$ resonances
have been reported by chambers and Wetherill (1998), Nagasawa et al (2000, 2002),
and can also be found in the simulations of the dynamical evolutions of test particle interior to the orbit
of Jupiter by Levison and Agnor (2003), and Nagasawa et al (2005). 

Similar outward motion was observed for the $\nu_{16}$ resonance as well. Morbidelli and Henrard (1991a \& b) 
examined the position of this resonance as a function of the orbital inclination, and showed that
it moves outward to a region between 2.1 AU and 2.9 AU for particle inclinations varying between
0 and $20^\circ$.

\begin{figure}
\centering{
\includegraphics[scale=0.34]{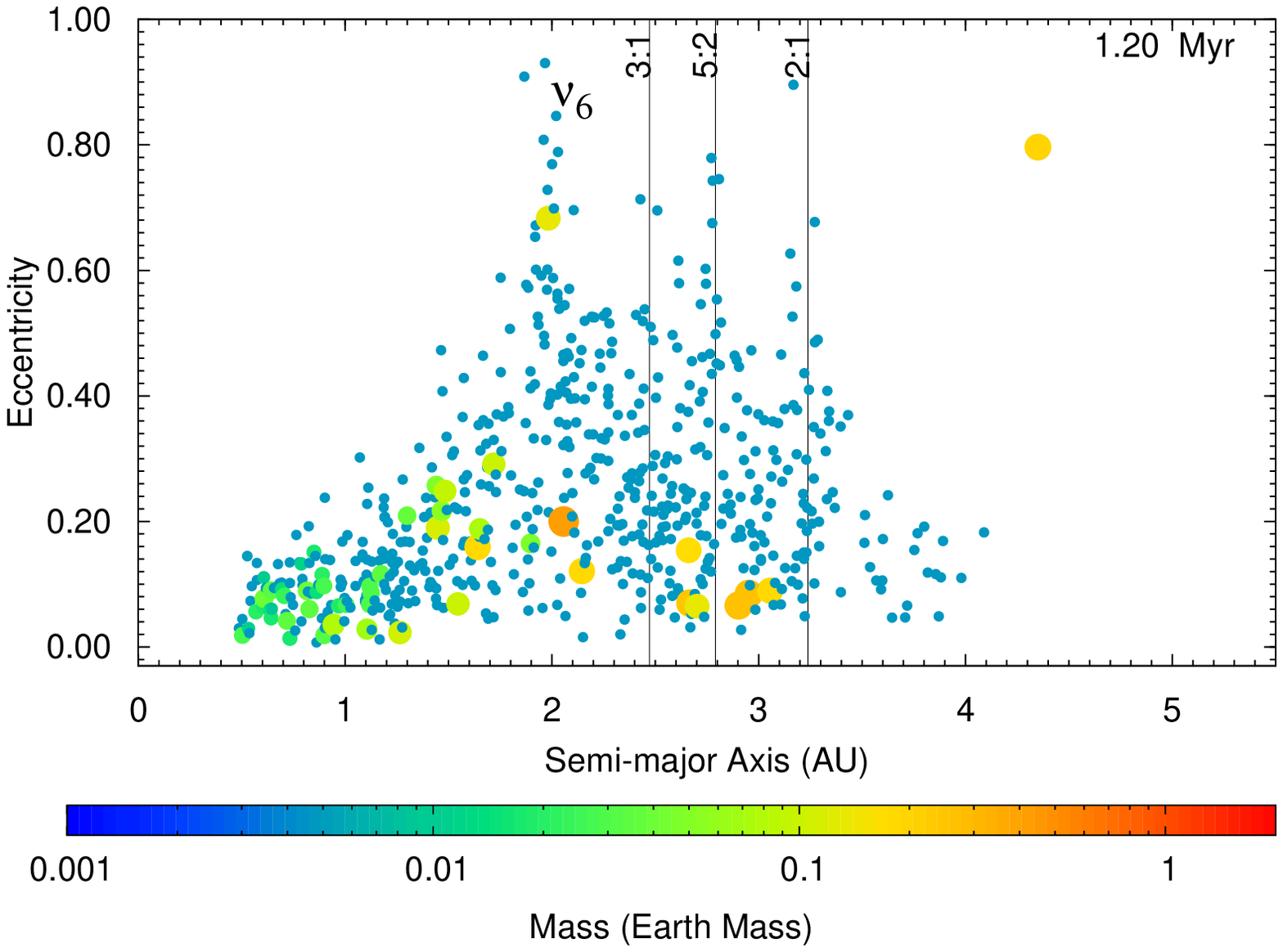}
\includegraphics[scale=0.34]{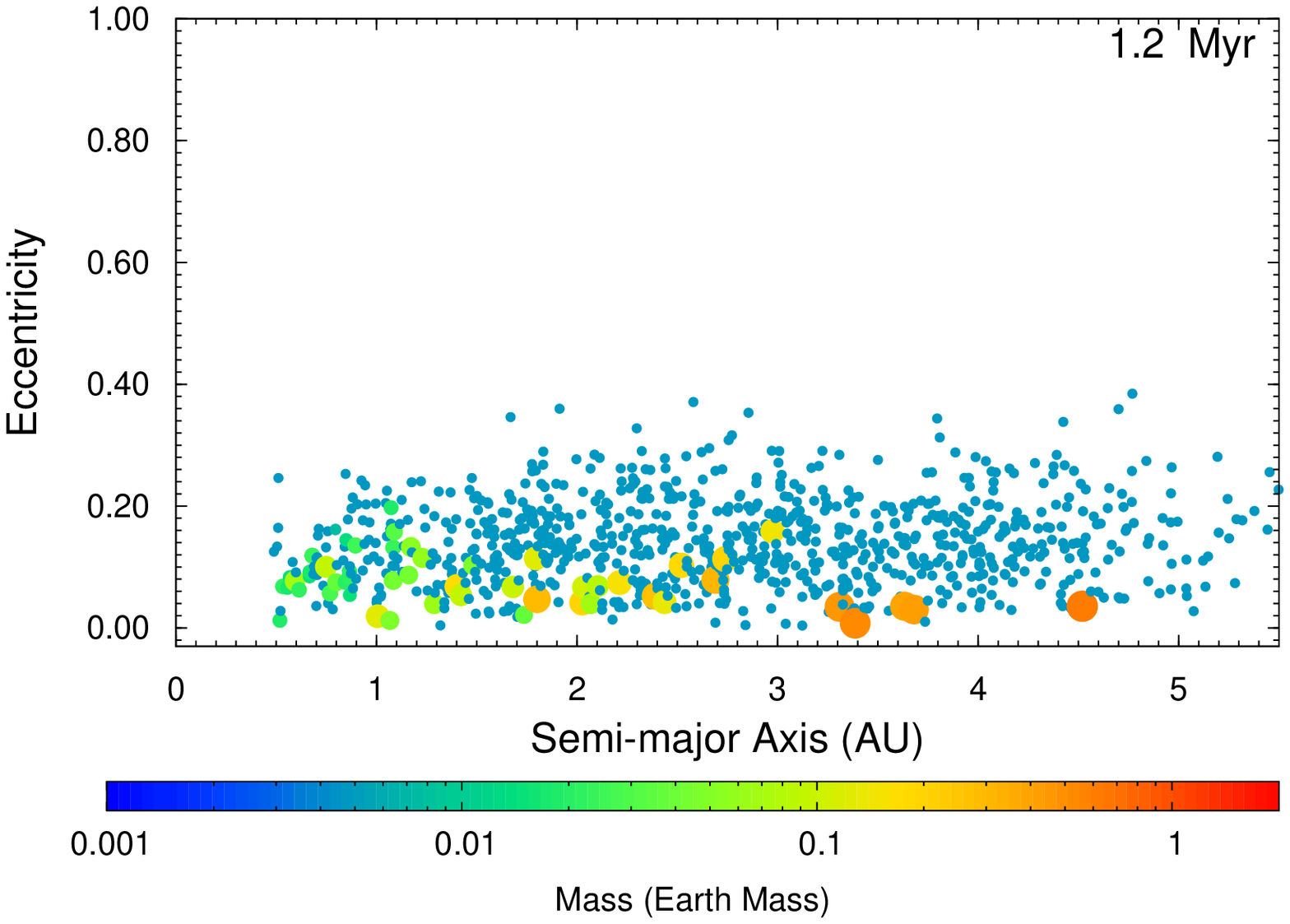}
\vskip 5pt
\includegraphics[scale=0.34]{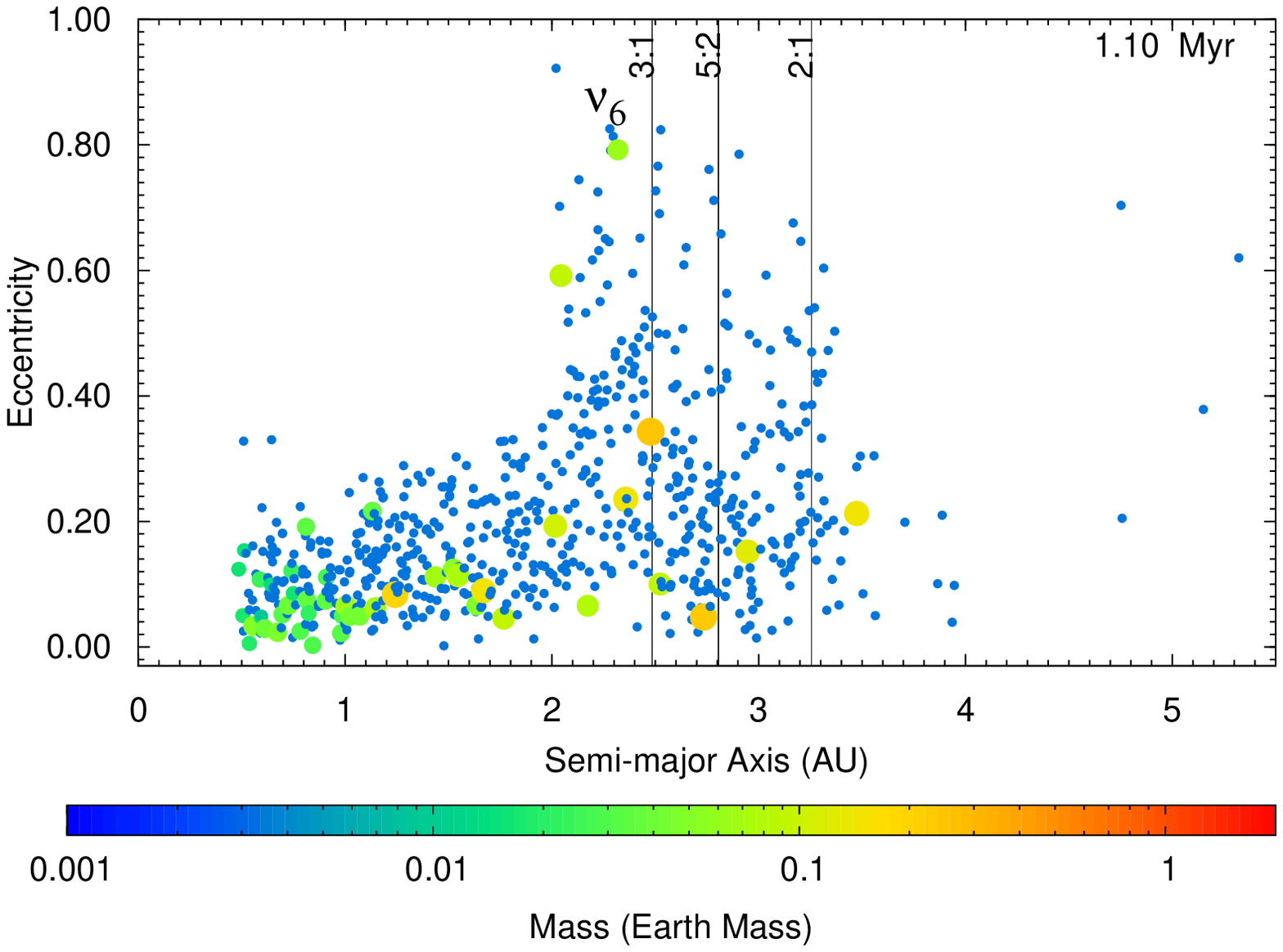}
\includegraphics[scale=0.34]{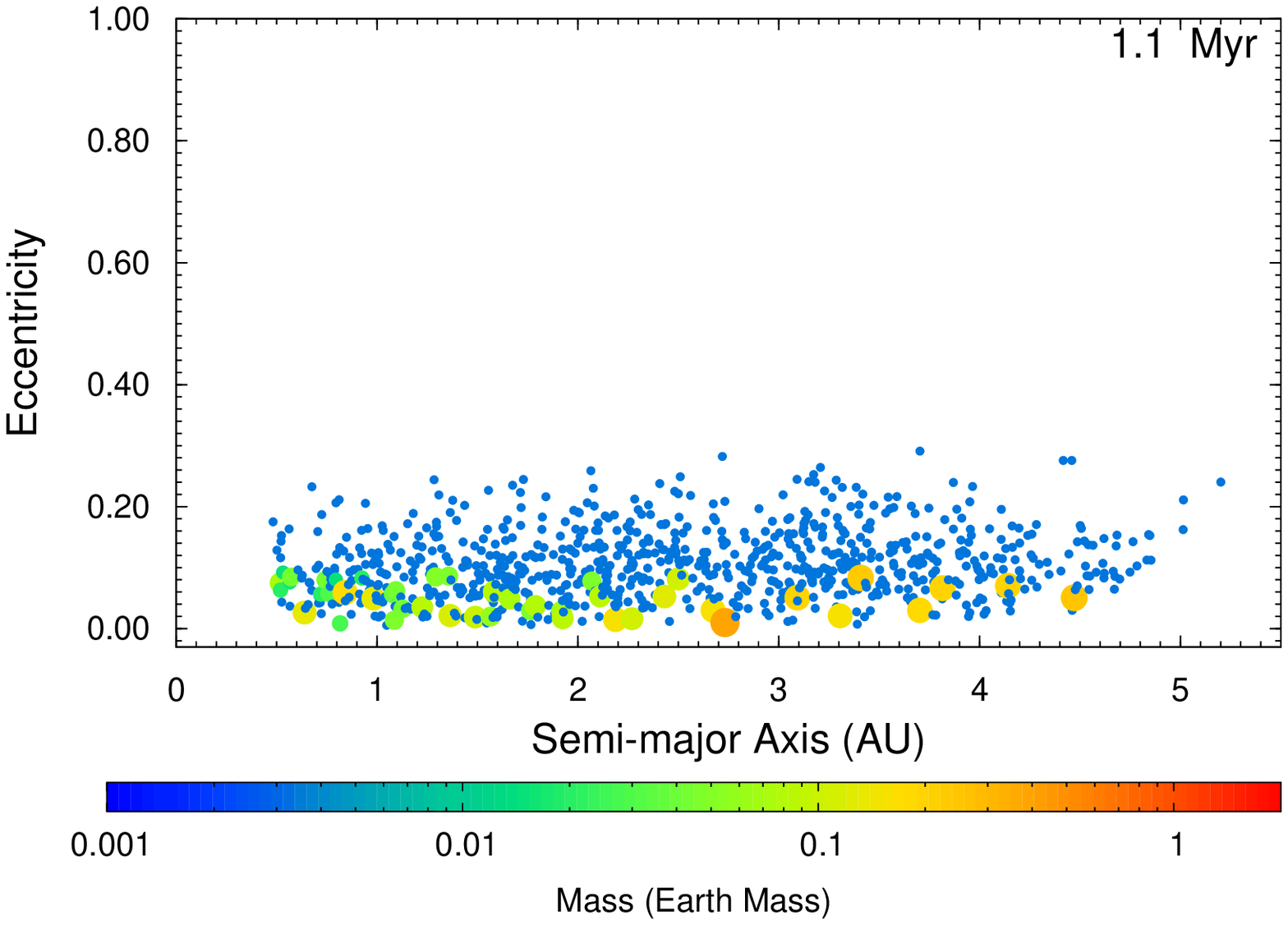}
\vskip 5pt
\includegraphics[scale=0.34]{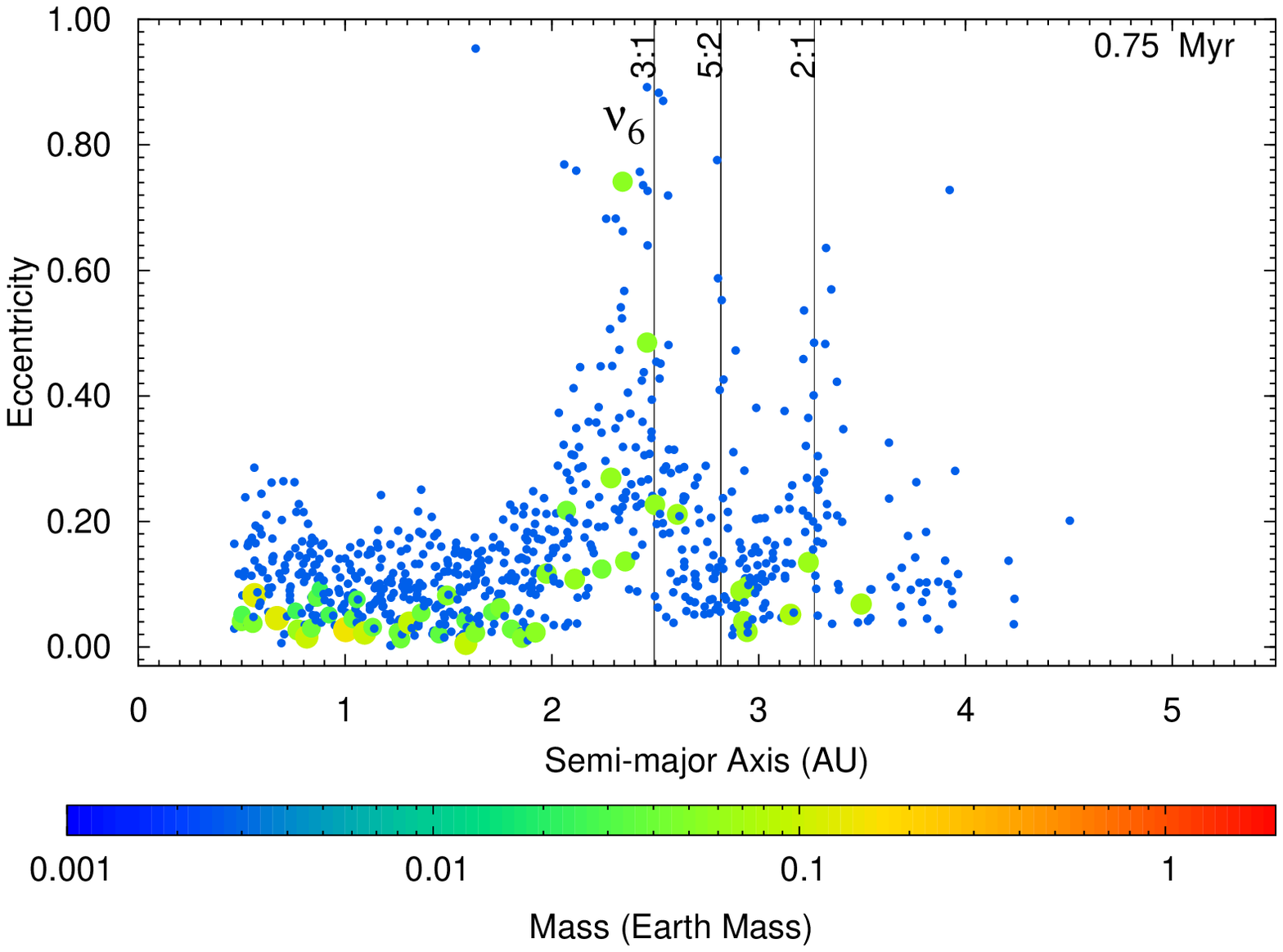}
\includegraphics[scale=0.34]{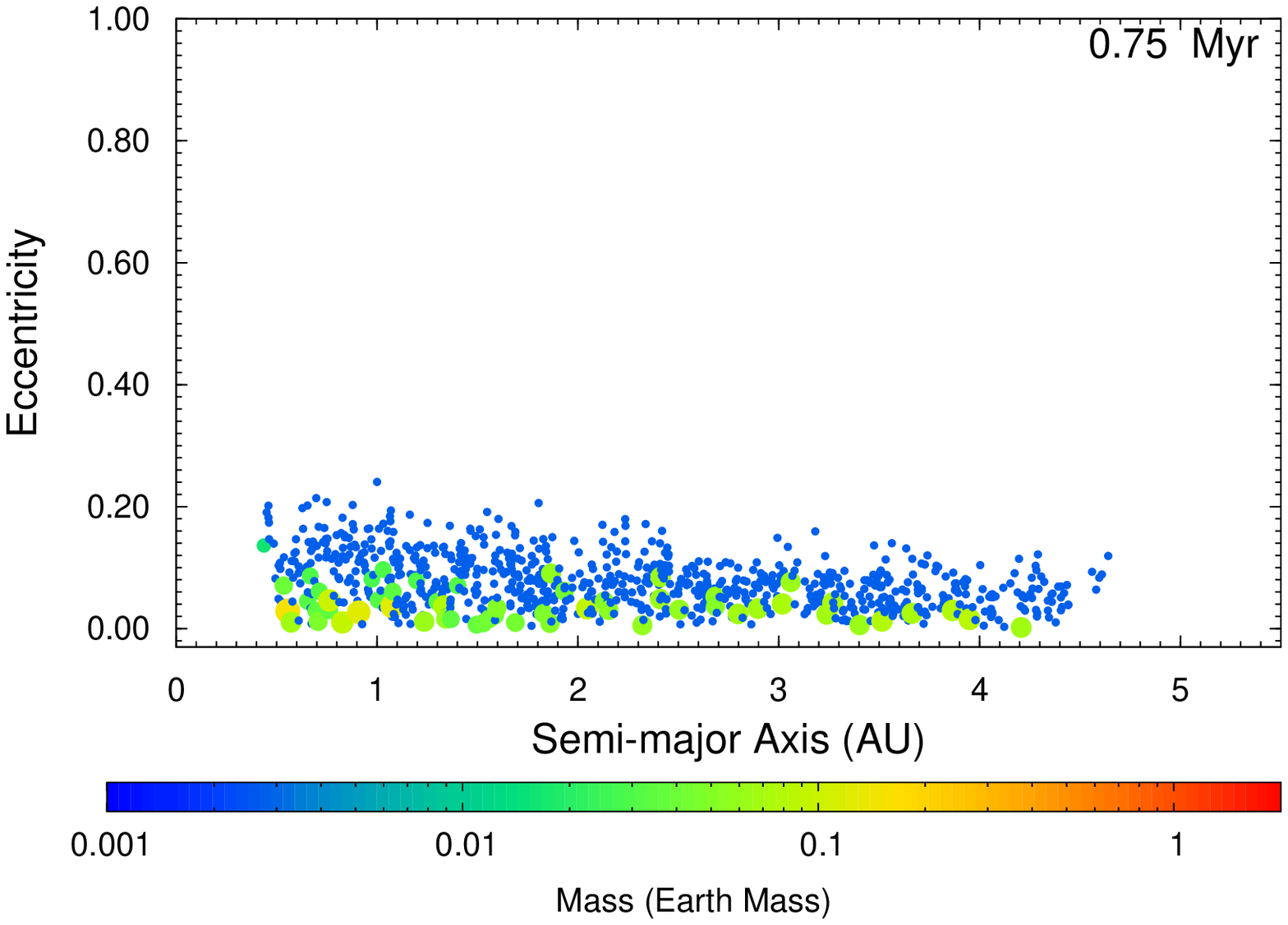}
}
\caption{Snapshots of the evolution of a disk of planetesimals and planetary embryos in a system
where Jupiter and Saturn are initially in their current orbits. From top to bottom, the disk's surface
density profile is proportional to $r^{-0.5}$, $r^{-1}$, and $r^{-1.5}$, respectively. 
The time of each panel corresponds to when the effect of $\nu_6$ was first enhanced.
The location of the $\nu_6$ resonance has been marked on left panels. The right panels show the 
state of the similar disks when the effects of giant planets are not included.}
\end{figure}

\begin{figure}
\centering{
\includegraphics[scale=0.5]{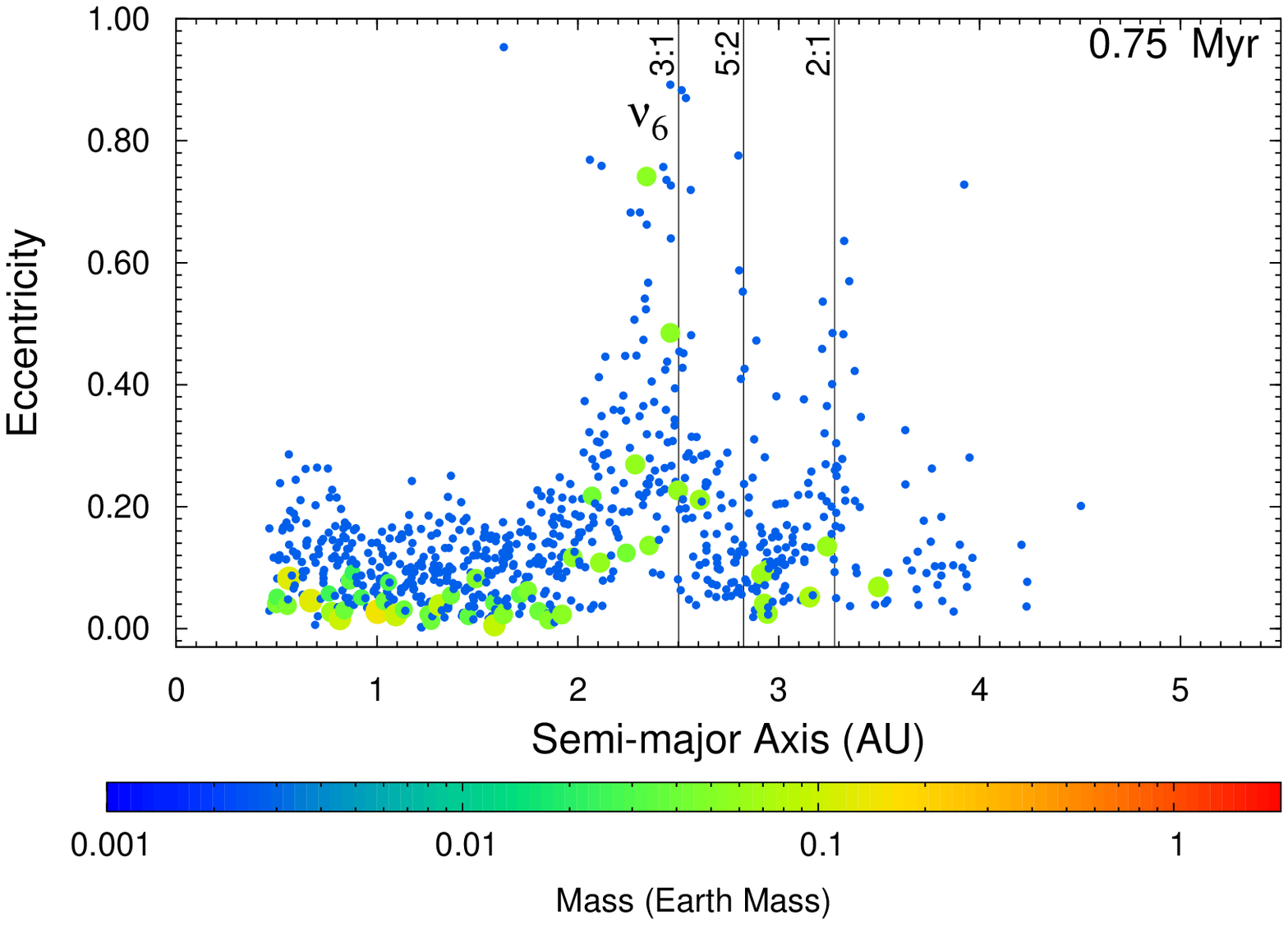}
\vskip 5pt
\includegraphics[scale=0.5]{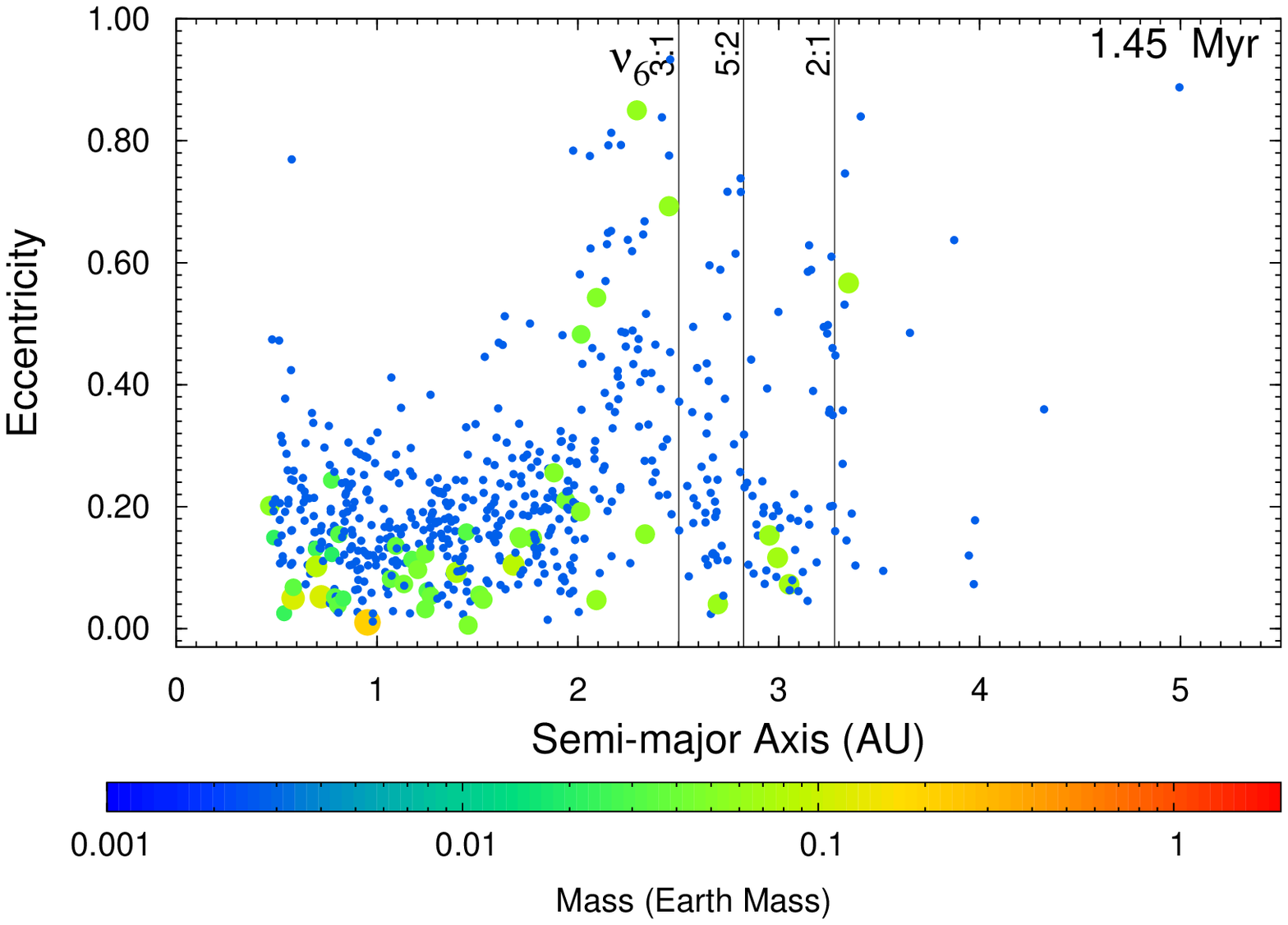}
\vskip 5pt
\includegraphics[scale=0.5]{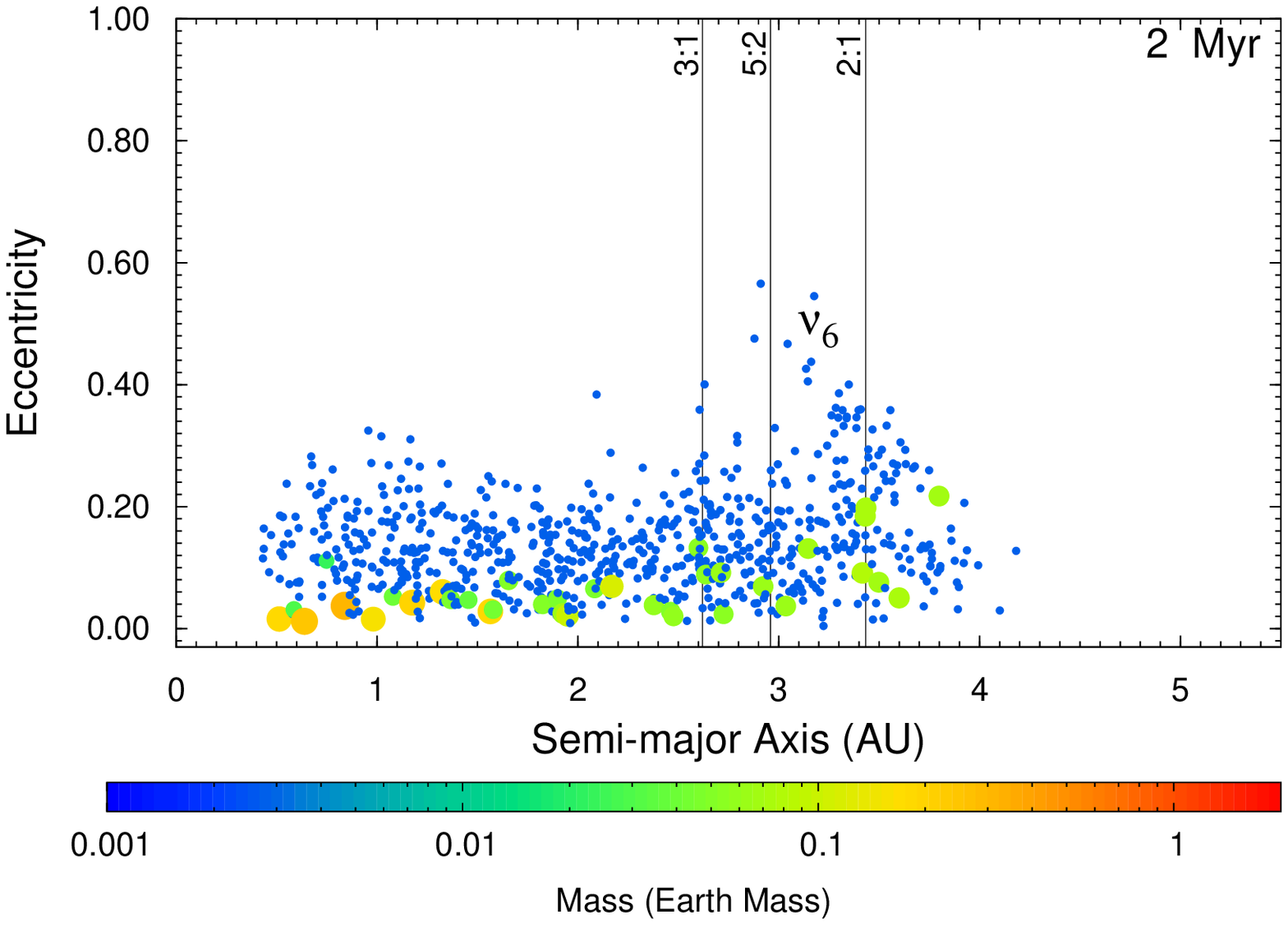}
}
\caption{Snapshots of the evolution of a disk of planetesimals and planetary embryos in a system
where the disk's surface density profile is proportional to $r^{-1.5}$. From top to bottom, the initial
orbits of Jupiter and Saturn are taken to be similar to their current orbits, have an eccentricity of 0.1,
and be similar to those in the Nice model. The time of each panel corresponds to when the effect of 
$\nu_6$ was first enhanced.}
\end{figure}

To determine the locations of secular resonances in our protoplanetary disk models,
we carried out simulations similar to those in Figures 2 and 3 where we also included massive planetary 
embryos. Figures 4 and 5 show snapshots of a sample of these simulations. The time in each panel
corresponds to when the effect of $\nu_6$ was first enhanced. In Figure 4, we show the results 
for disks with three different surface density profiles. The initial orbits of Jupiter and Saturn in these 
simulations were taken to be similar to their current orbits.
For the sake of comparison, we also show the state of the disk without the effects of giant planets on the right panels.
As shown here, from top to bottom, $\nu_6$ appears at $\sim 2$ AU, 2.3 AU, and 2.35 AU in disks with $\alpha=0.5, 1$ 
and 1.5, respectively. These results clearly show the initial outward displacement of $\nu_6$
compared to that in a disk of test particles (Figure 2) where the position of this resonance is at $\sim$2 AU.
In Figure 5, we show the results for a disk with $\alpha=1.5$ and for three different initial orbits of giant planets.
From top to bottom, the panels correspond to when the orbits of Jupiter and Saturn are taken to be similar to those
in their current orbits ($\nu_6$ at $\sim 2.3$ AU), carry an initial eccentricity of 0.1 ($\nu_6$ at $\sim 2.35$ AU), 
and be similar to those in the Nice model ($\nu_6$ at $\sim 3.35$ AU). 
The amount of the displacement of $\nu_6$ in Figures 4 and 5 is consistent with the findings of Nagasawa et al 
(2005)\footnote{According to the 
calculations by Nagasawa et al (2005), the outward displacement of $\nu_6$ in our disk models should be no more than 0.1 AU.
This outward displacement would be smaller in disks with less steep surface density profiles (i.e., lower values of $\alpha$). 
While in our simulations, the $\nu_6$ resonance did move outward, the amount of its displacement was
slightly larger than that suggested by Nagasawa et al (2005). For example, in a disk with $\alpha=1.5$ and 
considering Jupiter and Saturn in their current orbits, $\nu_6$ appeared around 2.35 AU, which corresponds to an 
outward displacement of 0.2-0.3 \rm AU compared to the location of this resonance in simulations considering test particles 
and same configuration for giant planets. Our results indicate that the calculations by Nagasawa et al (2005)  
underestimates the outward sweeping of secular resonances by not considering the disk gravitational potential.}. 

During the course of each simulation, we also examined the effect of $\nu_5$ resonance. We noticed
that this resonance never becomes strong as it is continuously countered by the mutual 
interactions of planetary embryos. 
We would like to emphasize that Figures 4 and 5 are only for illustrative purposes. These snapshots
are not intended to portray any correlation between the time and location of $\nu_6$ and $\nu_{16}$ with
the disk surface density profile. The proximity of the location of these resonances to the orbit of
giant planets depends on the initial orbital elements of embryos in each system.

\begin{figure}
\centering{
\includegraphics[scale=0.45]{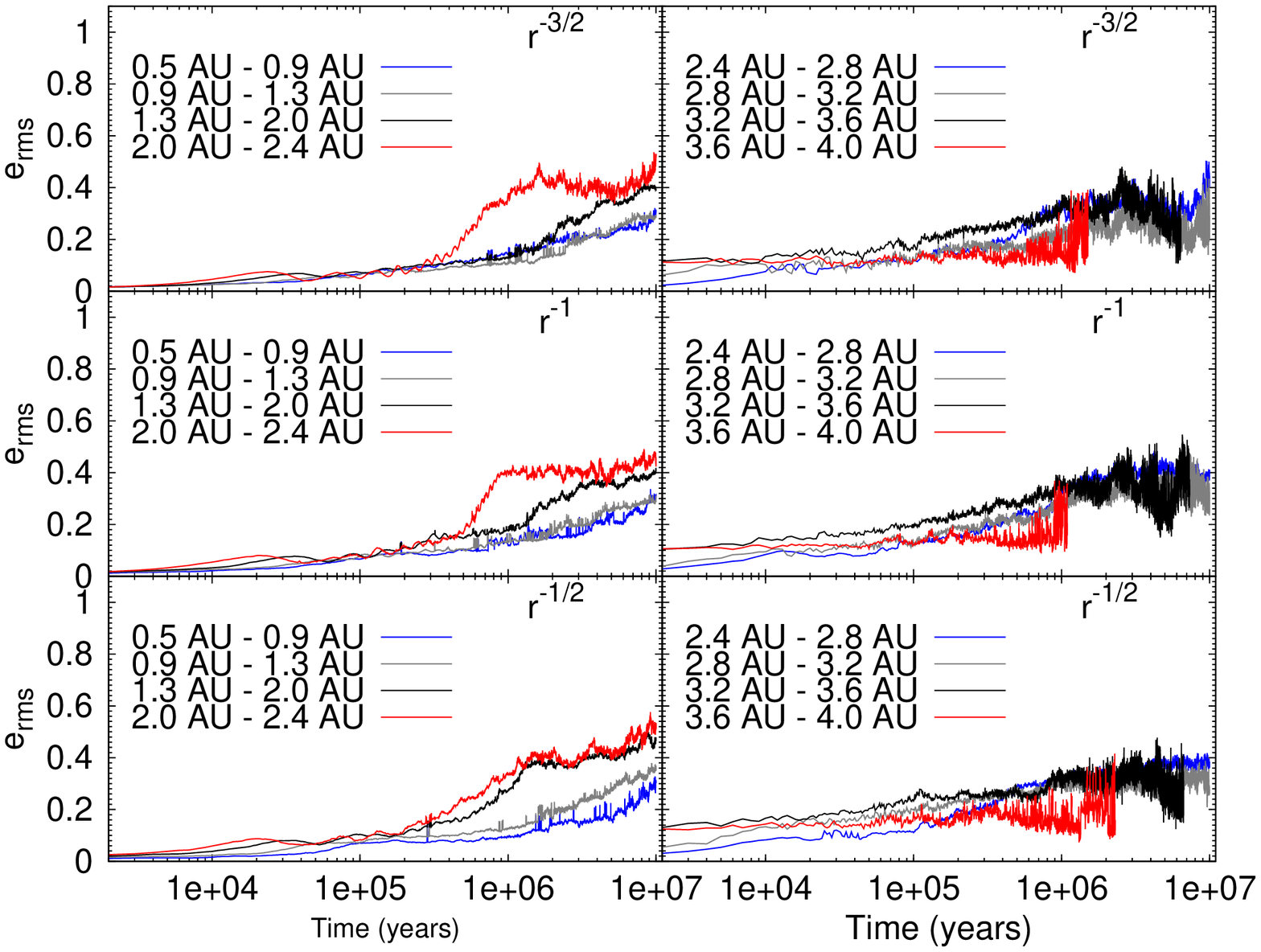}
\includegraphics[scale=0.45]{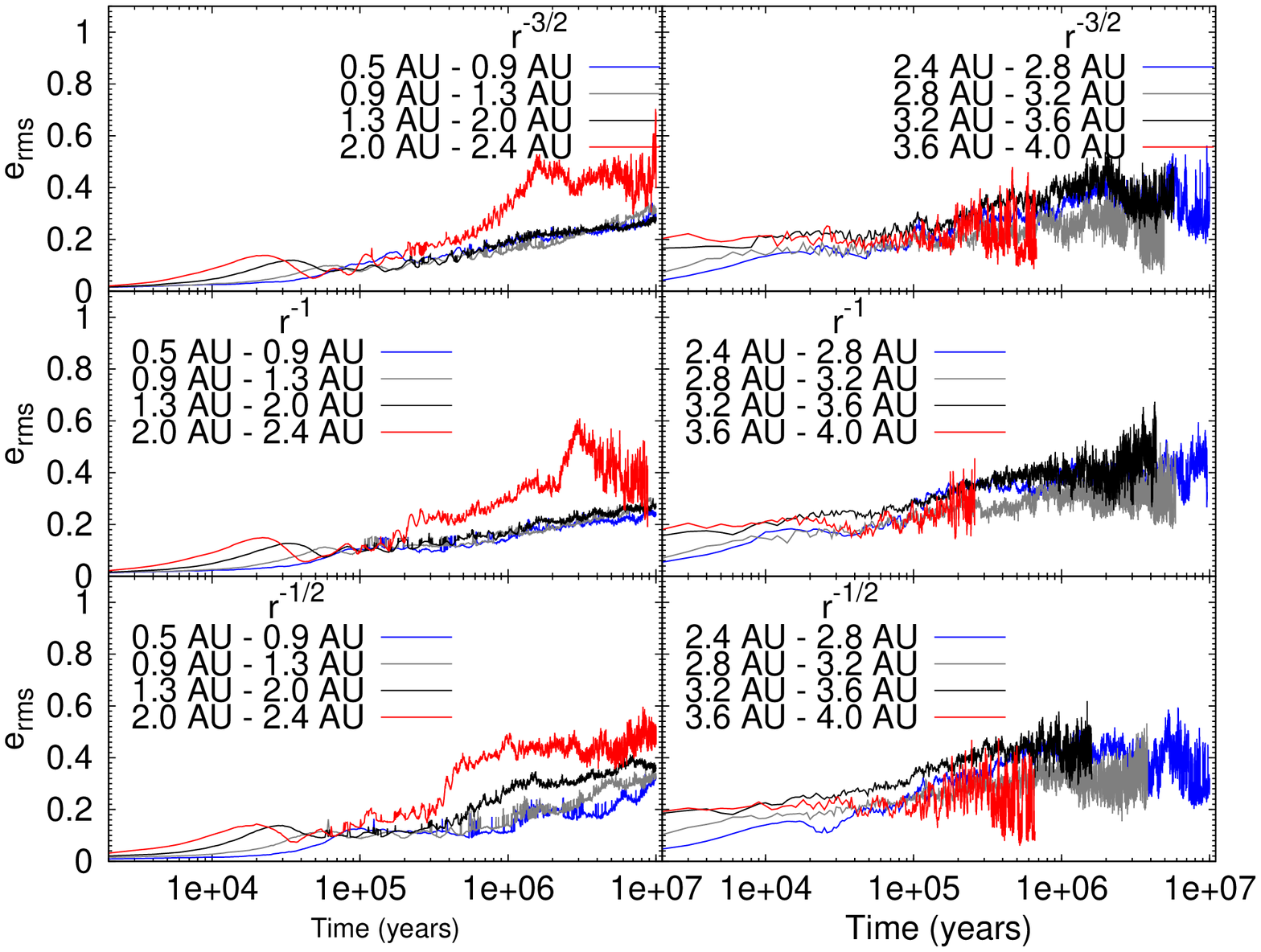}
\includegraphics[scale=0.45]{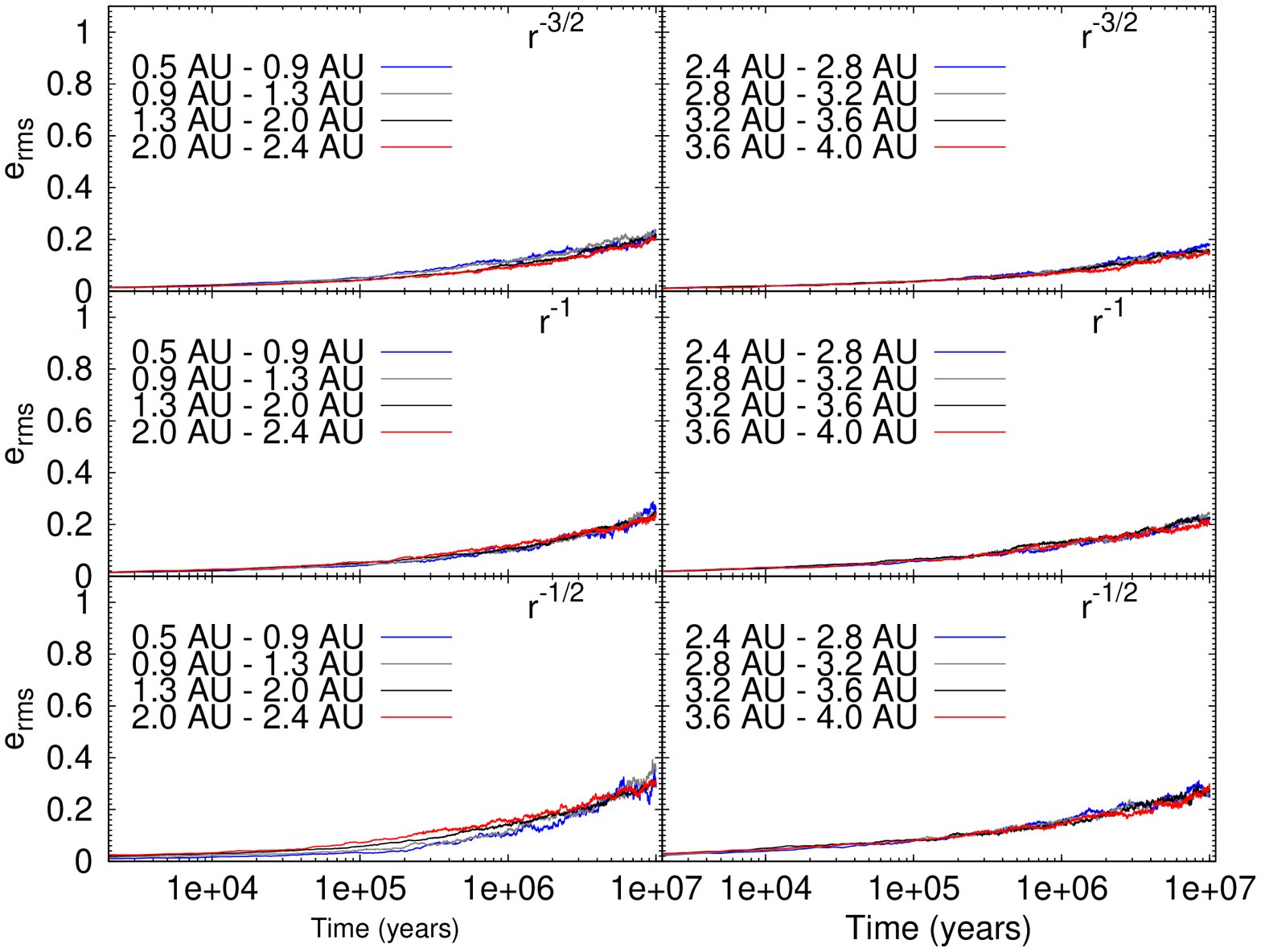}
}
\caption{Graphs of the root-mean-squares of the eccentricities
of planetesimals and planetary embryos in all three disk models for a system where Jupiter and
Saturn are in their current orbits (top), have an initial eccentricity of 0.1 (middle), 
and a system with no giant planets (bottom). The orbital excitation of disk objects at the location
of $\nu_6$ resonance can be seen in the top and middle panels for all three disk models.
The graph of eccentricities for the region of 3.6-4.0 AU has been stopped
at about 1 Myr because the variations of the eccentricity after this time
become so large that they will cover the other graphs.} 

\end{figure}

\begin{figure}
\centering{
\includegraphics[scale=0.45]{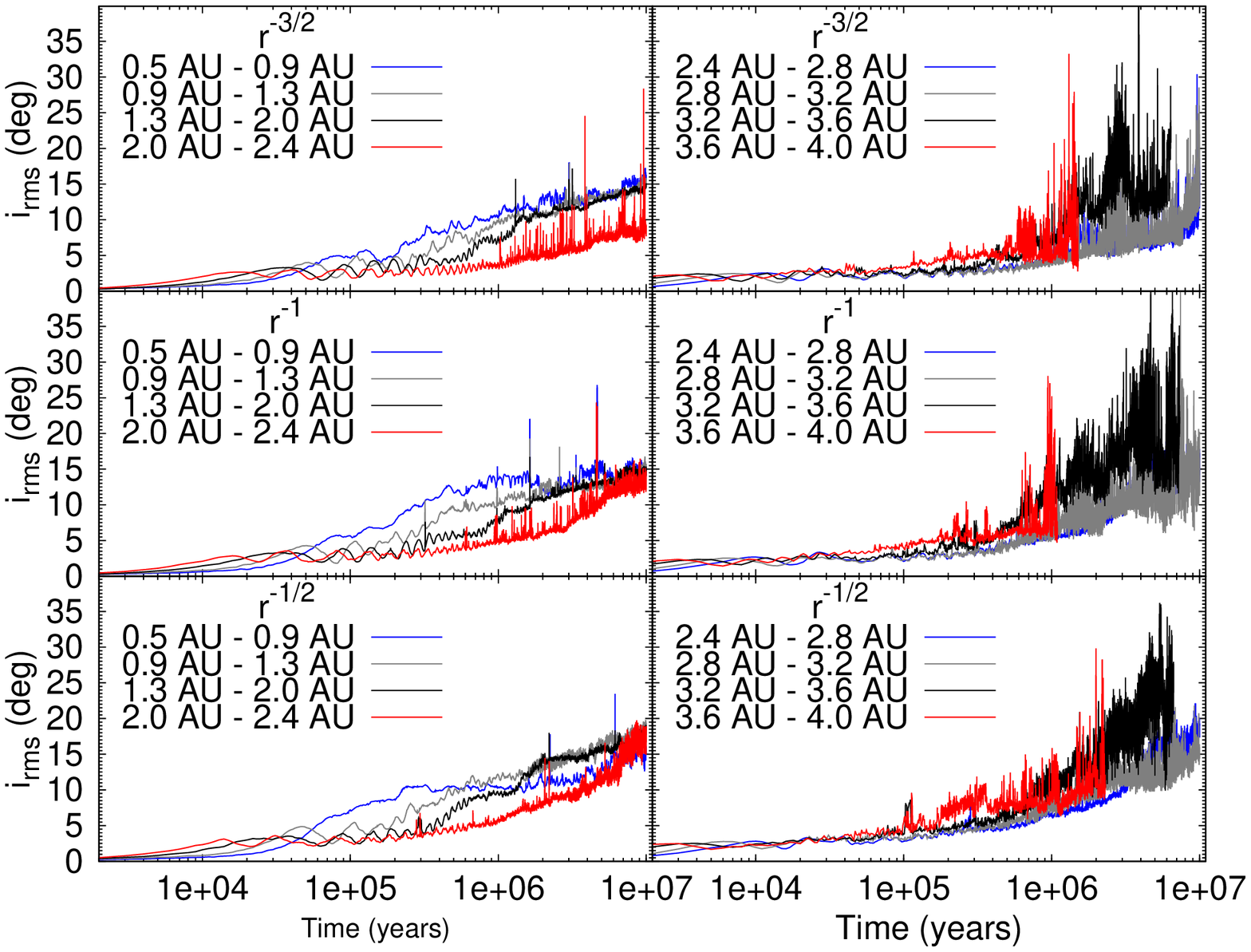}
\includegraphics[scale=0.45]{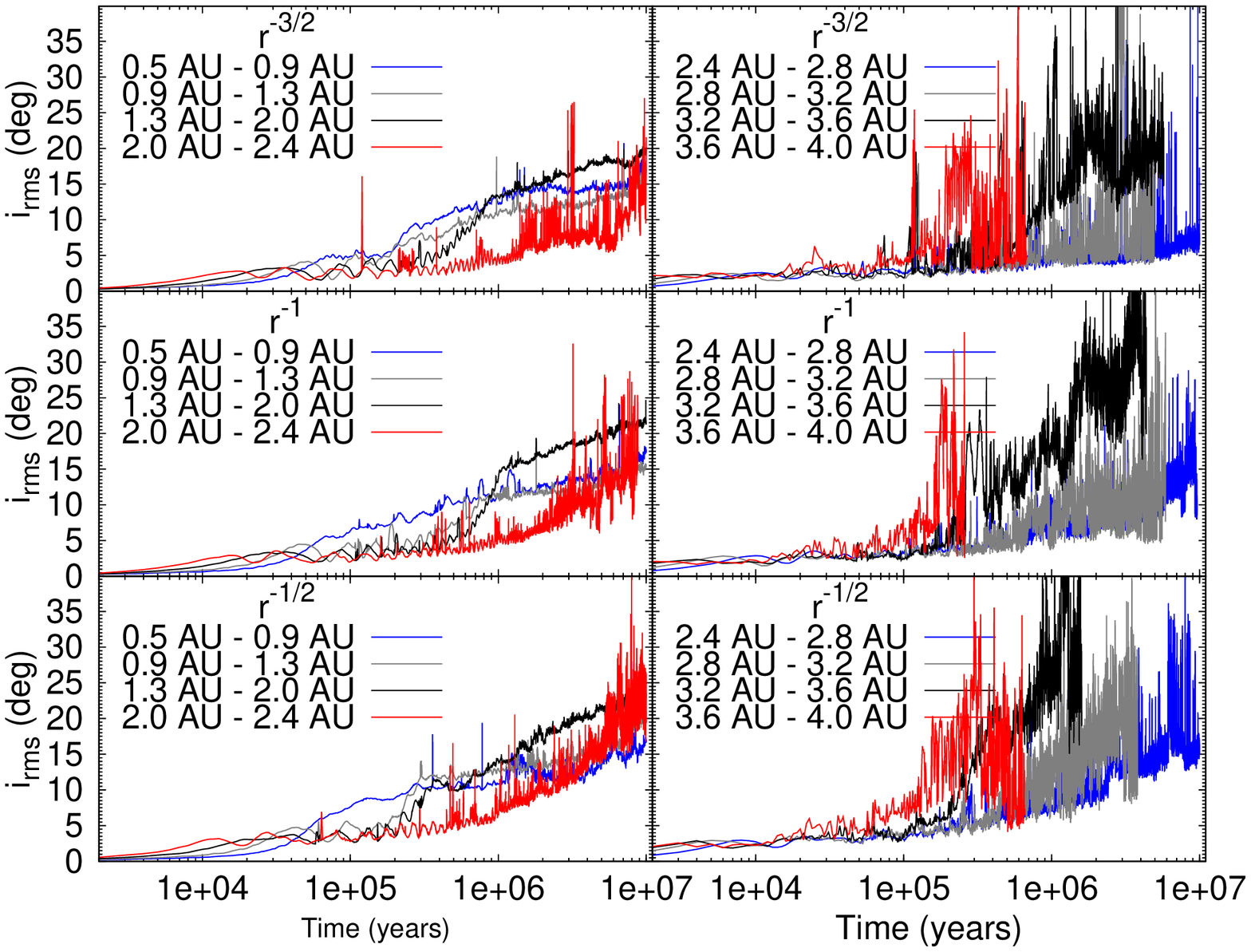}
\includegraphics[scale=0.45]{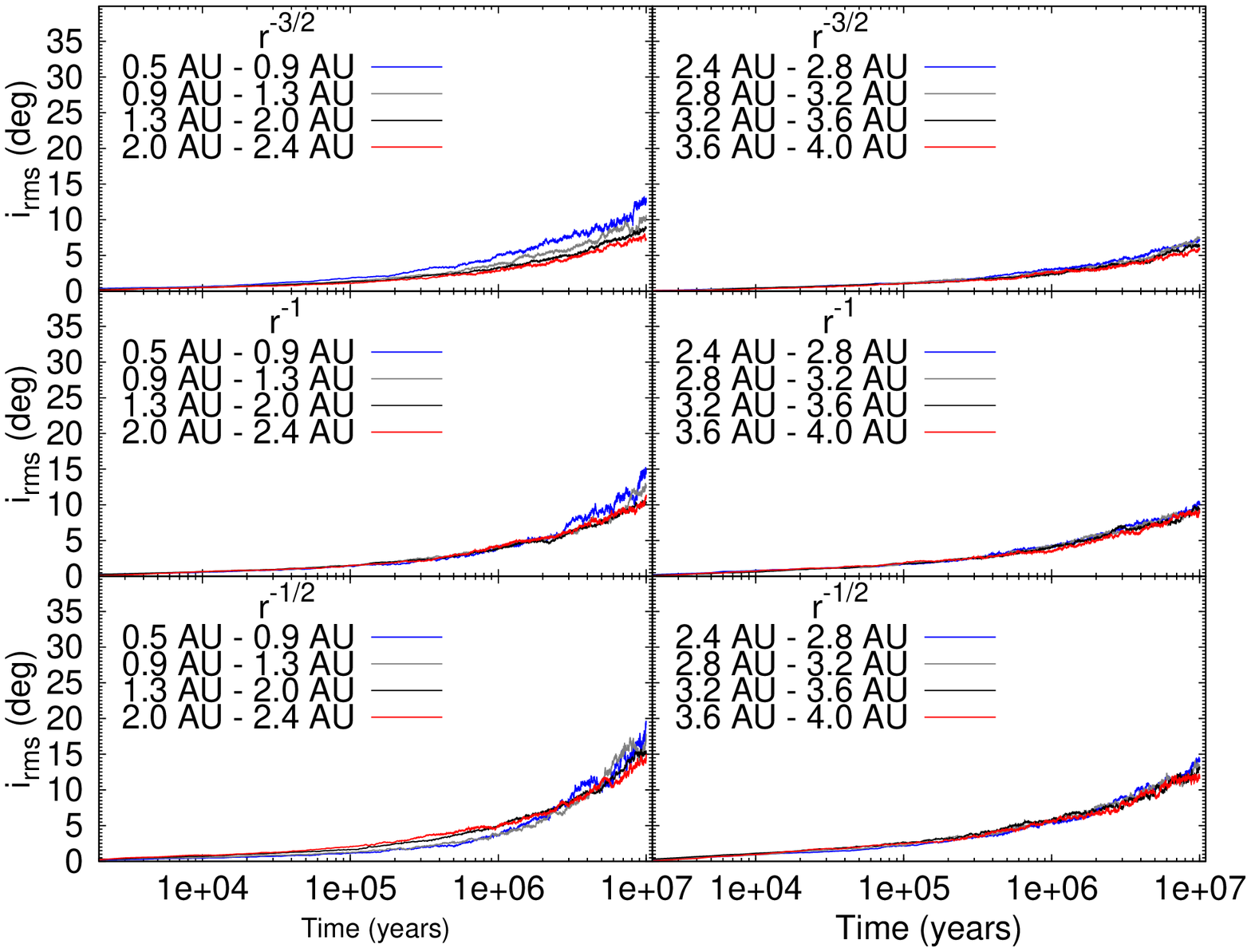}
}
\caption{Graphs of the root-mean-squares of the inclinations
of planetesimals and planetary embryos in all three disk models for a system where Jupiter and
Saturn are in their current orbits (top), have an initial eccentricity of 0.1 (middle), 
and a system with no giant planets (bottom). The orbital excitation of disk objects at the location
of $\nu_{16}$ resonance can be seen in the top and middle panels for all three disk models.
The graph of inclinations for the region of 3.6-4.0 AU has been stopped
at about 1 Myr because the variations of the inclination after this time
become so large that they will cover the other graphs.}
\end{figure}

To better quantify the effect of giant planets and mutual interactions of disk bodies in increasing the 
eccentricities and inclinations of the latter objects, we show in Figures 6 and 7 the 
root-mean-squares of their eccentricities and inclinations for our three disk surface density profiles. 
The top panels in these figures correspond to a system where Jupiter and Saturn are initially in their 
current orbits, the middle panels are for when they have an initial eccentricity of 0.1, and the bottom panels 
show the results for a system without giant planets.
As can be seen, objects are strongly excited in systems with giant planets. The 
orbital excitation is specifically pronounced in the outer region of the disk. 
The effects of the $\nu_6$ and $\nu_{16}$ resonances can also be seen in the
top and middle panels in the region of 2 AU -- 2.4 AU. Figures 6 and 7 also show that the orbital excitation 
of planetesimals and embryos in the region of 1.3 AU -- 2 AU is stronger for disks with less steep 
surface density profiles. This result can be attributed to fact that in such disks, the mutual 
interactions among planetesimals and planetary embryos increase with increasing the mass of 
the disk. Note that the graph of the region of 3.6 AU -- 4 AU shows the time evolution of 
$e_{\rm rms}$ only up to the point when less than 5 objects have survived. We do not show 
the rest of the graph because the variations in $e_{\rm rms}$ become quite large and cover 
those of the inner regions.

\vskip 20pt
{\bf \subsubsection{Disk-Mass Removal: Inward Sweeping of Secular Resonances}}

As a consequence of the interactions of giant planets with a massive disk, either through mean-motion 
and secular resonances, or because of the direct dynamical excitations of the objects at the outer part of 
the disk, the orbits of disk bodies become highly eccentric. This orbital excitation propagates 
throughout the disk via embryo-embryo and planetesimal-embryo interactions. The high eccentricities of
these objects causes many of them to either hit each other (and undergo breakage or fragmentation), 
collide with giant planets or central star, or leave the gravitational field of the system.  
In other words, the disk will lose mass.

\begin{figure}
\centering{
\includegraphics[width=8.7cm]{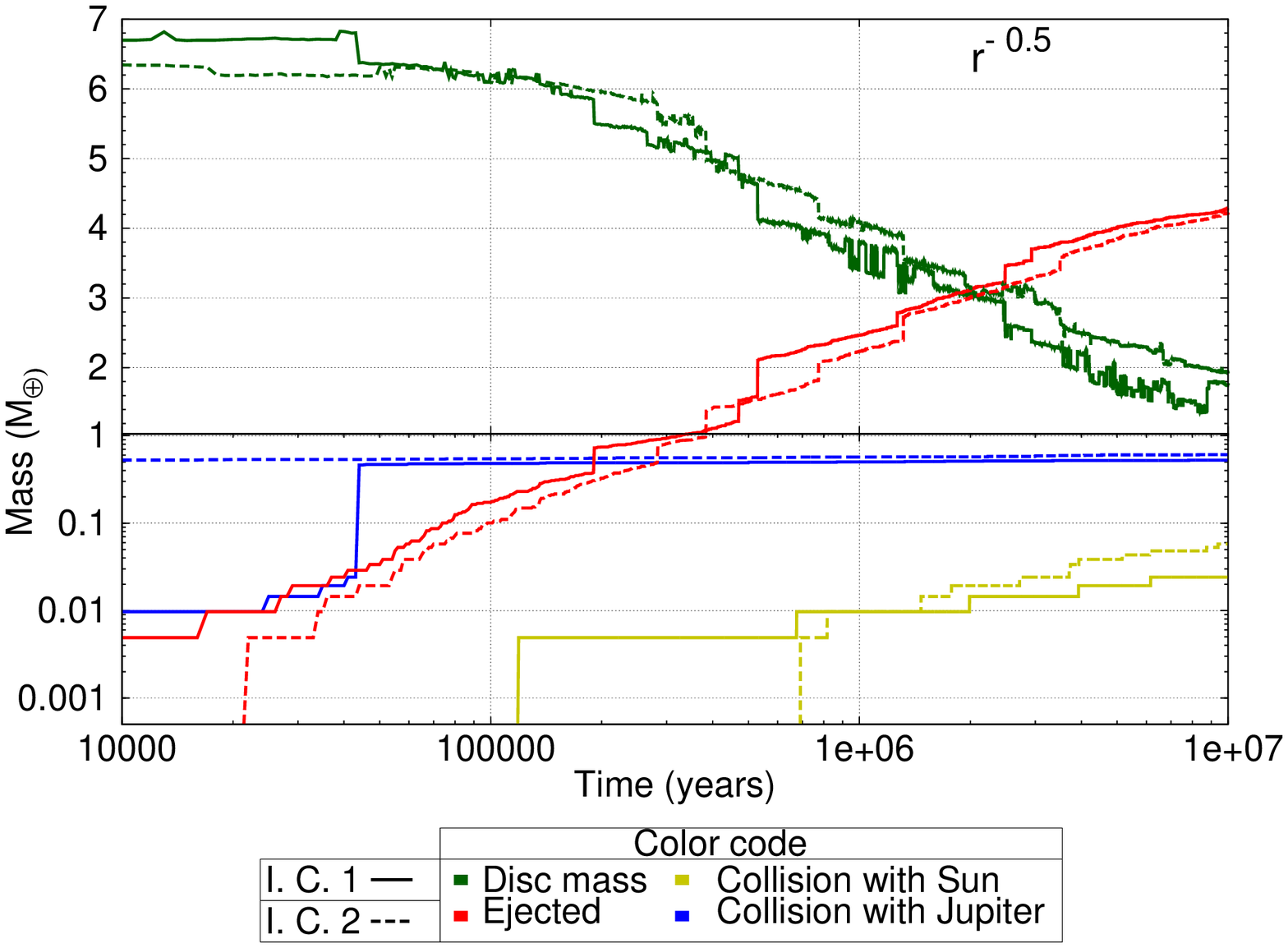}
\vskip 5pt
\includegraphics[width=8.7cm]{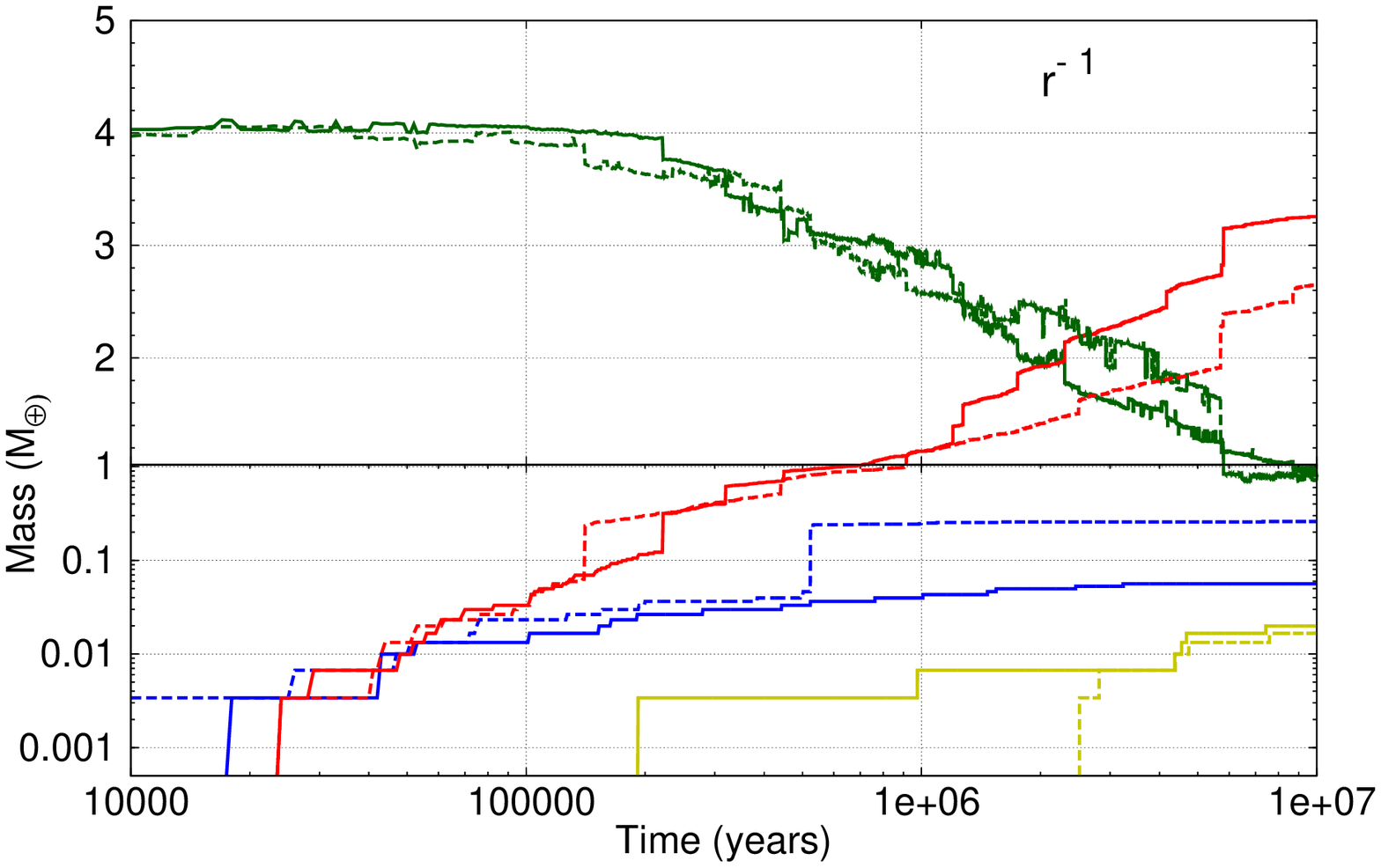}
\vskip 5pt
\includegraphics[width=8.7cm]{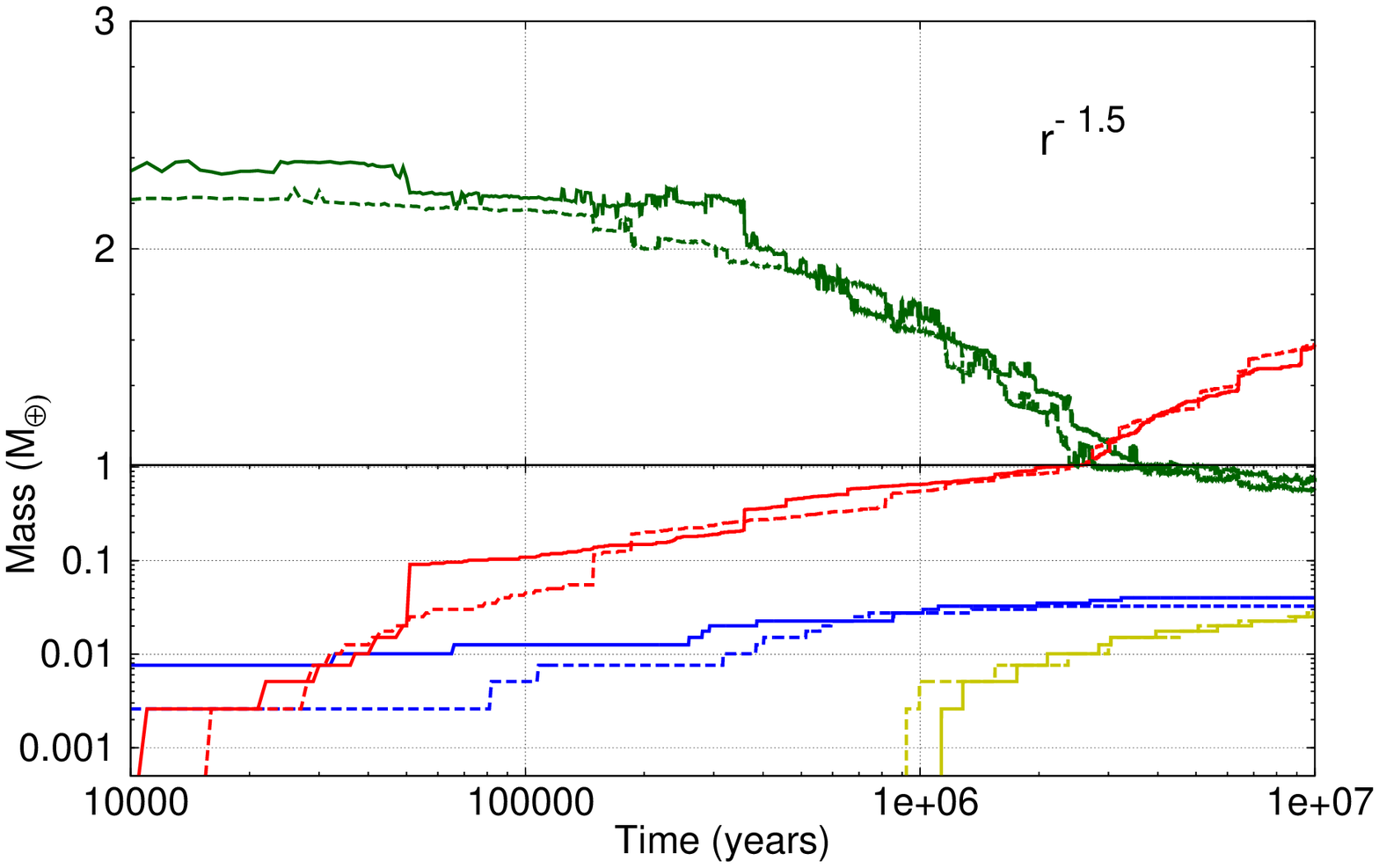}
}
\caption{Graphs of the evolution of the disk mass beyond 2 AU
in a system where Jupiter and Saturn are initially in their current orbits. From top to bottom,
the disk's surface density profile is proportional to $r^{-0.5}$, $r^{-1}$, and $r^{-1.5}$, respectively.
The green curve shows the decrease in the mass of the disk,
and red, yellow and blue correspond to the amount of the mass that is lost due to the ejection from the system,
collision with the Sun, and collision with Jupiter, respectively.
The notation “I.C. 1\&2” in the legend represents two different initial distribution for the disk material. 
Note that on the vertical axis, the scale is logarithmic for the values smaller than 1, and is linear 
for higher values.}
\end{figure}

Figure 8 shows sample graphs of the evolution of the mass of the disk in 
the region beyond 2 AU for different surface density profiles and for a system where 
Jupiter and Saturn are initially in their current orbits (similar results were obtained for other initial orbital 
configurations of the giant planets). The green curve shows the decrease in the mass of the disk,
and red, yellow and blue correspond to the amount of the mass that is lost due to the ejection from the system,
collision with the Sun, and collision with Jupiter, respectively. As shown here, mass-loss begins when the
orbits of planetary embryos become gradually excited. Once the orbital excitation of embryos intensifies 
and their orbital eccentricities reach high values, the rate of mass-loss is 
noticeably enhanced. 

As shown in Figure 8, the rate of mass-loss
varies with the radial profile of the disk. In a disk with a less 
steep surface density profile, the rate of mass-loss is higher. This is an expected result that can be 
explained noting that planetary embryos
are the main carriers of the mass, and that their masses increase as a function of their semimajor axes
(see section 2.1). As a result, a disk with more mass in its outer region (e.g., a disk with less steep
radial profile such as $r^{-0.5}$) will have larger embryos that 
are farther apart. In other words, such a disk will have fewer but more massive embryos in its outer region.
The latter causes the interactions between embryos to be stronger
which in turn causes the disk to become dynamically excited in a shorter time. When  
some of these embryos are scattered out of the system, because they
are more massive (compared to their counterparts in a disk with a more steep surface density profile),
they carry away more of the mass of the disk, increasing the rate at which the disk loses material. 

As the disk loses mass in its outer region, its regressing effect on the orbit of an object
weakens. This, combined with the inward migration of the giant planets due to the exchange of angular 
momentum with the disk causes secular resonances move slightly inward. Figure 9 shows this in a series 
of snapshots of the evolution of a disk with an $r^{-1}$ surface density profile. The giant planets
in this simulation were initially in the current orbits of Jupiter and Saturn. The locations of some
of the major mean-motion resonances with Jupiter are also shown.

\begin{figure}
\centering
\includegraphics[scale=.7]{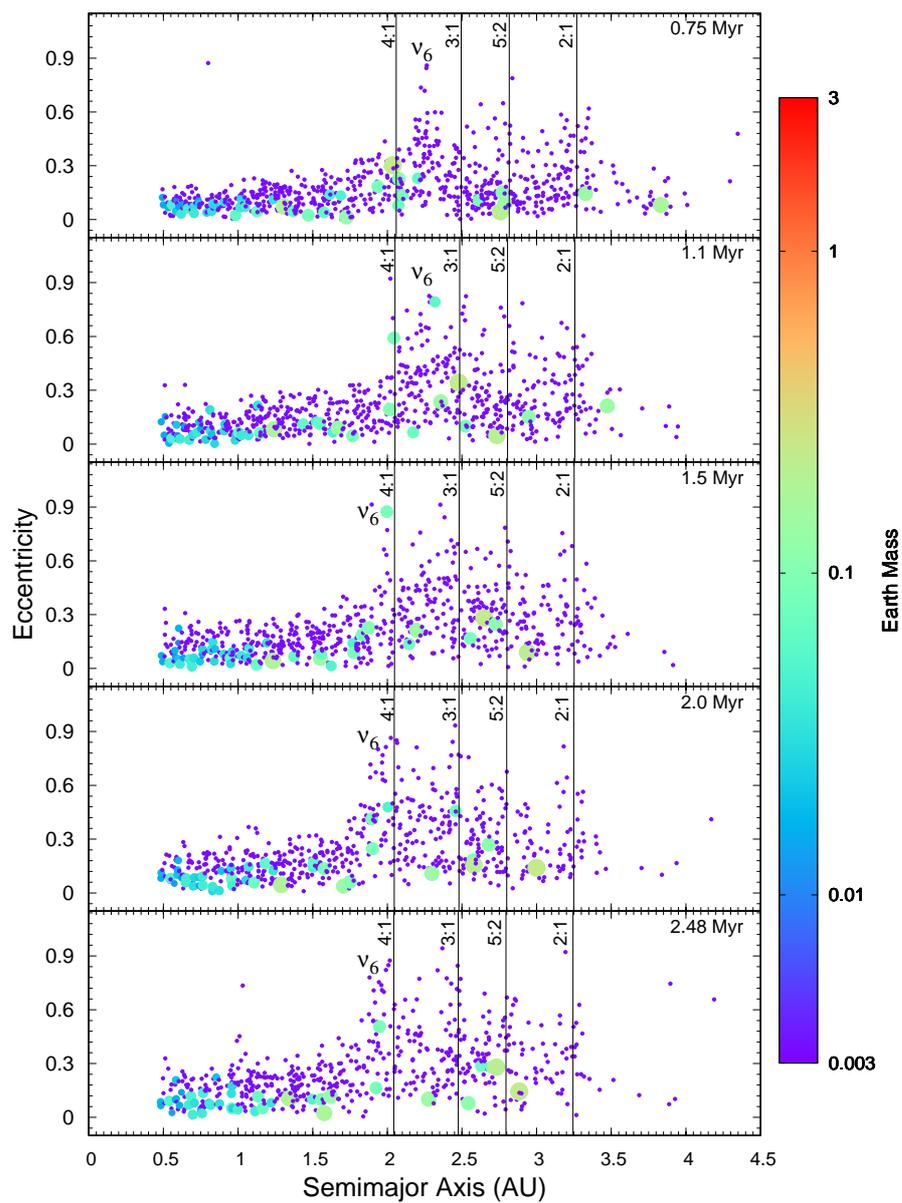}
\caption{Snapshots of the evolution of a disk of planetesimals and planetary embryos in a system
where Jupiter and Saturn are initially in their current orbits and the disk's surface
density profile is proportion to $r^{-1}$. The location of the $\nu_6$ resonance has been marked 
on the panels showing its shift inward. The main mean motion resonances with Jupiter are also shown.} 
\end{figure}

\begin{figure}
\centering{
\includegraphics[scale=0.34]{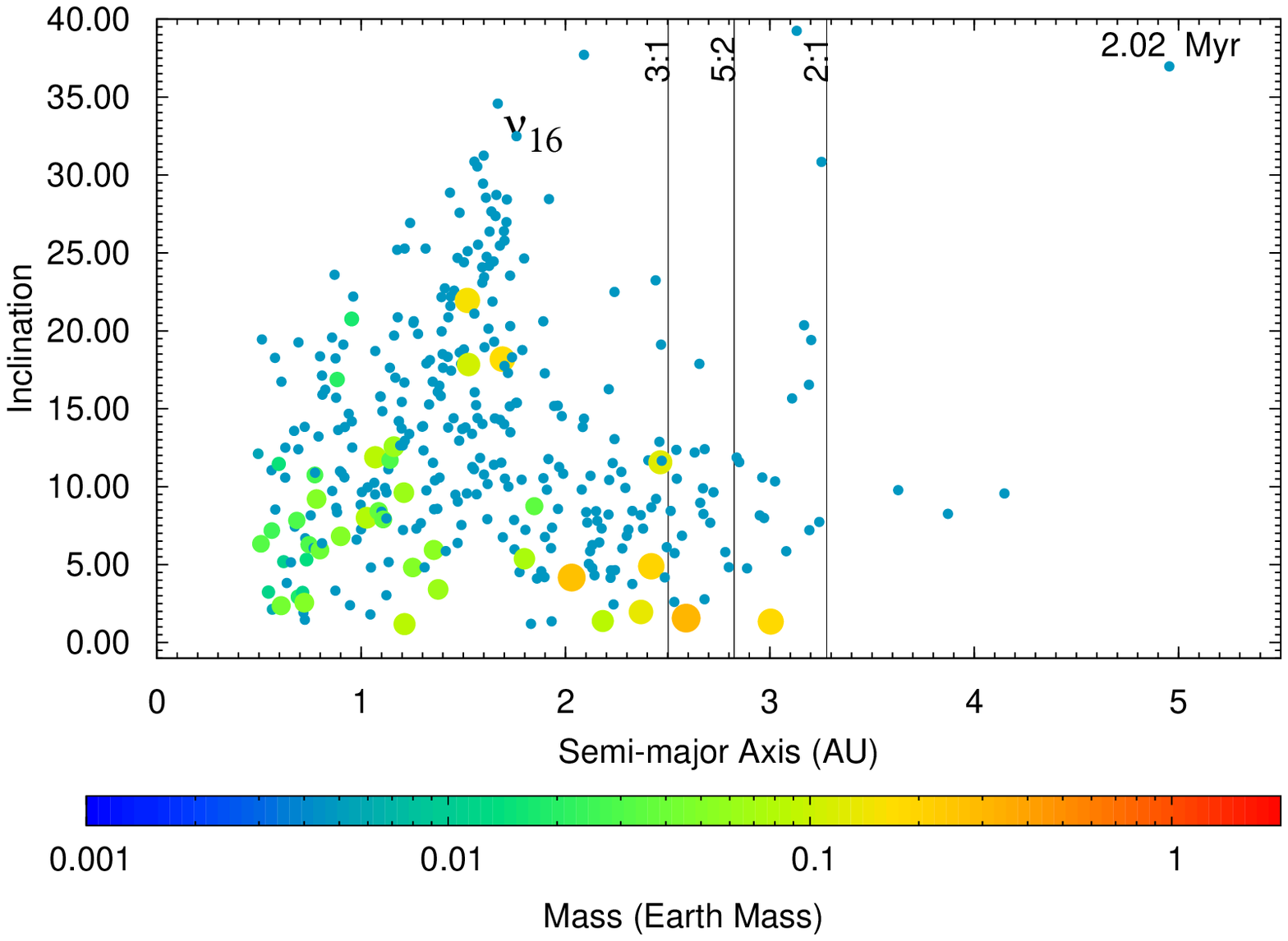}
\includegraphics[scale=0.34]{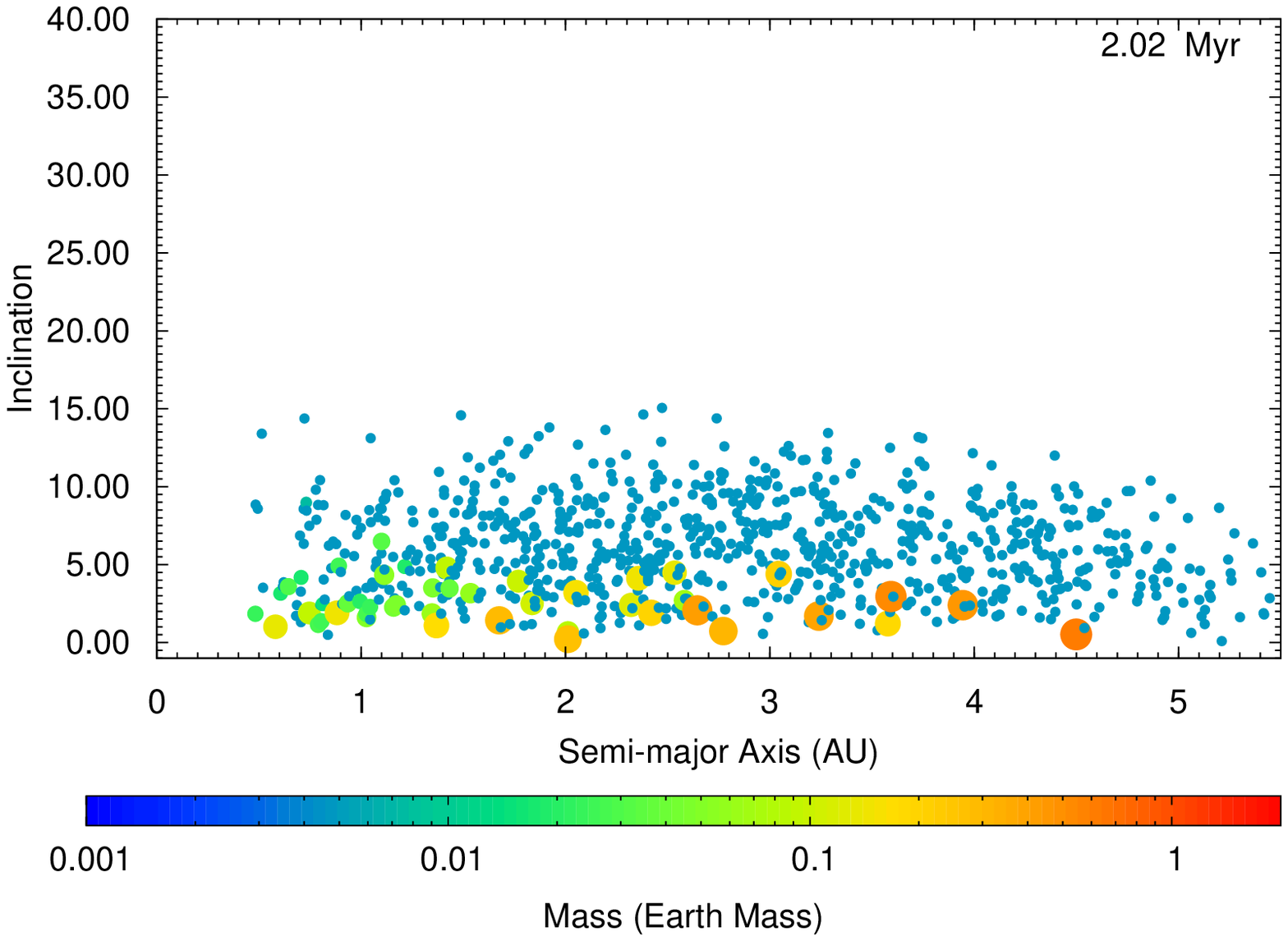}
\vskip 5pt
\includegraphics[scale=0.34]{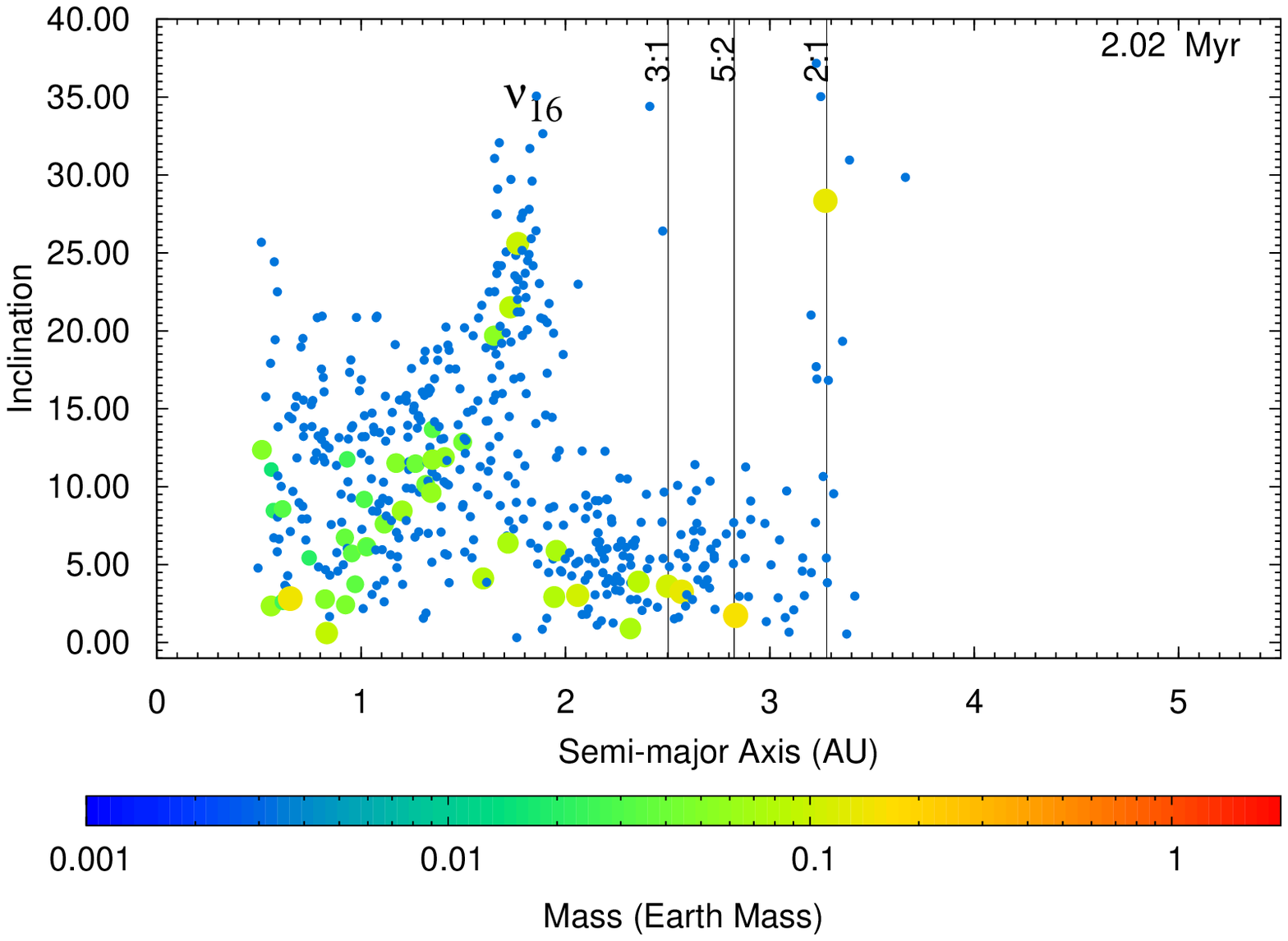}
\includegraphics[scale=0.34]{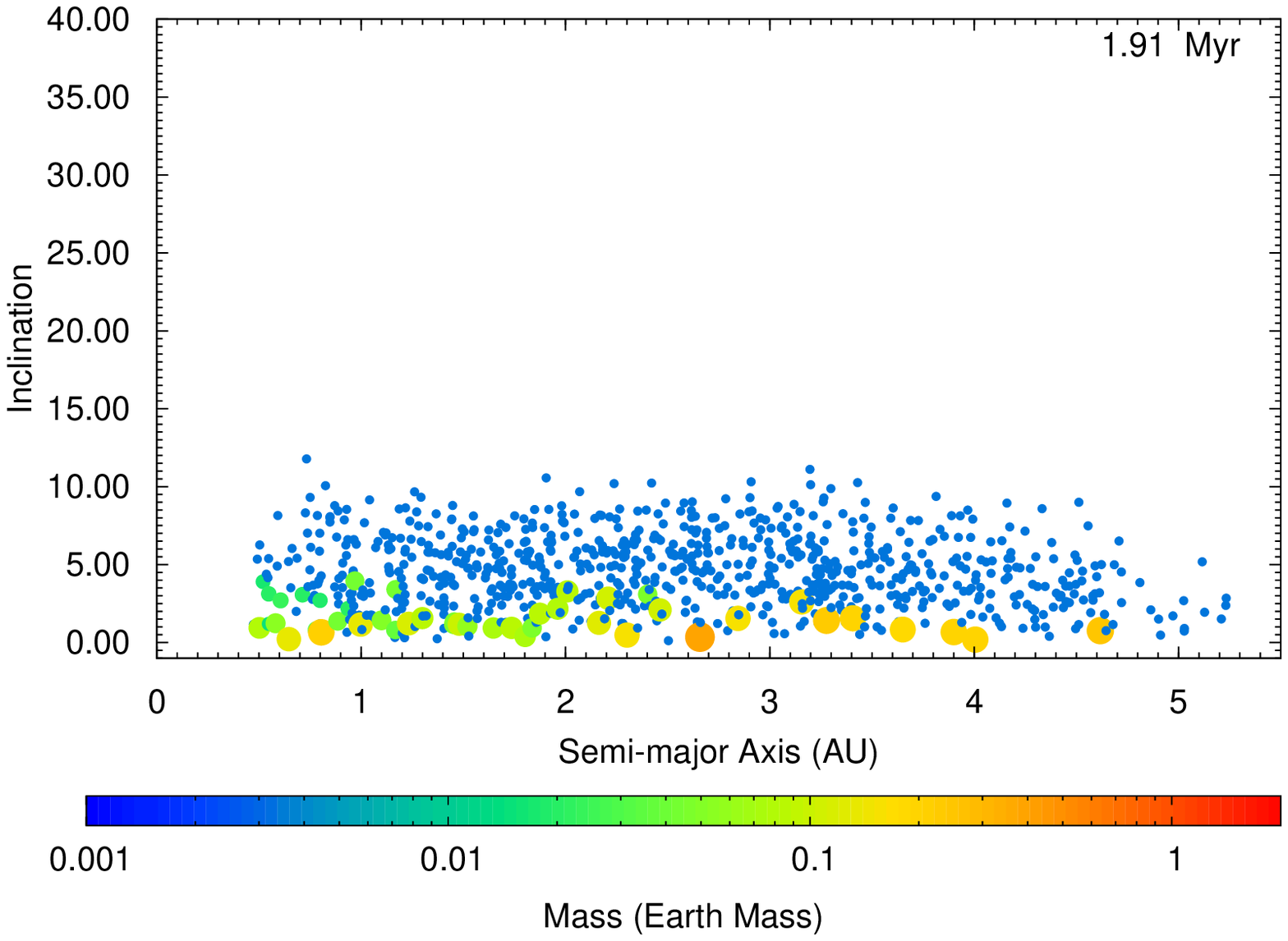}
\vskip 5pt
\includegraphics[scale=0.34]{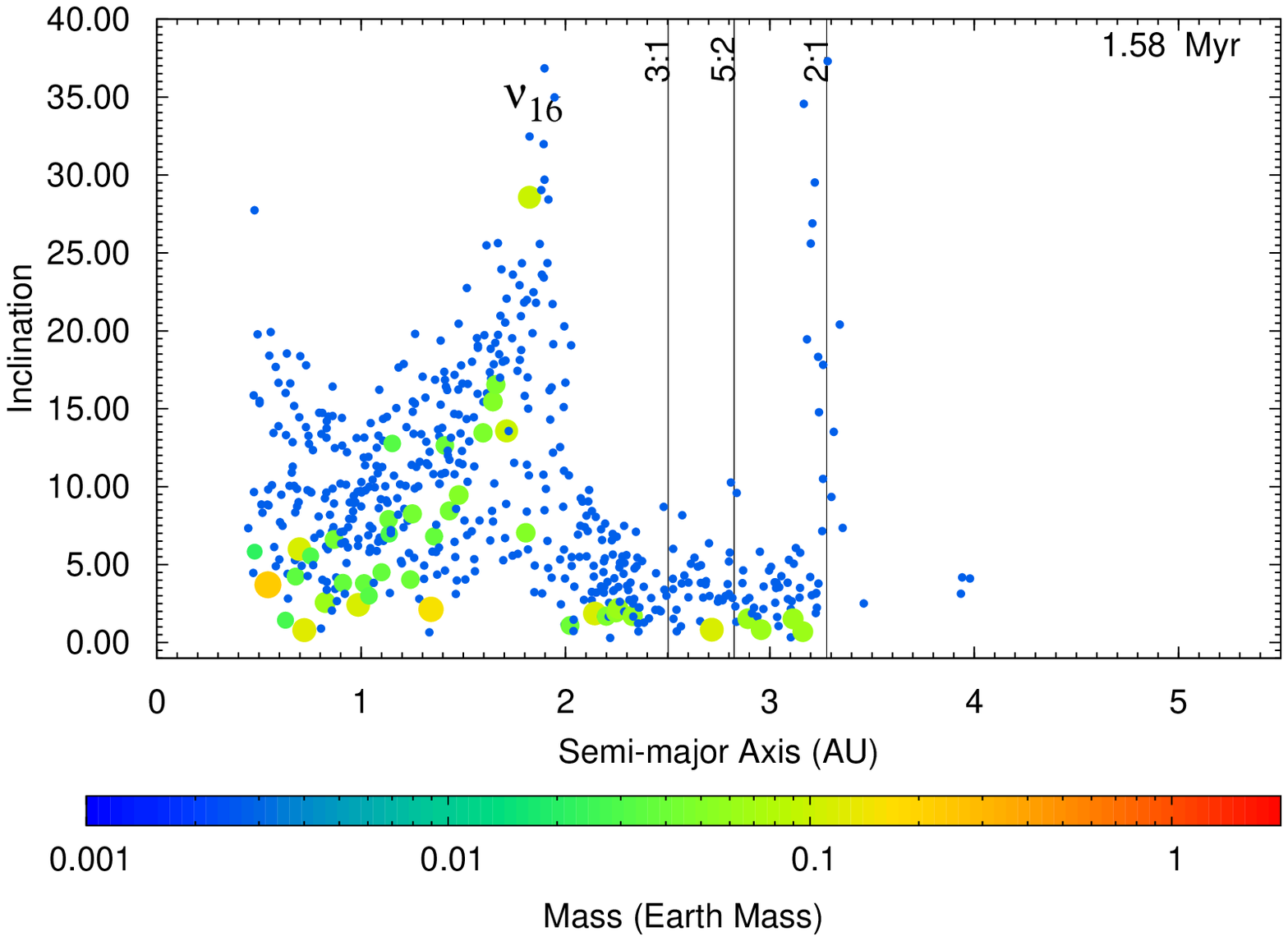}
\includegraphics[scale=0.34]{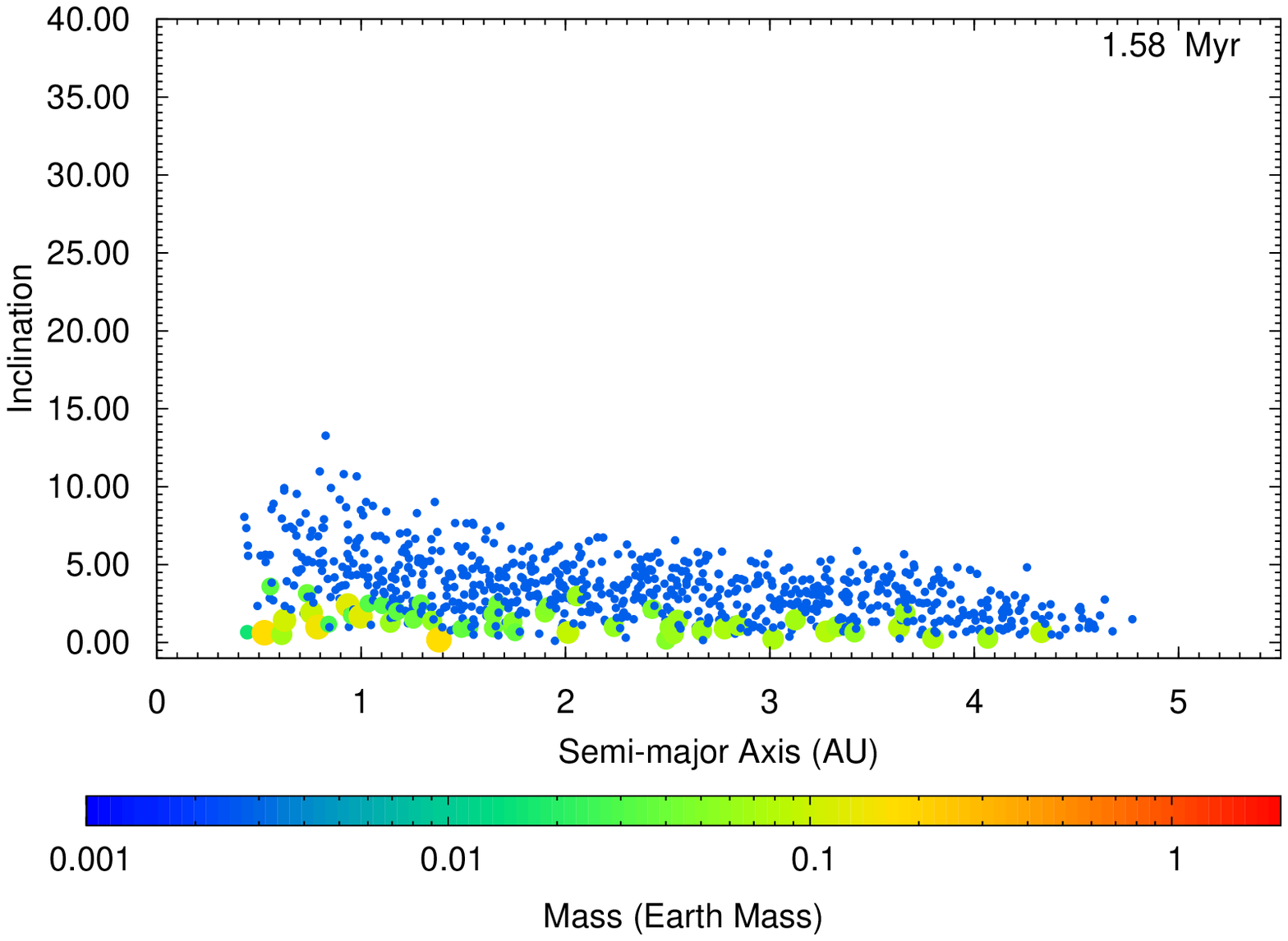}
}
\caption{Snapshots of the evolution of a disk of planetesimals and planetary embryos in a system
where Jupiter and Saturn are in their current orbits. From top to bottom, the disk's surface
density profile is proportional to $r^{-0.5}$, $r^{-1}$, and $r^{-1.5}$, respectively. 
The location of the $\nu_{16}$ resonance has been marked on the left panels. 
The time of each panel corresponds to when the effect of 
$\nu_{16}$ was first enhanced. The right panel shows
the state of the disk when the effects of the giant planets are not included. As shown in this panel, 
even in the absence of Jupiter and Saturn, the orbital inclinations of the disk objects are increased
due to their interactions with one another.}
\end{figure}

\begin{figure}
\centering{
\includegraphics[width=9cm]{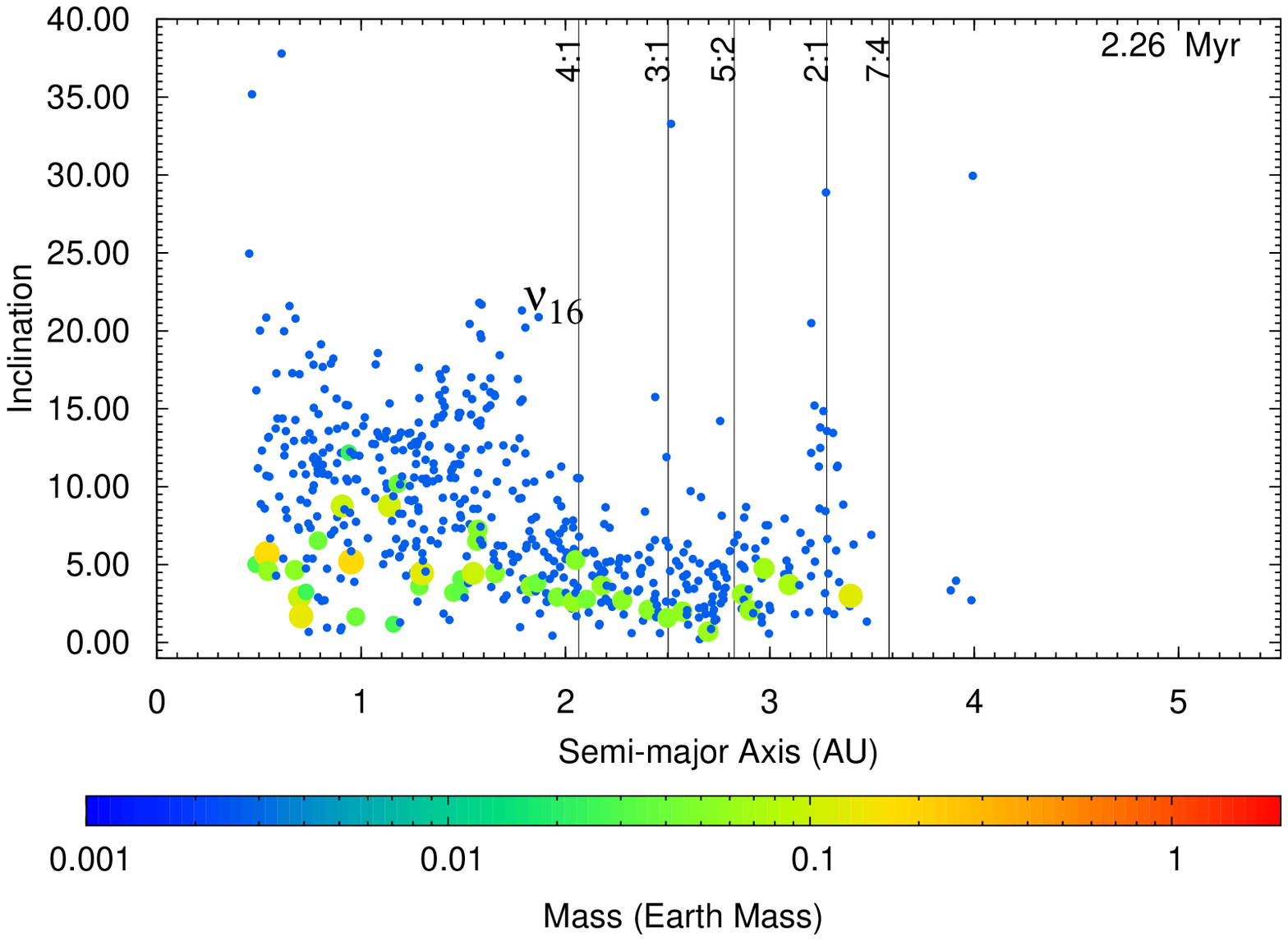}
\vskip -5pt
\includegraphics[width=9cm]{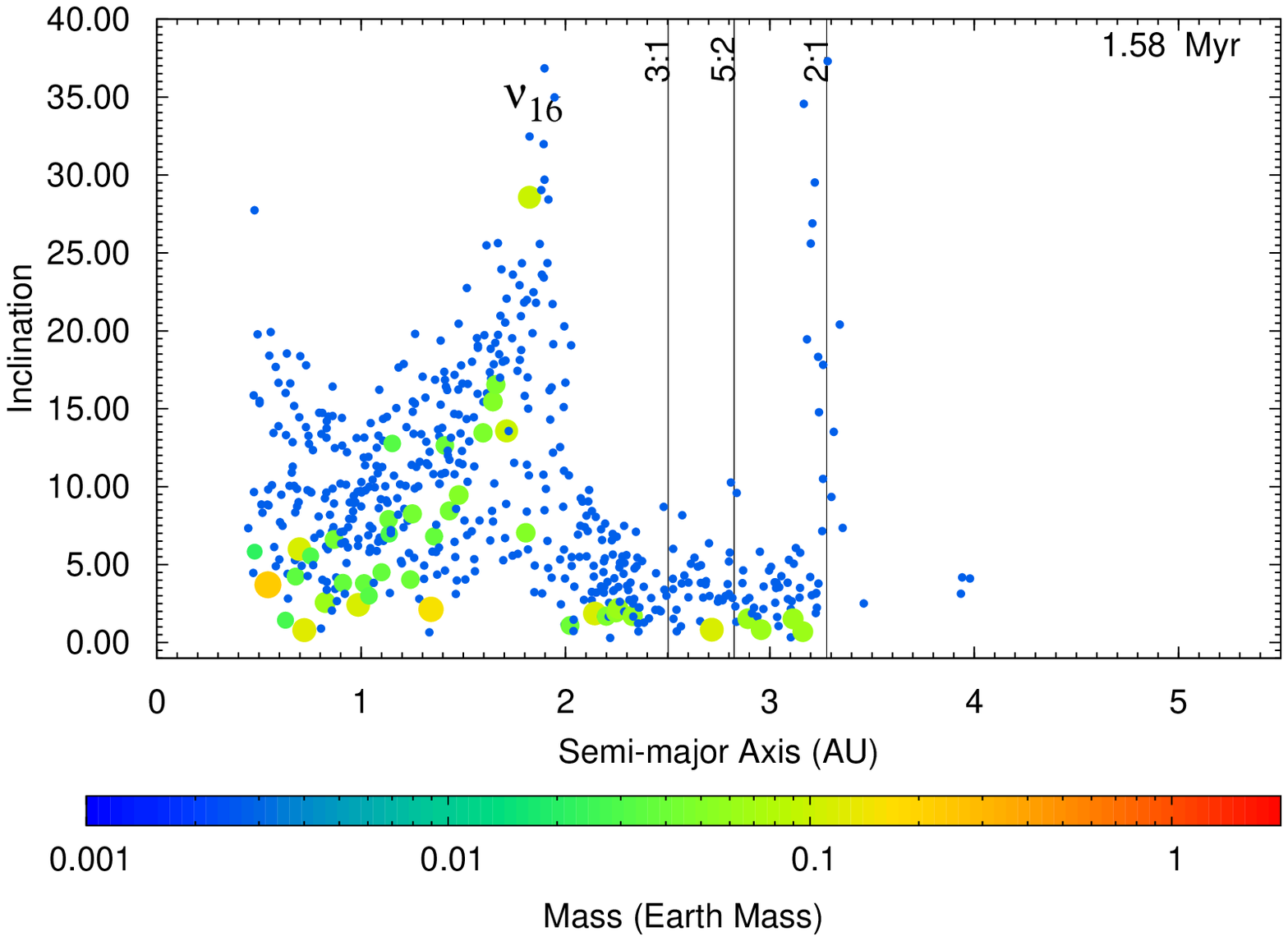}
\vskip -5pt
\includegraphics[width=9cm]{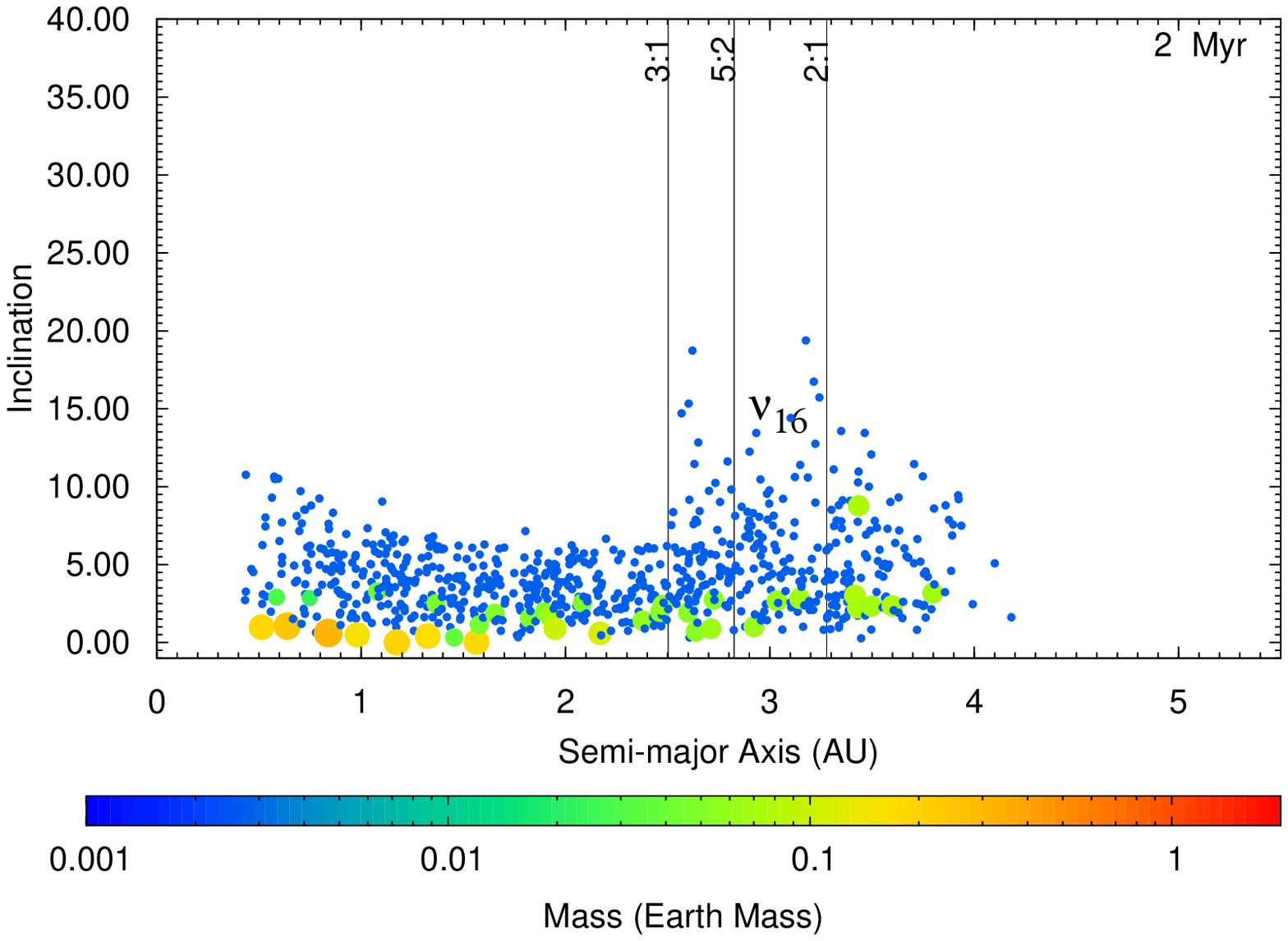}
}
\caption{Snapshots of the evolution of a disk of planetesimals and planetary embryos in a system
where the disk's surface density profile is proportion to $r^{-1.5}$. From top to bottom, the initial
orbits of Jupiter and Saturn are taken to be similar to their current orbits, have an eccentricity of 0.1,
and be similar to those in the Nice model. The time of each panel corresponds to when the effect of 
$\nu_{16}$ was first enhanced.} 
\end{figure}

Figures 10 and 11 show similar results for the inward displacement of $\nu_{16}$ resonance. 
The left panels in Figure 10 correspond 
to disks with $\alpha=0.5, 1$ and 1.5. As shown in this figure, $\nu_{16}$ appears at $\sim$ 1.5 AU, 
1.7 AU and 1.8 AU, respectively, which are interior to its test-particle-only position at 1.9 AU (Figure 3). 
The right panels in Figure 10 show the state of the same disks without the effects of the giant
planets. An important result shown here is that even when giant planets are not included in the simulations,
the mutual interactions among the disk bodies can increase the orbital inclinations of these objects.

Figure 11 shows the inward migration of the $\nu_{16}$ resonance in a disk with $\alpha=1.5$
and for three initial orbital configurations of giant planets. The inward displacement of $\nu_{16}$ 
compared to its test-particle-only position is clearly evident. For more details, we invite the 
reader to see the animations of our simulations in the electronic supplementary materials.

The results of our simulations indicate that the rate of the inward displacement of secular resonances
depends on the amount of the mass that is lost from the disk's 
outer region. In a disk with less steep surface density profile, outer embryos ($>1$ AU) are much 
larger and when one embryo is scattered out, the disk loses a considerable amount of mass. 
As a result, in such disks, the $\nu_6$ and $\nu_{16}$ resonances move inward in a very short time. 
For instance, in some of our simulations with a disk with an $r^{-0.5}$ profile,
the location of $\nu_6$ shifted from $\sim 2.3$ AU to 1.8 AU in only a few hundred thousand years. 
In contrast, when the disk in the same system was replaced with one with an $r^{-1.5}$ 
profile, this time increased to over 1 Myr.

\vskip 20pt
 
{\bf \subsubsection{Scattering into Terrestrial Planets Region}}

The displacement of $\nu_6$ and $\nu_{16}$ during the evolution of the system causes the disk to lose
most of its material in the region swept by these resonances.
In our simulations, this region was from 2.5 AU to 1.8 AU. 
Because we are interested in the contribution of these resonances to the scattering of material into
the terrestrial planet region, 
we calculated the amount of the mass that was scattered from this region (i.e. 2.5 - 1.8 AU) into the region 
interior to 1.5 AU. Table 1 shows the results for our three disk models and three initial orbits of giant planets. 
The columns in this table correspond to the amount of the mass that
was scattered into the region interior to 1.5 AU as a percentage of the final mass in that region.
As shown here, the amount of the scattered mass 
becomes larger as the disk's surface density profile becomes less steep. 
While this can be attributed to the large sizes of planetary embryos in such disks,
it must be noted that the scattering of  material is not solely due to the effect of the $\nu_6$ and $\nu_{16}$
resonances. The mutual interactions of planetary embryos also contribute. This can be seen in
the bottom row of Table 1 where similar pattern has been reported for simulations with no 
giant planets (implying that the inward scattering has been solely due to embryo-embryo interaction).
A comparison between the values of the mass scattered into 1.5 AU in this case with those in cases where 
giant planets exist indicates that $\nu_6$ and $\nu_{16}$ are much more effective in {\it removing} material than 
contributing to their radial mixing. In other words, \textit {radial mixing is primarily driven by
the interactions between planetary embryos}. Similar results have also been reported by Quintana and Lissauer (2014)
who ran lower resolution simulations of terrestrial planet formation.

\begin{table}
\centering
\vskip 35pt
 \caption{Amount of the mass scattered from the region 2.5 - 1.8 AU into the region 0.5 - 1.5 AU.
The values in each column show the amount of scattered mass as a percentage of the final mass of the disk
at 0.5 - 1.5 AU.}
  \begin{tabular}{lccccc}
  \hline
 System       &  $\alpha=1.5$ && $\alpha=1$  && $\alpha=0.5$ \\
  \hline
Jupiter/Saturn current orbits &  1.41  & &  6.92 && 8.81   \\
Jupiter/Saturn $(e=0.1)$ & 0.31   && 1.95   & & 5.26  \\
No giant planet        & 2.16  & & 7.20   &&  11.27     \\
\hline
\end{tabular}
\label{tab5}
\end{table}

\vskip 20pt
\section{Secular Resonances in Non-Homogeneous Disks}

In this section, we study the effects of secular resonances in a non-homogeneous disk of protoplanetary
bodies. As we are interested in the contribution of these resonances to the process of the formation of Mars 
and other terrestrial planets, we use the results of the simulations by Izidoro et al (2014), and study the 
effects of $\nu_5$, $\nu_6$ and $\nu_{16}$ resonances in a sub-set of those simulations where a Mars-analog formed. 

We begin by considering a protoplanetary disk extending from 0.5 AU to 4 AU, with a bi-modal 
distribution of planetesimals and planetary embryos similar to that in section 2.1. Following Izidoro et al (2014),
we divide the total mass of the disk equally between the total mass of planetesimals 
and the total mass of planetary embryos. 
We introduce non-uniformities to the disk in the form of local mass depletion 
with the surface density of the disk given by

\begin{equation}
\Sigma(r)=
\left\{
\begin{array}{lll}
8\,({\rm g\> cm^{-2}})\,({r}/{1\, {\rm AU}})^{-3/2} \qquad\qquad\quad; 
\hspace{0.4cm} {\rm outside\,\, the\,\, depleted\,\, region,}\\  \\ 
8\,({\rm g\> cm^{-2}})\, (1-\beta) \, ({r}/{1\, {\rm AU}})^{-3/2 } \qquad; 
\hspace{.3cm} {\rm inside\,\, the\,\, depleted\,\, region.}
\end{array}
\right.
\end{equation}

\vskip 5pt
\noindent
The quantity $0\leq \beta \leq 1$ in equation (2) represents the fraction of the initial
mass that is removed from the disk in order to create a non-homogeneity.  
Each individual planetesimal is assigned a mass equal to 0.0025 Earth-masses and, similar to the disk in 
section 2.1, the masses of the planetary embryos are set to scale with the number of 
their mutual Hill radii $(\Delta)$ as $M\sim r^{3/4}\Delta^{3/2}$ 
(Kokubo \& Ida 2000; Raymond et al. 2005, 2009; Izidoro et al 2013, 2014). 
Planetary embryos are spaced from one another at distances of 3 to 6 mutual Hill radii, and
the embryo-to-planetesimal mass-ratio is taken to be $\sim 8$ around 1.5 AU (Raymond et al. 2009).

{\bf \subsection{Numerical Simulations and Results}}

In Izidoro et al (2014), we carried out 84 simulations considering different initial distributions for
planetesimals and planetary embryos, different locations and spatial extents for the disk's
mass-depleted region, and two sets of initial orbital configurations for Jupiter and Saturn 
(their current orbits and the Nice model). Similar to the 
simulations presented in section 2, the initial orbital inclinations 
of all bodies were chosen randomly from the range of $10^{-4}$ to $10^{-3}$ deg., and their mean
anomalies 
were taken to be between 0 and $360^\circ$. The arguments of periastrons and longitudes of 
ascending nodes of all objects were initially set to zero. The time-steps of simulations were set to 6 days.

In general, simulations by Izidoro et al (2014) pointed to two distinct results.
\vskip 2pt
\noindent
1) Similar to previous studies (O'Brien et al. 2006; Raymond et al. 2009), when giant planets were initially in 
circular orbits, simulations were systematically unsuccessful in forming planetary systems with a Mars-analog 
at 1.5 AU. These results suggested that if Jupiter and Saturn were initially in circular orbits, as in the Nice 
model (Tsiganis et al. 2005) or the model by Levison et al. (2011), a disk with a local mass-depletion 
may be able to form a Mars-sized planet around 1.5 AU only if an additional mechanism removes material from 
the disk in the region of the asteroid belt.

\vskip 5pt
\noindent
2) When Jupiter and Saturn were initially in their current orbits, a significant portion
of the disk material 
was removed from its outer regions causing less material to be scattered
into the mass-depleted region. The latter created a favorable condition for Mars-analogs to form
around 1.5 AU especially when the depletion factor was between 50\% and 75\%. 
In these simulations, Mars formed as an embryo in a massive part of the disk and was scattered
into the mass-depleted region. These results suggested that systems in which Jupiter and Saturn were initially
in orbits with low to moderate eccentricities, have a better chance for forming Mars.

In the rest of this section, we explore the effect of the secular resonances of Jupiter and Saturn in both cases 
above.

\vskip 20pt
{\bf \subsubsection{Locations of $\nu_5$, $\nu_6$, and $\nu_{16}$ Resonances}}

As explained in section 2, to obtain the precise locations of secular resonances, the evolution of the
system has to be numerically simulated and  
the mass of the disk and its interaction with giant planets have to be taken into account.
Figure 12 shows the locations of $\nu_5$, $\nu_6$, and $\nu_{16}$ in three
simulations from Izidoro et al (2014) in which a Mars-analog successfully formed. The region of mass-depletion
in these simulations extents from 1.3 AU to 2.0 AU, and the amount of mass removed from this region (i.e., the
quantity $\beta$ in equation 2) is equal to 50\% in the top panel and 75\% in the middle and bottom panels.
The times of 
the panels on the left correspond to when the ${\nu_6}$ resonance was first enhanced. The times on the right
panels correspond to the first enhancement of the $\nu_{16}$ resonance. 
A comparison between these results and the results shown in the top panels of Figures 5 and 11, where the 
disk surface density profile is proportional to $r^{-1.5}$ and Jupiter and Saturn were initially in their current
orbits, indicates that despite the lower mass of 
the disk in these simulations (see equation 2), the locations of $\nu_5$, $\nu_6$ and $\nu_{16}$ resonances 
do not seem to have changed significantly. This can also be seen in Figure 13 where we show the 
root-mean-squares of the eccentricities and inclinations of planetesimals and planetary embryos 
for the three systems of Figure 12.
The left panels of the figure on the top show that the effect of $\nu_6$ resonance is more pronounced 
in the region between 2 AU and 2.4 AU. This is similar to the results shown in the top panel of Figure 6.
The bottom figure shows the small effect of $\nu_{16}$ resonance appearing in the region of 1.3 AU
to 2 AU. This is also similar to the results shown in the top panel of figure 7.

The above-mentioned similarities between the dynamical excitations of disks with and without non-uniformities
suggest that the inhomogeneities that facilitate the formation of Mars do not seem to have significant effects on 
the location and evolution of secular resonances. This is mainly due to the fact that in general, the amount
of the mass that is locally removed from the disk (i.e., $\beta$) is much smaller
than the total mass of the disk and as a result its effect in changing the locations of secular resonances 
is considerably small.

\vskip 20pt
{\bf \subsubsection{Disk-Mass Removal}}

Given that in a non-homogeneous, Mars-forming disk, secular resonances appear to be almost at the same 
location as those in a disk with no non-uniformity, and show similar dynamical behavior, it would be important to
determine to what extent the removal of the disk mass by these resonances affects the process
of Mars formation and radial mixing among planetesimals and protoplanetary bodies.
We use, as examples, the three systems of Figure 12. However, the results of our analysis are general
and can be extended to all Mars-forming simulations in Izidoro et al (2014).

As expected, similar to the classical model (section 2.2.3), the $\nu_5$ resonance does not have a
significant contribution in removing mass from its surrounding and the radial mixing of
disk material. The scattering of objects in the region around this resonance is primarily
due to their interactions with planetary embryos (Bottke et al 2006; Haghighipour and Scott 2012).
Results of our simulations show that as Earth- and Venus-analogs form and accrete protoplanetary bodies,
the effect of this resonance becomes weaker till it finally disappears
when Earth and Venus are fully formed.

The $\nu_6$ and $\nu_{16}$ resonances, on the other hand, play important roles in removing material 
and scattering of planetary embryos to other parts of the disk. Figure 14 shows the evolution of the
mass of the disk in the three systems of Figure 12. As shown here, during the evolution of the disk,
the mass of the depleted region (shown in purple) does not change considerably (it slightly increases
as some of the material in its outer region are scattered into it). The overall mass of the disk (shown in green), 
however, decreases drastically as the disk loses material from the region of the asteroid belt. This
is an important process that is necessary to ensure that too much material would not be scattered 
into the disk's mass-depleted region so that Mars can maintain its low mass as well as its long-term orbital
stability. It is important to note that in systems where the location of the mass-depletion coincides
with the nominal locations of $\nu_6$ and $\nu_{16}$ resonances, the absence of mass will have significant
effects on the orbital precession of bodies near these resonances. We will discuss these effects in future
studies.

\vskip 20pt
\noindent
{\bf PART 2: PLANETS PROPERTIES AND DISK SURFACE DENSITY}
\vskip 10pt
\noindent
In this part of the paper, we present the results of our study of the final properties
of terrestrial planets for disks with different surface density profiles. Because in models with initially non-uniform and
partially depleted disks, the subsequent radial mixing of planetesimals and planetary embryos 
after the first few million years of the evolution of the system causes the disk to behave similar to disks in the
classical model, and

\clearpage

\begin{figure}
\centering{
\vskip 50pt
\includegraphics[scale=0.34]{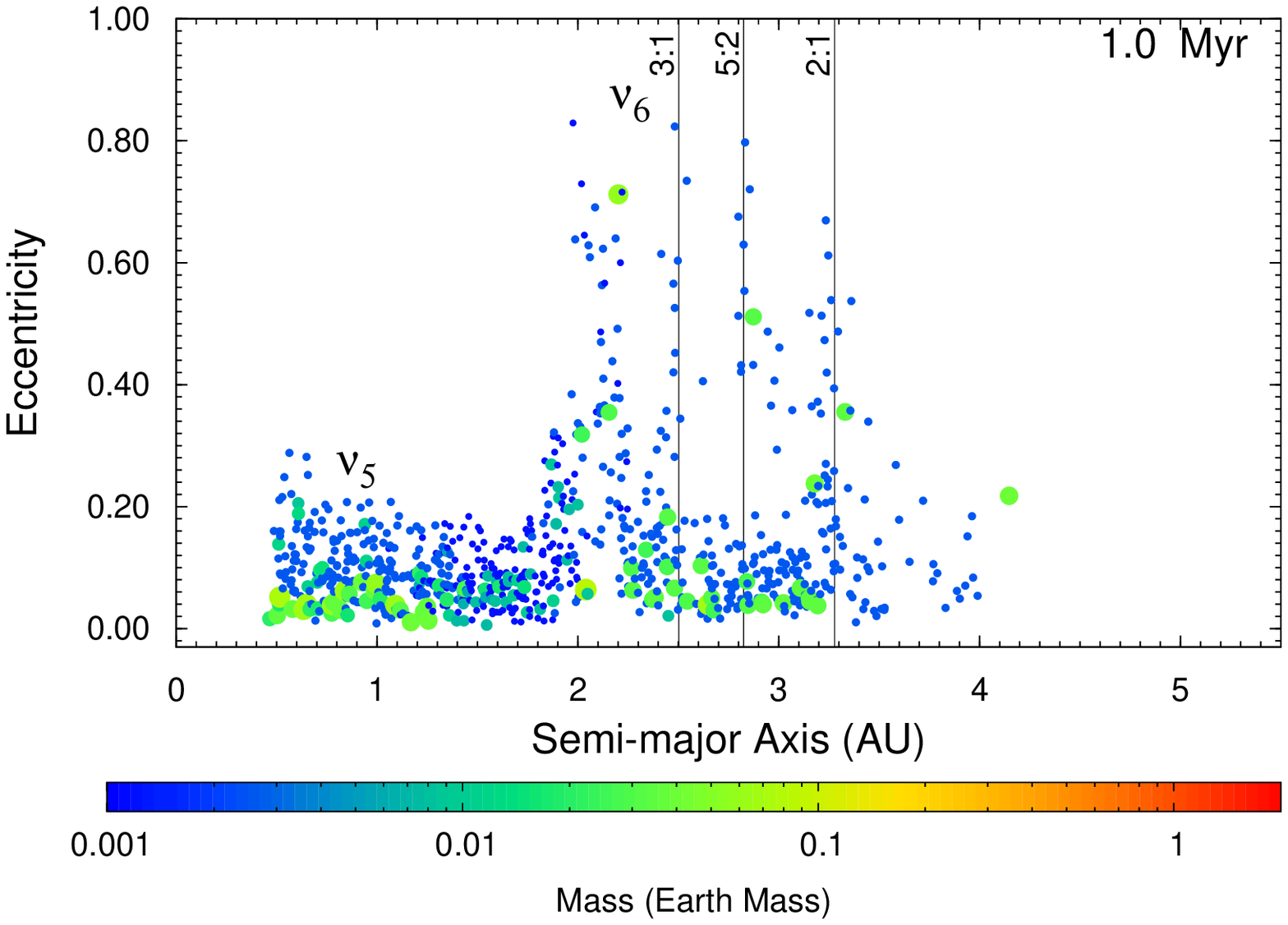}
\includegraphics[scale=0.34]{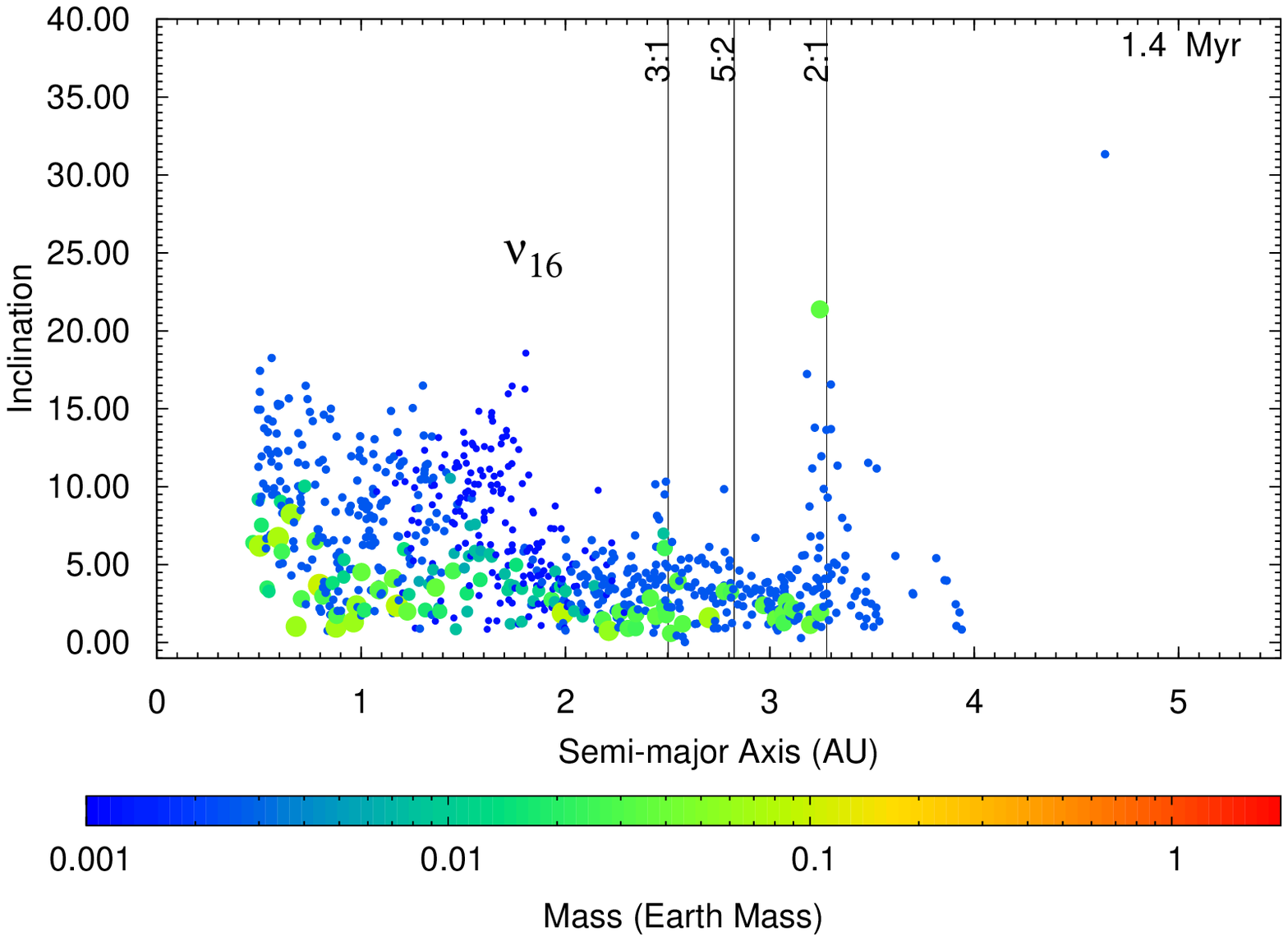}
\vskip 5pt
\includegraphics[scale=0.34]{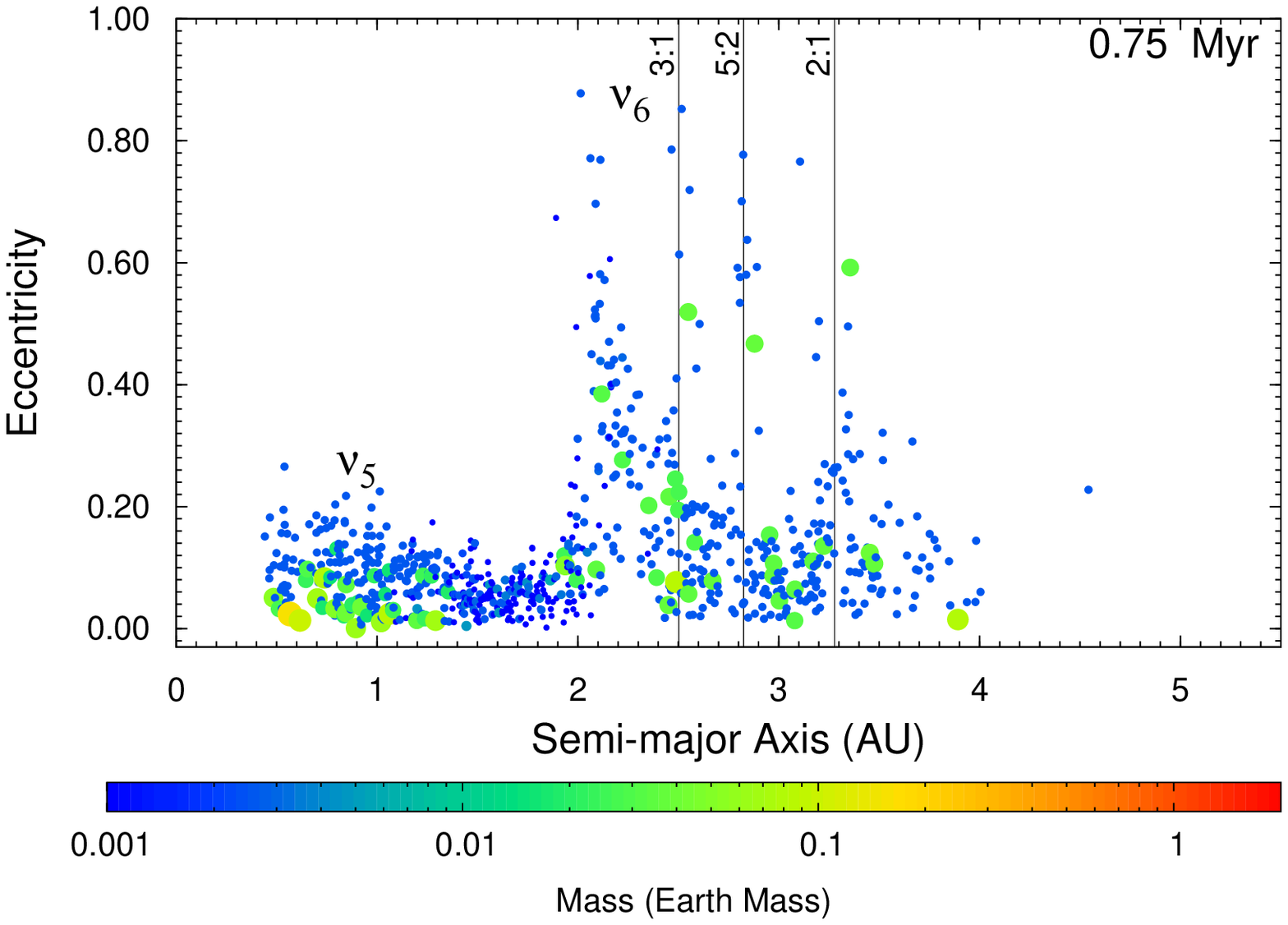}
\includegraphics[scale=0.34]{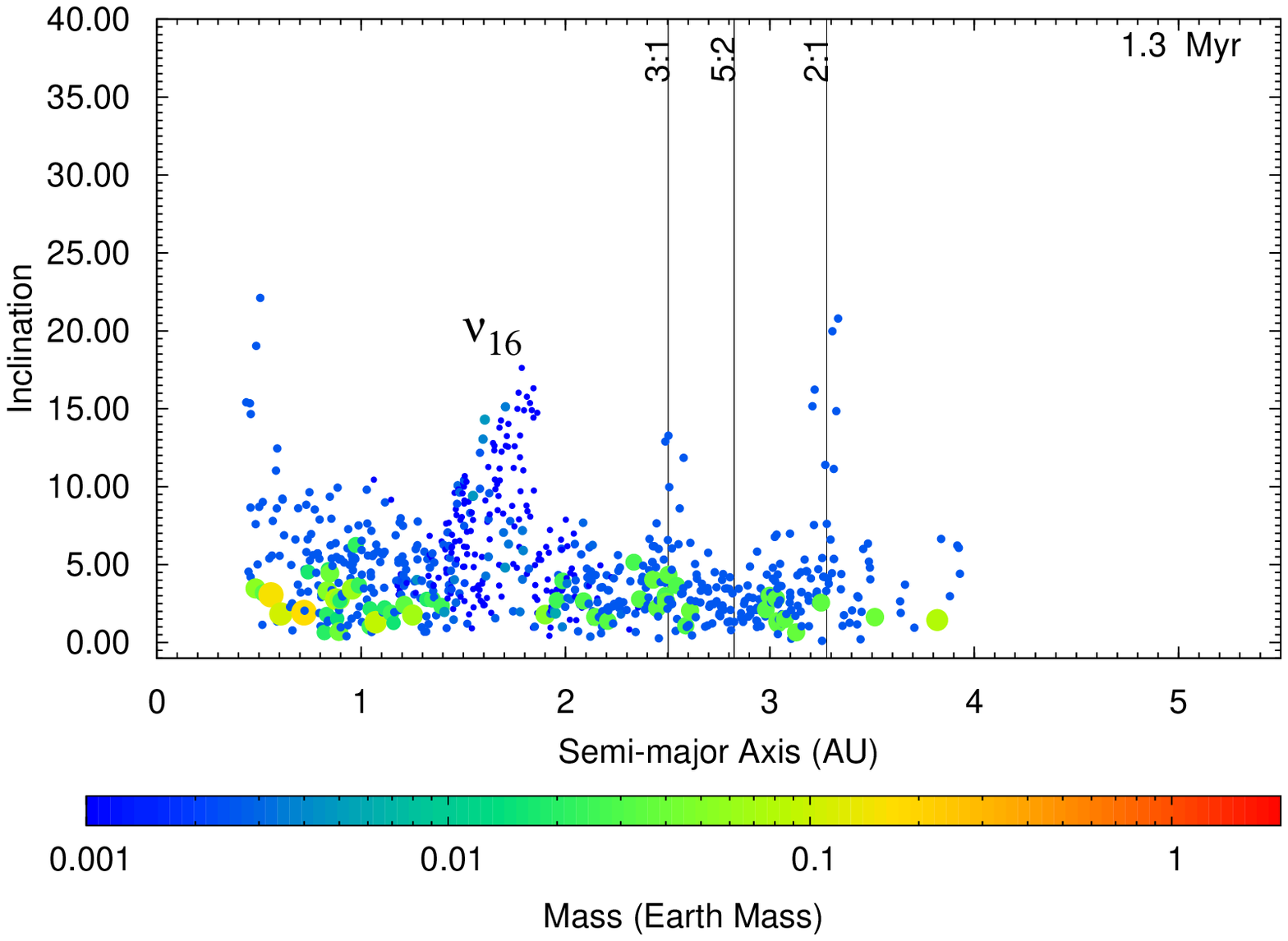}
\vskip 5pt
\includegraphics[scale=0.34]{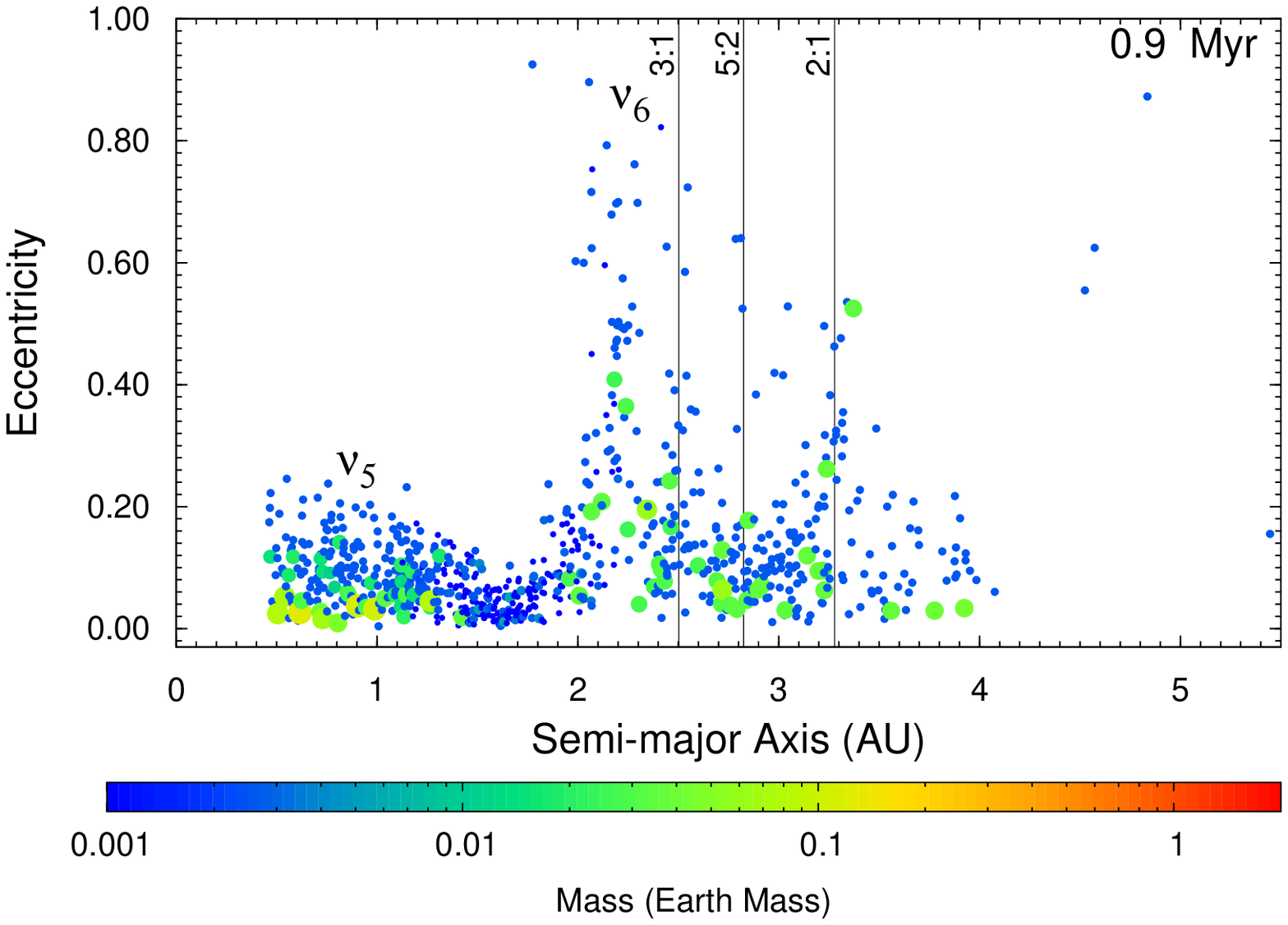}
\includegraphics[scale=0.34]{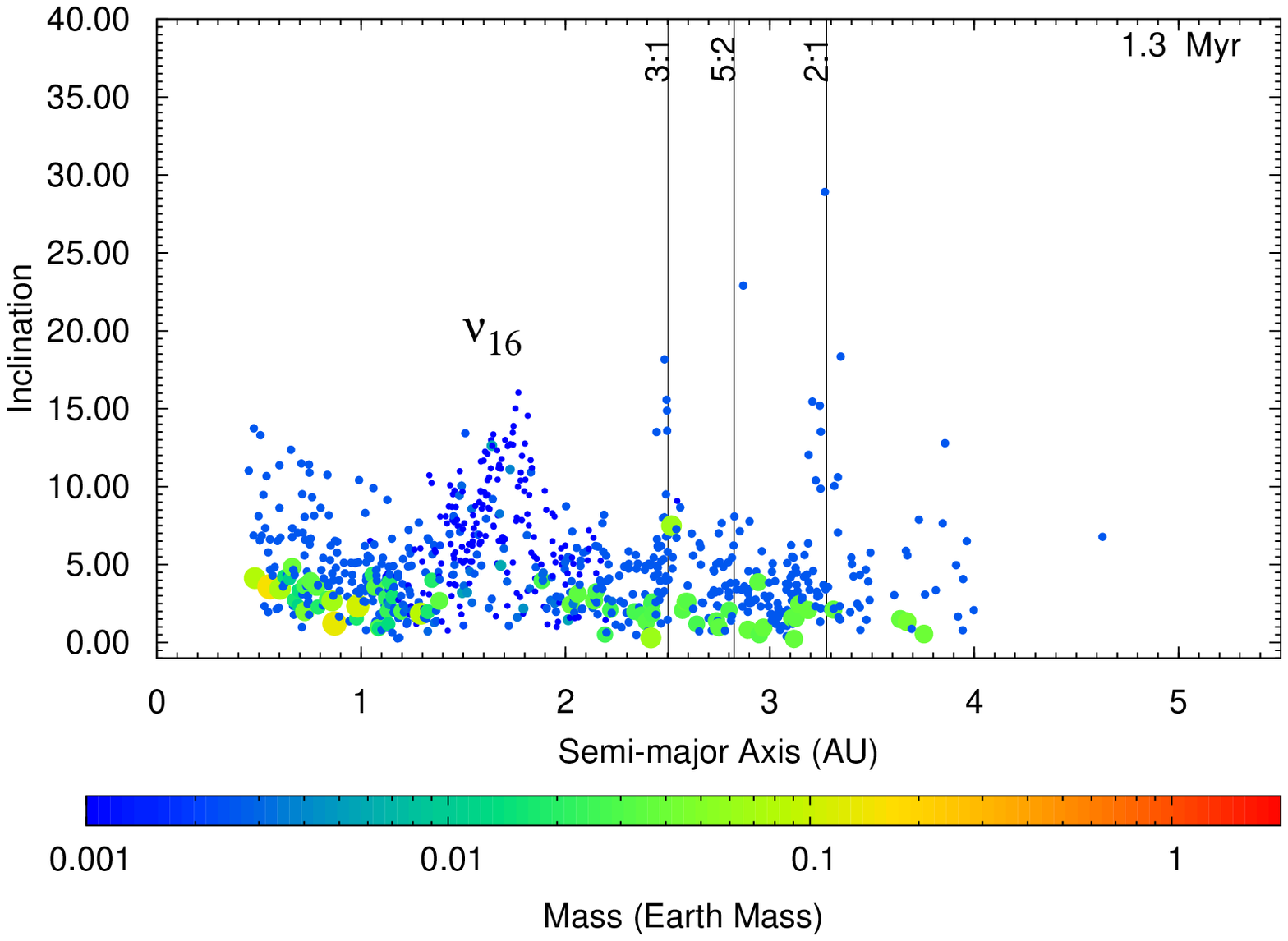}
}
\caption{Graphs of the the locations of $\nu_5$, $\nu_6$, and $\nu_{16}$ in three
simulations from Izidoro et al (2014) in which a Mars-analog successfully formed. The region of mass-depletion
in these simulations extends from 1.3 AU to 2.0 AU. The amount of mass removed from this region is equal to 50\% 
in the top panel and 75\% in the middle and bottom panels. The times of 
the panels on the left correspond to when the ${\nu_6}$ resonance was first enhanced. The times on the right
panels correspond to the first enhancement of the $\nu_{16}$ resonance.}
\end{figure}

\begin{figure}
\centering{
\vskip 50pt
\includegraphics[width=10cm]{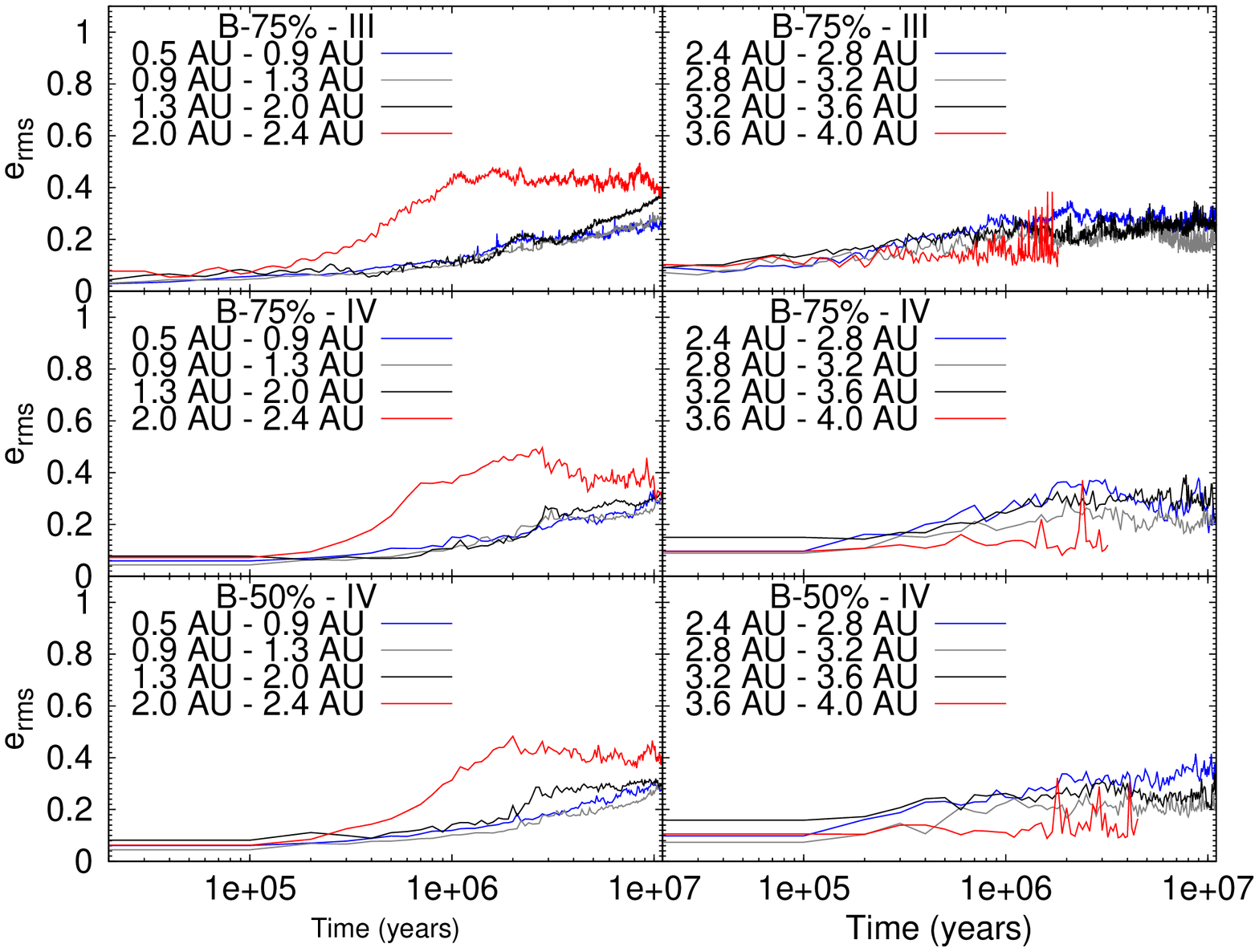}
\vskip 15pt
\includegraphics[width=10cm]{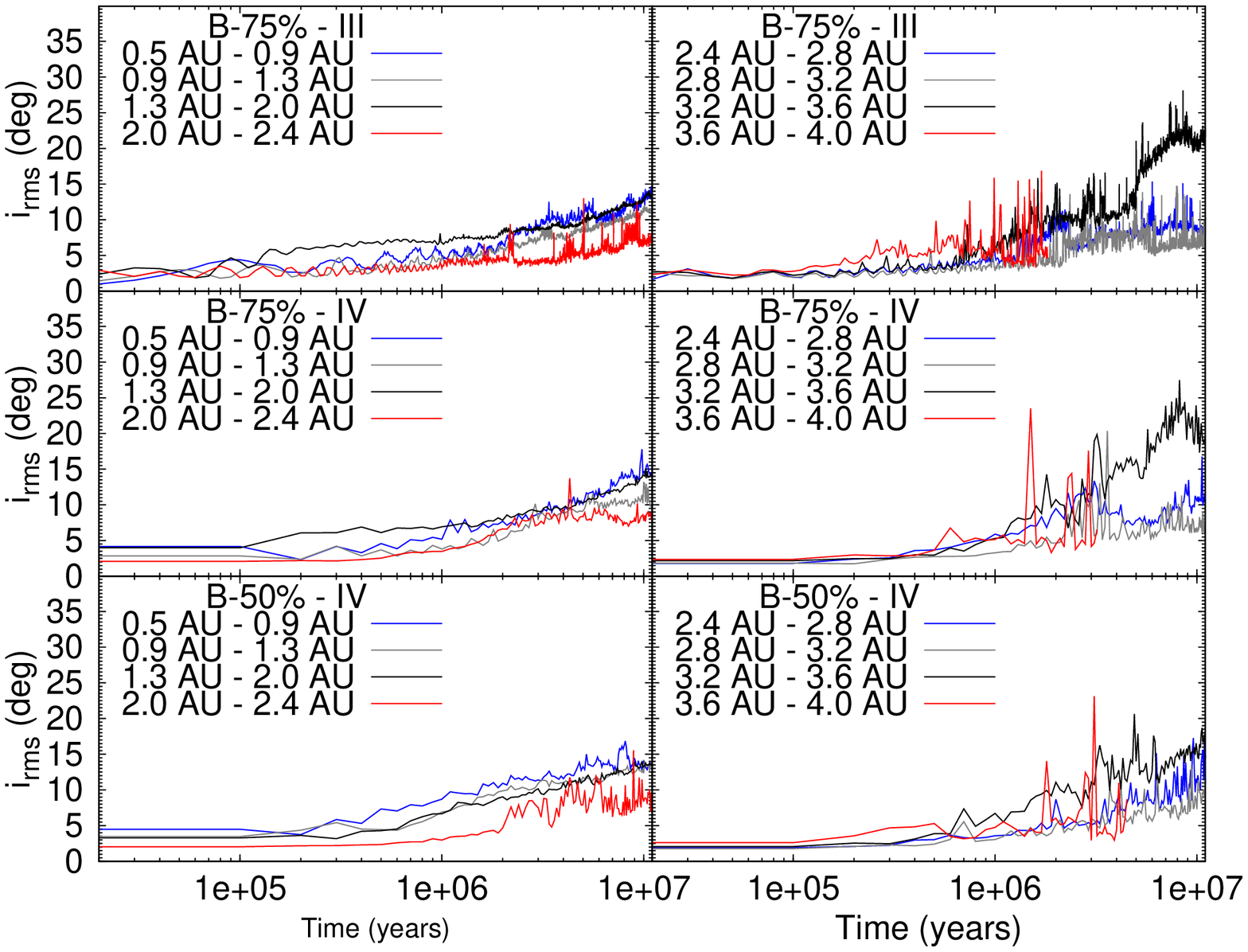}
}
\caption{Graphs of the root-mean-squares of the eccentricities and inclinations of
planetesimals and planetary embryos in the simulations of Figure 12. A comparison between
these results and those shown in Figures 6 and 7 indicate that despite the lower mass of
the disk in these simulations, the locations of $\nu_6$ and $\nu_{16}$ resonances did not change
significantly.} 
\end{figure}

\clearpage

\begin{figure}
\centering{
\includegraphics[width=12.5cm]{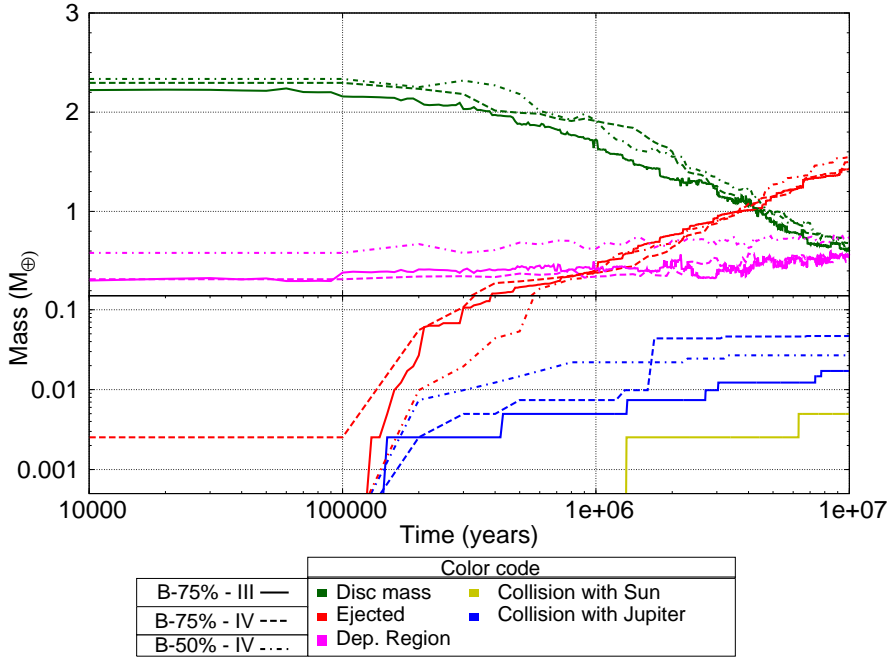}
}
\caption{Graph of the evolution of the mass of the disk in the simulations of Figure 12.
The green curve shows the decrease in the mass of the disk,
and red, yellow and blue correspond to the amount of the mass that is lost due to the ejection from the system,
collision with the Sun, and collision with Jupiter, respectively. The color purple shows the mass of the
disk in the depleted area (1.3-2.0 AU).
Note that on the vertical axis, the scale is logarithmic for the values smaller than 1, and is linear 
for higher values.}
\end{figure}

\noindent
because the time of the formation of terrestrial planets is an order of magnitude longer,
we ignore the non-uniformities and focus our study on the disks in the classical model. In section 4, we present 
a detailed discussion of the final properties of planets and their connection to the initial surface density of the 
protoplanetary disk.

\section{Properties of the Final Planetary Systems}

As mentioned in the introduction, the properties of the final planetary systems, in addition to the effect
of giant planets (i.e., mean-motion and secular resonances), depend also on the spatial distribution of
planetesimals and planetary embryos. Raymond et al (2005b) and Quintana and Lissauer (2014) 
studied these correlations by simulating 
the late stage of terrestrial planet formation in disks with different surface density profiles and in the 
context of the classical model. Considering systems with equal-mass disks of 10 Earth-masses and a Jupiter-mass
planet in a circular orbit at 5.5 AU, Raymond et al (2005b) found that terrestrial planet formation occurs in  
shorter times in disks with steeper surface density profiles (high values of $\alpha$). 
In these systems, the number of the final 
planets is larger, they are more massive, and they have shorter orbital periods and lower water-mass fractions.
Quintana and Lissauer (2014) performed similar simulations but with lower resolution, and showed that 
although a planet in the orbit of Jupiter can play an important role in the final assembly of terrestrial planets, 
its presence it not necessary to ensure radial mixing among the disk objects.

Raymond et al (2005b) and Quintana and Lissauer (2014) carried out their studies on disks with similar masses, and considered only
one giant planet. However, as shown in section 3, the dynamics of planetesimals and planetary embryos
is strongly affected by their mutual interactions as well as the perturbation of Saturn and the initial orbital
configuration of the two giant planets. It would, therefore, 
be important to examine how the results and properties of the final planets change with
the mass of the disk and the number of giant planets. 

We recall that, as 
mentioned in section 2.1, our simulations of the classical model included three different models of the 
protoplanetary disk, each with a different surface density profile. The surface densities of these disks were chosen
such that their values at 1 AU would be equal to 8 g/cm$^2$. As a result, each disk carried a different amount of mass
(see section 2.1). 
In the rest of this section, we refer to these disks as disk model 1 (i.e., identical values at 1 AU, but different 
total masses). We will also consider a second set of disks (hereafter referred to as disk model 2) 
where the total
masses of disks are identical and equal to 7 Earth-masses, however, their values at 1 AU will be different.
For disk radial profiles with $\alpha = 0.5, 1$ and 1.5, this value of disk mass corresponds to a 
surface density of 5.8, 8.5 and 11.5 g/cm$^2$  at 1 AU.

In order to be able to compare our results with those of Raymond et al (2005b) and Quintana and Lissauer (2014), 
we carried out simulations with 
one and two giant planets. When using disk model 1, we first simulated the system with only one Jupiter-mass planet
and then repeated the simulations with both Jupiter and Saturn. With disk model 2, however, we only considered
two giant planets. Similar to the simulations in the previous
section, we considered planets to be either in their current orbits, have an initial eccentricity of 0.1, or be in orbits
similar to those in the Nice model. To determine the water contents of the final
planets, we considered the embryos and planetesimals inside 2 AU to be dry, those between 2 AU and 2.5 AU to carry
1\% water, and beyond 2.5 AU, the water content of the disk material to be 5\% (Raymond et al. 2004, Izidoro et al. 2014). 

Simulations were carried out for 300 Myr. Table 2 summarizes the results. From left to right the columns in this 
table represent 
giant planets' initial configurations, disk surface density profile and its value at 1 AU, 
mean number of the final planets (we define a planet as an object with a mass larger 0.05 Earth-masses 
and with a semimajor axis smaller than 2 AU), mean value of the semimajor axis of the innermost planet, 
mean number of large planets, mean value of 
planetary mass relative to that of the disk, mean value of the mass of the largest planet, 
mean value of the time of the formation
of the final planetary system, mean value of the water-mass fraction in the system, mean orbital eccentricity,
and the number of Mars-analogs (we define a Mars-analog as a planet with a mass of 1--3 Mars-masses
formed between 1.3 AU and 2 AU).
The quantities shown by an asterisk refer to their corresponding values for planets interior to 1 AU.

In the following, we present an analysis of these results and compare any possible trend with those reported by 
Raymond et al (2005b). Table 3 shows a summary of these comparisons where a check mark indicates compatibility.
Following trends suggested by Raymond et al (2005b), from left to right the columns in this table refer to 
giant planet(s) configuration,
whether simulations formed a large number of planets, if they formed in a short time, whether they are close
to their host stars, if they are massive, and whether they have large water contents.

{\bf \subsection{Correlation Between the Surface Density of the Disk and the Number, Location, and Mass of the Final Planets}}

Although our results showed very good agreement with those of Quintana and Lissauer (2014) on the number
of the inner planets, we did not find a correlation between the mean number of the final planets 
and the disk surface density profile or the orbital configuration of Jupiter and Saturn
in any of the simulations of disk models 1 and 2, either with one or two giant planets.
For instance, as suggested by Raymond et al (2005b), in simulations with only one giant planet,
more planets were expected to form for higher values of $\alpha$. However, we did not see such a trend
in our simulations. No correlation was observed with the properties of the innermost planet as well
(Table 2).

Raymond et al (2005b) also suggested that disks with steeper density profiles tend to produce planets closer 
to the star. In our simulations, the disk model 2 seemed to show a slight correlation of this kind.
However, this trend breaks down in simulations using the disk model 1. In fact the latter
simulations suggest that the properties of the final innermost planets vary with the total mass 
of the disk and the initial orbital configurations of giant planets. 

Our simulations also show that the total mass of the disk is 
a strong factor in determining the final masses of the planets. Results of the simulations with disk model 1 
show no trend between the masses of the planets and the value of $\alpha$ and instead suggest that
the most important factor affecting the masses of the final planets is the value of the disk
surface density at 1 AU. Only in simulations where disks with similar masses were used, 
the results show similar trends as those in Raymond et al (2005b).

\vskip 10pt
{\bf \subsection{Correlation Between the Surface Density of the Disk and the Time of the Formation of the Final Planets}}

Raymond  et al (2005b) suggested that in disks with steeper surface density profiles, 
planets form more quickly. However, such a pattern was not observed in our simulations of disk model 1
(where both the radial profile of the disk surface density and the disk's total mass vary)
independent of the giant planets' orbital configurations. This is a direct consequence of considering
disks with different total masses. Only when disks with similar masses were used, that is, same total
disk mass was considered for all radial profiles of the disk surface density (e.g., in disk model 2), 
simulations showed shorter time of planet formation for higher values of $\alpha$ as suggested by 
Raymond et al (2005b). For a given surface 
density profile, however, the time of formation was strongly influenced by the orbital architecture of 
giant planets.

\begin{table}
\scriptsize
\renewcommand{\arraystretch}{1.4}
\setlength{\tabcolsep}{5pt}
\caption{Final results of the simulations of terrestrial planet formation in disks with different masses and surface density profiles.
From left to right the columns represent giant planet(s) configuration (Conf.),  slope of the density profile $(\alpha)$, 
surface density profile at 1 AU $(\Sigma_1)$, mean number of planets $(N)$, mean semimajor axis of the innermost planet $(a_{\rm min})$, 
mean value of the mass of the largest planet in Earth-masses $(M_{\rm L})$, mean ratio of the mass of planets to the mass of the disk
$({M_{\rm p}}/{M_{\rm disk}})$, mean value of planetary mass in Earth-masses $(M_{\rm p})$, mean value of the time of formation $(T_{\rm Form})$, 
mean value of water-mass fraction (WMF), mean value of planetary eccentricity $(e_{\rm p})$, 
Number of Mars analogs $(N_{\rm Mars})$,
mean number of planets interior to 1 AU $({N^*})$, 
mean value of the mass of the largest planet interior to 1 AU $({M_{\rm L}^\ast})$, Mean value of the time of formation interior to 1 AU $({T_{\rm Form}^\ast})$, 
and mean value of water-mass fraction interior to 1 AU (WMF$^\ast$). In the column for giant planets configuration, J=Jupiter, S=Saturn, C=current orbit, 
E=eccentric orbit $(e=0.1)$, and NM=Nice Model.}
\hskip -60pt
\begin{tabular}{@{}lccccccccccccccc@{}}
  \hline
  \\
Conf. &  $\alpha$       & $\Sigma_1$      & $\rm N$       & $a_{\rm min }$  &  $M_{\rm L} (M_{\oplus})$  & $M_{\rm p}/M_{\rm disk}$   &   ${M_{\rm p}} (M_{\oplus})$   & $T_{\rm Form}$ (Myr) & $\rm WMF$   & $e_{\rm p}$    &  $N_{\rm Mars}$ & $N^*$   &  $M_{\rm L}^\ast (M_{\oplus})$    &   ${T_{\rm Form}^\ast}$ (Myr) &   WMF$^\ast$ \\
\hline\hline
& & & & & & & Disk Model 1 & & & & & \\  
\hline\hline
 JCE  &   0.5  & 8      &   3.33 &          0.45  &       1.14   &      0.26 &   0.75 &        36.08  &           3.5373e-03  &      0.04 &   0 &      2.00   &         0.68   &      38.16  &      1.4088e-03  \\ 
JC    & 0.5    &  8     &   2.67 &          0.56  &       2.08   &      0.34  & 1.24 &       37.37  &           1.2700e-02  &      0.03 &    0 &      1.33   &         0.71   &      38.25  &      1.4729e-03  \\ 
 JNM   &  0.5    &   8    &   2.00 &          0.66  &       2.24   &      0.36  & 1.77 &     108.82  &           1.7613e-02  &     0.11 &    0 &      1.00   &         1.31   &     103.77  &      1.0834e-02  \\ 
  JSCE  &  0.5    & 8      &   2.33 &          0.63  &       1.29   &      0.21  & 0.82  &      32.76  &           8.0769e-03  &     0.07 &  1 &        1.67   &         1.02   &      35.20  &      3.5852e-04  \\ 
JSC    &  0.5     &   8    &   2.67 &          0.65  &       1.43   &      0.31  &  1.10  &    88.99  &           1.4957e-02  &    0.10 &    0 &      1.33   &         1.05   &     105.62  &      4.3333e-03  \\ 
JSNM    &   0.5   &   8    &   3.00 &          0.62  &       1.73   &       0.38  &   1.22&     82.01  &           1.5150e-02   &  0.05    &  0 &        1.33   &         0.80   &     101.16  &      3.1367e-03  \\ 
 \hline
JCE     & 1    &  8     &   2.33 &          0.66  &       1.31   &      0.32  & 0.92  &      42.46  &           3.3006e-04  &      0.14 &   0 &       1.00   &         1.31   &      54.01  &      3.0821e-04  \\ 
JC    &   1  & 8      &   3.00 &          0.59  &       1.11   &       0.38  &  0.85  &   69.60  &           7.8981e-03  &      0.07 &      0 &    1.67   &         0.87   &      41.42  &      3.1071e-03  \\ 
JNM    & 1    &  8     &   1.67 &          0.40  &       1.35   &      0.33  &  1.32  &     49.53  &           9.4547e-03  &      0.03 &    0 &      1.00   &         0.86   &      46.84  &      2.6023e-03  \\ 
 JSCE   &  1   &  8     &   2.67 &          0.55  &       0.90   &   0.25     &  0.62   &   37.13  &           6.6529e-05  &      0.14 &    1 &     1.67   &         0.74   &      37.41  &      9.6197e-06  \\ 
JSC    & 1    &  8     &   2.67 &          0.58  &       1.31   &   0.35    &   0.89  &   78.63  &           7.4890e-03  &      0.08 &      0 &    1.33   &         0.87   &      56.30  &      5.4693e-04  \\ 
JSNM    & 1    &  8     &   2.67 &          0.61  &       1.52   &   0.42   &  1.05  &    59.78  &           1.0986e-02  &      0.09 &      0 &    1.33   &         0.95   &      64.76  &      7.2642e-03  \\ 
    \hline
 JCE        & 1.5    &  8     &   3.00 &          0.55  &       1.16   &      0.42  &  0.69 &      38.83  &           4.3831e-04  &      0.08 &   2 &      1.67   &         0.83   &      46.38  &      5.4021e-04  \\ 
 JC   &   1.5  &   8    &   4.33 &          0.54  &       0.86   &  0.50      &   0.57  &       52.53  &           2.0525e-03  &      0.06 &      0 &    2.33   &         0.63   &      53.44  &      1.0514e-03  \\ 
  JNM   &  1.5   &  8     &   2.33 &          0.35  &       0.90   &       0.41  &  0.86    &   85.89  &           8.5457e-03  &      0.10 &      0 &    1.33   &         0.93   &      88.41  &      5.5836e-03  \\ 
 JSCE   &  1.5   &  8     &   2.00 &          0.63  &       1.05   &     0.33  &   0.81&     56.58  &           3.9374e-05  &      0.13 &      0 &    1.33   &         0.94   &      64.72  &      3.8087e-06  \\ 
JSC    & 1.5    & 8      &   2.67 &          0.61  &       1.36   &      0.42  &   0.78   &  78.34  &           4.2284e-04  &      0.08 &      1 &    1.00   &         1.19   &      95.85  &      8.1467e-04  \\ 
JSNM    & 1.5    &  8     &   2.67 &          0.57  &       1.51   &      0.60  &   1.10   &   96.61  &           6.6732e-03  &      0.12 &    0 &      1.33   &         0.95   &      97.12  &      4.9686e-03  \\ 
   \hline\hline
  
& & & & & & & Disk Model 2 & & & & & \\

   \hline \hline
JSCE & 0.5    & 5.8      &   2.67 &          0.59  &       0.85   &       0.22  &   0.58  &   98.29  &           2.7475e-03  &      0.14 &    1 &      1.33   &         0.69   &     124.66  &      2.2161e-03  \\ 
JSC    & 0.5    & 5.8       &   2.67 &          0.64  &       1.35   &      0.31  &  0.82   &   96.28  &           1.0064e-02  &      0.08 &    0 &      1.33   &         1.04   &     117.89  &      5.4368e-03  \\ 
 JSNM   & 0.5     &   5.8    &   2.67 &          0.72  &       1.20   &       0.35  & 0.9   &     90.39  &           1.7050e-02  &      0.08 &  0 &        1.00   &         1.02   &     120.67  &      1.0484e-02  \\  
    \hline
JSCE &  1   &  8.5     &   2.00 &          0.59  &       1.21   &      0.27  &  0.96  &     56.14  &           3.5222e-04  &      0.20 &    0 &      1.00   &         1.06   &      51.98  &      3.7514e-04  \\ 
JSC    &  1   & 8.5      &   3.00 &          0.53  &       1.17   &       0.38  &  0.86   &   62.29  &           3.3493e-03  &      0.05 &  0 &        1.67   &         0.88   &      42.95  &      1.4891e-03  \\ 
 JSNM   & 1    &   8.5    &   2.67 &          0.68  &       1.57   &      0.45  &   1.17  &    82.86  &           1.2246e-02  &      0.08 &  0 &        1.00   &         1.33   &      95.16  &      5.5143e-03  \\ 
\hline
 JSCE   &  1.5   &  11.5     &   3.00 &          0.55  &       1.20   &      0.39  &    0.92 &    33.25  &           5.2152e-04  &      0.09 &  1 &        1.67   &         1.09   &      33.49  &      4.4628e-05  \\ 
JSC    &  1.5   &  11.5     &   2.33 &          0.55  &       1.62   &       0.42  &   1.27  &   32.66  &           1.4428e-03  &      0.08 &    0 &      1.67   &         1.18   &      30.56  &      1.2641e-03  \\ 
 JSNM   & 1.5    &  11.5     &   3.00 &          0.57  &       1.88   &       0.54  &  1.27 &      41.36  &           1.0394e-02  &      0.14 &  0 &        1.33   &         1.58   &      45.09  &      4.1647e-03  \\     
    \hline\hline
    
\end{tabular}
\end{table}

\vskip 10pt
{\bf \subsection{Correlation Between the Surface Density of the Disk and Final Planets' Water Contents}}

Our simulations indicated an interesting trend in the water contents of the final planets.  
In disks with $\alpha$=1 and 1.5, the inclusion of Saturn resulted in dryer planets. For instance, as shown in Table 2, 
the mean water-mass fraction of the final planets in simulations using the disk model 1 with $\alpha=1.5$ and only Jupiter 
in an eccentric orbit is $\sim 4.38831\times 10^{-4}$. However, when in the same simulations, Saturn is 
also included and is considered to be initially 
in an eccentric orbit, the water-mass fraction reduces to $\sim 3.9375\times 10^{-5}$. This trend can be attributed to
the strong effect of the $\nu_6$ resonance which, as explained in the previous section, scatters objects from the outer
part of the disk (where bodies carry more water) to regions outside of terrestrial planets' accretion zones.

The effect of the $\nu_6$ resonance in scattering water-carrying objects becomes weaker
in disks with $\alpha=0.5$. In many cases, the final planets 
in these systems contained more water compared to those from disks with $\alpha=1$ and 1.5. Table 2 shows this in more detail.
As mentioned in section 2, disks with $\alpha=0.5$ are very massive in their outer parts. The mass of the disk
in this region is carried primarily by a few large embryos that have high water contents (5\% by mass). 
As these embryos are very
massive, their strong mutual interactions with other objects play an efficient role in the radial mixing of bodies,
countering the effect of $\nu_6$ resonance, and delivering more water to the inner part of the disk. 
This efficiency in water delivery can be observed even in systems where Jupiter and Saturn are initially in 
slightly eccentric orbits ($e=0.1$, see Table 1).
The water-mass fractions of the planets formed inside 1 AU in the latter systems were comparable to the expected 
value of Earth's water-mass fraction, $\sim 5 \times 10^{-4}$ (Raymond et al, 2009).

\begin{table}
\scriptsize
\renewcommand{\arraystretch}{1.4}
\setlength{\tabcolsep}{5pt}
\caption{Comparing our results with the trends reported by Raymond et al (2005b). The first column represents
the initial orbital configuration of the giant planets where J=Jupiter, S=Saturn, C=current orbit, E=eccentric
orbit $(e=0.1)$, and NM=Nice Model. The three columns under each item show the comparison of the results 
for two disks with different values of $\alpha$. The left column corresponds to a comparison between 
a disk with $\alpha$=0.5 and a disk with $\alpha$=1. The middle column is for a disk with $\alpha$=1 comparing
with a disk with $\alpha$=1.5, and the right column is for comparing a disk with $\alpha$=0.5 with
a disk with $\alpha$=1.5. A $\checkmark$ indicates that a trend observed in our simulations is similar to that
reported by Raymond et al (2005b) and a $\times$ means otherwise.
.}
   \centering 
\begin{tabular}{@{}lccccc@{}}
Conf. &   More numerous         &   Formed more quickly      & Formed closer to the star  &  More massive   
& Lower water content \\
\hline \hline
      &        &        & Disk Model 1  &  &  \\
 \hline \hline
JCE   & $\times$ $\checkmark$ $\times$ & $\times$ $\checkmark$ $\times$  & $\times$ $\checkmark$ $\times$ &$\checkmark$ $\times$ $\times$ & $\checkmark$ $\checkmark$ $\checkmark$ \\

JC    & $\checkmark$ $\checkmark$ $\checkmark$ & $\times$ $\checkmark$ $\times$  & $\times$ $\checkmark$ $\checkmark$ &$\times$ $\times$ $\times$ & $\checkmark$ $\checkmark$ $\checkmark$ \\

JNM    & $\times$ $\checkmark$ $\checkmark$ & $\checkmark$ $\times$ $\checkmark$  & $\checkmark$ $\checkmark$ $\checkmark$ &$\times$ $\times$ $\times$ & $\checkmark$ $\checkmark$ $\checkmark$ \\

JSCE    & $\checkmark$ $\times$ $\times$ & $\times$ $\times$ $\times$  & $\checkmark$ $\times$ $\times$ &$\times$ $\checkmark$ $\times$ & $\checkmark$ $\checkmark$ $\checkmark$ \\

JSC    & $\times$ $\times$ $\times$ & $\checkmark$ $\checkmark$ $\checkmark$ & $\checkmark$ $\times$ $\checkmark$ &$\times$ $\times$ $\times$ & $\checkmark$ $\checkmark$ $\checkmark$ \\

JSNM    & $\times$ $\times$ $\times$ & $\checkmark$ $\times$ $\times$ & $\checkmark$ $\checkmark$ $\checkmark$ & $\times$ $\checkmark$ $\times$ & $\checkmark$ $\checkmark$ $\checkmark$ \\ 
 \hline \hline
      &        &        & Disk Model 2  &  &  \\
 \hline \hline
JSCE    & $\times$ $\checkmark$ $\checkmark$ & $\checkmark$ $\checkmark$ $\checkmark$  & $\times$ $\checkmark$ $\checkmark$ & $\checkmark$ $\times$ $\checkmark$ & $\checkmark$ $\checkmark$ $\checkmark$ \\

JSC    & $\checkmark$ $\times$ $\times$ & $\checkmark$ $\checkmark$ $\checkmark$  & $\checkmark$ $\times$ $\checkmark$ & $\checkmark$ $\checkmark$ $\checkmark$ & $\checkmark$ $\checkmark$ $\checkmark$ \\ 
 
JSNM    & $\times$ $\checkmark$ $\checkmark$ & $\checkmark$ $\checkmark$ $\checkmark$  & $\checkmark$ $\checkmark$ $\checkmark$ & $\checkmark$ $\checkmark$ $\checkmark$ & $\checkmark$ $\checkmark$ $\checkmark$ \\ 
      
\hline
\end{tabular}
\end{table}

\vskip 20pt

\section{Conclusions}

We carried out extensive numerical simulations of the final stage of 
terrestrial planet formation in disks with different surface density profiles and for different initial
orbital configurations of Jupiter and Saturn. Simulations were carried out in the context of the classical 
model, and also for non-uniform protoplanetary disks as in the recent depleted-disk model of the formation Mars by
Izidoro et al (2014). Our goal was to determine the role that the secular resonances of giant planets,
and the mass and surface density of the disk play in the final assembly of terrestrial planets as well as their
physical properties. We were especially interested in these effects on the formation of Mars in partially 
depleted disks (Izidoro et al 2014).
Results of our simulations indicated that, irrespective of the non-uniformities in the disk, $\nu_5$ does not 
have significant effect on the dynamics of planetary embryos in its vicinity, whereas $\nu_6$ and $\nu_{16}$ play 
important roles in depleting their affecting areas, particularly in disks with less steep surface density profiles. 
Similar results were obtained in non-uniform disks that succeeded to form Mars as proposed by Izidoro et al (2014).
Our simulations showed that while in partially depleted, Mars-forming disks secular resonances do not alter the orbits 
of planetesimals and planetary embryos in the regions where Earth, Venus, and Mars form, their clearing effect in 
the asteroid belt becomes essential to ensuring that objects from this region
will not be scattered into the accretion zone of Mars so that Mars can maintain its mass and orbital
stability. 

Simulations also showed that although a small fraction of the material scattered by the $\nu_6$ resonance reaches
the terrestrial planet region, interactions among planetary embryos has a dominant effect on their radial mixing
and final mass and elemental contents of terrestrial planets. Similar results have also been reported by Quintana and Lissauer (2014).
Despite the latter, our results suggested that 
in disks with less steep surface density profile, where terrestrial planets may take longer 
to form and may be smaller, embryo scattering due to the $\nu_6$ resonance reaches closer distances and may account
for the higher water contents of terrestrial planets. In disks with surface density profiles proportional to
$r^{-1.5}$ (commonly used in models of terrestrial planet formation), $\nu_6$ causes final planets to be
drier compared to simulations in which only Jupiter is included (e.g., Raymond et al 2005b).

We also searched for trends between the orbital and physical properties of the final terrestrial
planets, and the initial mass and surface density of the protoplanetary disk, in the results of our long-term (300 Myr) simulations. 
Results suggested some correlations with the choices of the initial parameters. 
For instance, in all our simulations of disk models 1 and 2, where the latter is similar to 
the model adopted in Raymond et al (2005b), steeper surface density profiles tend to produce drier planets. 
Or, when Saturn was added to the simulations, the water contents of the final planets decreased. However,
in general, results of our simulations indicated that unlike previous studies (e.g., Raymond et al 2005b),
when integrations are considered to be more general (i.e., the mass of the disk,
its surface density profile, and the number and orbital configurations of giant planets are varied), in most cases
no specific trend seems to exist. We caution that the lack of correlation between the properties of the
final planets and the choices of the initial conditions in our simulations may disappear, and stronger correlations
may be found if more integrations are carried out for larger ranges of disk mass and surface density profile,
as well as different orbital placements of planetesimals and planetary embryos (we recall that in this study, we carried
out a total of 63 simulations). The latter integrations are currently in progress.

It is important to note that in this study, we assumed that giant planets had been fully
formed, and did not consider their effects on the dynamics of the disk during their formation.
We refer the reader to Haghighipour and Scott (2008, 2012) where these authors present the
results of a study on the effect of a forming giant planet in the orbit of Jupiter on the 
dynamics of planetesimals and planetary embryos. As shown by these authors, the perturbing
effect of the planet begins to appear when its mass exceeds
10 Earth-masses. Although we do not expect the results, as illustrated in this study, 
to change conceptually, we note that a more comprehensive
model needs to include the growth (and possible simultaneously migration) of giant planets, as well.

\begin{acknowledgements}

We are indebted to A. Izidoro for his help with making the figures and analysis of the results,
and for his invaluable comments that  greatly improved our manuscript. We are also thankful to the referees for
their constructive suggestions and recommendations. 
NH acknowledges support from the NASA ADAP program under grant NNX13AF20G, NASA PAST program under grant NNX14AJ38G, 
HST grant HST-GO-12548.06-A, and NASA Astrobiology Institute under Cooperative 
Agreement NNA09DA77A at the Institute for Astronomy, University of Hawaii.
OCW acknowledges support from FAPESP (PROC. 2011/08171-3) and CNPq (proc. 312813/2013-9).
Support for program HST-GO-12548.06-A was provided by NASA through a grant from the Space Telescope
Science Institute, which is operated by the Association of Universities for Research in Astronomy
Incorporated, under NASA grant NAS5-26555. 
NH would also like to thank the Alexander von Humboldt Foundation, and the Kavli Institute for Theoretical
Physics at the University of California-Santa Barbara (KITP) for their kind hospitality
during the final stage of this project. KITP visiting program is supported in part by the 
National Science Foundation under Grant No. NSF PHY11-25915.

\end{acknowledgements}


\begin{thebibliography}{}


\bibitem{Agnor99}
Agnor, C. B., Canup, R. M., and Levison, H. F.: On the Character and Consequences of Large 
Impacts in the Late Stage of Terrestrial Planet Formation, Icarus, {\bf 142}, 219-237 (1999)

\bibitem{Agnor912}
Agnor, C. B., and Lin, D. N. C.: On the Migration of Jupiter and Saturn: Constraints from
Linear Models of Secular Resonant Coupling with the Terrestrial Planets, ApJ, {\bf 745}, 143 (2012)

\bibitem{Bottke06}
Bottke, W. F., Nesvorn\'y, D., Grimm, R. E., Morbidelli, A., O'Brien, D. P.:
Iron meteorites as remnants of planetesimals formed in the terrestrial planet region,
Nature, {\bf 439}, 821-824 (2006)

\bibitem{Chambers98}
Chambers, J. E. and Wetherill, G. W.: 1998, Making the Terrestrial Planets: N-Body Integrations 
of Planetary Embryos in Three Dimensions, Icarus, {\bf 136}, 304-327 (1998)

\bibitem{Chambers99}
Chambers, J. E.: A Hybrid Symplectic Integrator that Permits Close Encounters Between Massive Bodies, 
MNRAS, {\bf 304}, 793-799 (1999)

\bibitem{Chambers01}
Chambers, J. E.: Making More Terrestrial Planets, Icarus, {\bf 152}, 205-224 (2001)

\bibitem{Chambers02}
Chambers, J. E. and Cassen, P.: The Effect of Surface Density Profile and Giant Planet Eccentricities 
on Planetary Accretion in the Inner Solar System, MAPS, {\bf 37}, 1523-1540 (2002)

\bibitem{Chambers14}
Chambers, J. E.: Forming Terrestrial Planets, Science, {\bf 344}, 479-480 (2014)

\bibitem{Dauphs11}
Dauphas, N. and Pourmand, A.: Hf-W-Th evidence for rapid growth of Mars and its status as a planetary 
embryo, Nature, {\bf 473}, 489-492 (2011)

\bibitem{Gomes05}
Gomes, R. S., Levison, H. F., Tsiganis, K. and Morbidelli, A.: Origin of the Cataclysmic Late 
Heavy Bombardment Period of the terrestrial Planets, Nature, {\bf 435}, 466-469 (2005)

\bibitem{Haghighipour08}
Haghighipour, N. and Scott, E. R. D.: Meteorite Constraints on the Early Stages of Planetary Growth in 
the Inner Solar System, Lunar Planet. Sci., {\bf39}, 1679 (2008)

\bibitem{Haghighipour12}
Haghighipour, N. and Scott, E. R. D.: On The Effect of Giant Planets on the Scattering of Parent Bodies 
of Iron Meteorite from the Terrestrial Planet Region into the Asteroid Belt: A Concept Study, ApJ, 
{\bf 749}, 113 (2012)

\bibitem{Hansen09}
Hansen, B. M. S.: Formation of the Terrestrial Planets from a Narrow Annulus, 
ApJ, {\bf 703}, 1131 (2009)

\bibitem{Izidoro13}
Izidoro, A., Torres, K. S., Winter, O. C., and Haghighipour, N.:
A Compound Model for the Origin of Earth's Water, ApJ, {\bf 767}, 54 (2013)

\bibitem{Izidoro14}
Izidoro, A., Haghighipour, N., Winter, O. C., and Tsuchida, M.: 
Terrestrial Planet Formation in a Protoplanetary Disk with a Local Mass Depletion: 
A Successful Scenario for the Formation of Mars, ApJ, {\bf 782}, 31 (2014)

\bibitem{Izidoro15}
Izidoro, A., Raymond, S. N., Morbidelli, A., and Winter, O. C.: Terrestrial Planet Formation 
Constrained by Mars and the Structure of the Asteroid Belt, MNRAS, {\bf 453}, 3620 (2015)

\bibitem{Izidoro16}
Izidoro, A., Haghighipour, N., Gomes, R. S., and Winter, O.C.: 
Simulating Terrestrial Planet Formation Using Mercury Integrator: A Bug Report, 
in preparation for submission to ApJS (2016)

\bibitem{Kaib15}
Kaib, N. A., and Cowan, N. B.: The Feeding Zones of Terrestrial Planets and Insight into Moon Formation,
Icarus, {\bf 252}, 161-174 (2015)

\bibitem{Kaib16}
Kaib, N. A., and Chambers, J. E.: The Fragility of the Terrestrial Planets During a Giant Planet Instability,
MNRAS, in press (arXiv:1510.08448) 

\bibitem{Kokubo98}
Kokubo, E., and Ida, S.: Oligarchic Growth of Protoplanets, Icarus, {\bf 131}, 171-178 (1998)

\bibitem{Kokubo00}
Kokubo, E., and Ida, S.: Formation of Protoplanets from Planetesimals in the Solar Nebula, 
Icarus, {\bf 143}, 15-27 (2000)

\bibitem{Levison03}
Levison, H. F. and Agnor, C.: The Role of Giant Planets in Terrestrial Planet Formation,
AJ, {\bf 125}, 2692-2713 (2003) 

\bibitem{Levison11}
Levison, H. F., Morbidelli, A., Tsiganis, K., Nesvorn\'y, D., and Gomes, R. : 
Late Orbital Instabilities in the Outer Planets Induced by Interaction with a 
Self-gravitating Planetesimal Disk, ApJ, {\bf 142}, 152, (2011)

\bibitem{Lykawka13}
Lykawka, P. S., and Ito, T.: Terrestrial Planet Formation During the Migration and Resonance
Crossing of Giant Planets, ApJ, {\bf 773}, 65 (2013)

\bibitem{Milani90}
Milani, A. and   Kne\u znevi\' c, Z.: 
Secular Perturbation Theory and Computation of Asteroid Proper Elements,
CeMDA, {\bf 49}, 347-411 (1990)

\bibitem{Minton09}
Minton, D. A. and Malhotra, R.: A Record of Planet Migration in The Main Asteroid Belt, Nature, 
{\bf 457}, 1109-1111 (2009)

\bibitem{Minton11}
Minton, D. A. and Malhotra, R.: Secular Resonance Sweeping of the Main Asteroid Belt During 
Planet Migration, ApJ, {\bf 732}, article id.53 (2011)

\bibitem{Morbidelli91a}
Morbidelli, A., Henrard, J.: Secular resonances in the asteroid belt - Theoretical perturbation 
approach and the problem of their location, CMeDA, {\bf 51}, 131-167 (1991a)

\bibitem{Morbidelli91b}
Morbidelli, A., Henrard, J.: The main secular resonances nu6, nu5 and nu16 in the asteroid belt, 
CMeDA, {\bf 51}, 169-197 (1991b)

\bibitem{Morbidelli05}
Morbidelli, A., Lavison, H. F., Tsiganis K. and Gomes, R. S.: Chaotic Capture of Jupiter's 
Trojan Asteroids in the Early Solar System, Nature, {\bf 435}, 462-465 (2005)

\bibitem{Morishima08}
Morishima, R.,  Schmidt, M. W., Stadel, J., Moore, B.: 
Formation and Accretion History of Terrestrial Planets from Runaway Growth through 
to Late Time: Implications for Orbital Eccentricity, ApJ, {\bf 685}, 1247-1261 (2008)

\bibitem{Nagasawa00}
Nagasawa, M., Tanaka, H., and Ida, S.: 
Orbital Evolution of Asteroids during Depletion of the Solar Nebula, 
ApJ, {\bf 119}, 1480-1497 (2000)

\bibitem{Nagasawa02}
Nagasawa, M., Ida, S., Tanaka, H.: Excitation of Orbital Inclinations of Asteroids during 
Depletion of a Protoplanetary Disk: Dependence on the Disk Configuration,
Icarus, {\bf 159}, 322-327 (2002)
 
\bibitem{Nagasawa05}
Nagasawa, M., Lin, D. N. C., and  Thommes, E.: Dynamical Shake-up of Planetary Systems. 
I. Embryo Trapping and Induced Collisions by the Sweeping Secular Resonance and Embryo-Disk Tidal Interaction,
ApJ, {\bf 635}, 578-598 (2005)


\bibitem{Nesvorny98}
Nesvorn\'y, D. and Morbidelli, A.: Three-Body Mean Motion Resonances and Chaotic Structure 
of the Asteroid Belt, AJ, {\bf 116}, 3029-3037 (1998)

\bibitem{Nimmo07}
Nimmo, E. and Kleine, T.: How rapidly did Mars accrete? Uncertainties in the Hf-W 
timing of core formation, Icarus, {\bf 191}, 497-504 (2007)

\bibitem{OBrien06}
O'Brien, D. P., Morbidelli, A., and  Levison, H. F.: 
Terrestrial planet formation with strong dynamical friction, Icarus, {\bf 184}, 39-58 (2006)

\bibitem{Quintana14}
Quintana, E. V., and Lissauer, J. J.:
The Effect of Planets Beyond the Ice Line on the Accretion of Volatiles by Habitable-Zone Rocky Planets,
ApJ, {\bf 786}, 33 (2014)

\bibitem{Raymond04}
Raymond, S. N., Quinn, T. and Lunine, J. I.: Making Other Earths: Dynamical Simulations of 
Terrestrial Planet Formation and Water Delivery, Icarus, {\bf 168}, 1-17 (2004)

\bibitem{Raymond05a}
Raymond, S. N., Quinn, T. and Lunine, J. I.: The Formation and Habitability of Terrestrial 
Planets in the Presence of Close-in Giant Planets, Icarus, {\bf 177}, 256-263 (2005a)

\bibitem{Raymond05b}
Raymond, S. N., Quinn, T. and Lunine, J. I.:Terrestrial Planet Formation in Disks with
Varying Surface Density Profiles, ApJ, {\bf 632}, 670-676 (2005b)

\bibitem{Raymond06}
Raymond, S. N., Quinn, T. and Lunine, J. I.: High Resolution Simulations of the Final Assembly 
of Earth-like Planets 1: Terrestrial Accretion and Dynamics, Icarus, {\bf 183}, 265-282 (2006)

\bibitem{Raymond07}
Raymond, S. N., Quinn, T. and Lunine, J. I.:  High Resolution Simulations of the Final Assembly 
of Earth-like Planets 2: Water Delivery and planetary Habitability, Astrobio. J., {\bf 7}, 66-84 (2007)

\bibitem{Raymond09}
Raymond, S. N., O'Brien, D. P., Morbidelli, A. and Kaib, N. A.: 
Building the Terrestrial Planets: Constrained Accretion in the Inner Solar System,
Icarus, {\bf 203}, 644-662 (2009)

\bibitem{Tsiganis05}
Tsiganis, K., Gomes, R. S., Morbidelli, A. and Lavison, H. F.: Origin of the Architecture of 
the Giant Planets of the Solar System, Nature, {\bf 435}, 459-461 (2005)

\bibitem{Walsh11}
Walsh, K. J., Morbidelli, A., Raymond, S. N., O'Brien, D. P. and Mandell, A. M.:, A low mass for
Mars from Jupiter's early gas-driven migration, Nature, {\bf 475}, 206-209 (2011)

\bibitem{Ward81}
Ward, W. R.: Solar Nebula Dispersal and the Stability of Planetary Systems I. Scanning Secular 
Resonance Theory, Icarus, {\bf 47}, 234-264 (1981)

\bibitem{Wetherill90a}
Wetherill, G. W.: Comparison of Analytical and Physical Modeling of Planetesimal Accumulation,
Icarus, {\bf 88}, 336-354 (1990a)

\bibitem{Wetherill90b}
Wetherill, G. W.: Formation of Earth, AREPS, {\bf 18}, 205-256 (1990b)

\bibitem{Wetherill94}
Wetherill, G. W.: Provenance of the terrestrial planets, 
Geochimica et Cosmochimica Acta, {\bf 58}, 4513-4520 (1994) 

\bibitem{Wetherill96}
Wetherill, G. W.: The Formation and Habitability of Extra-Solar Planets,
Icarus, {\bf 119}, 219-238 (1996)

\bibitem{Wetherill96}
Wetherill, G. W.: Contemplation of Things Past, 
AREPS, {\bf 26}, 1-22 (1998)

\end{thebibliography}
\end{document}